\documentclass[%
 reprint,
 onecolumn,
 letter,
superscriptaddress,
showkeys,
 amsmath,amssymb,
 aps,
 keywords,
 prfluids,
floatfix,
]{revtex4-2}

\usepackage{dcolumn}% Align table columns on decimal point
\usepackage{bm}% bold math
\usepackage{graphicx}% Include figure files
\usepackage{subcaption}

\usepackage{booktabs}

\usepackage{xcolor}
\usepackage{xfrac}
\usepackage{float}

\usepackage{mathtools}

\usepackage{comment}

\usepackage{siunitx}

\usepackage[ruled]{algorithm2e}

\usepackage{hyperref}

\UseRawInputEncoding

\renewcommand{\selectlanguage}[1]{}

\begin{document}

\title{On the low drag regime of flatback airfoils}

\author{Konstantinos Kellaris}
\email{kkellaris@mail.ntua.gr}
\affiliation{School of Mechanical Engineering, National Technical University of Athens, Greece}

\author{George Papadakis}
\affiliation{ School of Naval Architecture \& Marine Engineering, National Technical University of Athens, Greece}

\author{Miguel A. Mendez}
\affiliation{Environmental and Applied Fluid Dynamics Department, von Karman Institute for Fluid Dynamics, Belgium}

\author{Marinos Manolesos}
\affiliation{School of Mechanical Engineering, National Technical University of Athens, Greece}

\date{October 14, 2024}

\begin{abstract}
Flatback airfoils, characterized by a blunt trailing edge, are used at the root of large wind turbine blades. A low-drag pocket has recently been identified in the flow past these airfoils at high angles of attack, potentially offering opportunities for enhanced energy extraction.
This study uses three-dimensional Detached Eddy Simulations (DES) combined with statistical and data-driven modal analysis techniques to explore the aerodynamics and coherent structures of a flatback airfoil in these conditions.
Two angles of attack -- one inside $\left(12^{\circ}\right)$ and one outside $\left(0^{\circ}\right)$ of the low-drag pocket -- are examined more thoroughly. The spanwise correlation length of secondary instability is analyzed in terms of autocorrelation of the $\Gamma_{1}$ vortex identification criterion, while coherent structures were extracted via the multiscale Proper Orthogonal Decomposition (mPOD). The results show increased base pressure, BL thickness, vortex formation
length, and more organized wake structures inside the low-drag regime. While the primary instability (B\'enard-von K\'arm\'an vortex street) dominates in both cases, the secondary instability is distinguishable only for the $12^{\circ}$ case and is identified as a Mode S$^{\prime}$ instability.
\end{abstract}


\maketitle

\section{Introduction}

As wind turbine (WT) rotor diameters increase to reduce the
Levelized Cost of Energy (LCOE), flatback (FB) airfoils -- i.e. airfoils
with a blunt trailing edge (TE)-- have gained traction in recent years.
FB airfoils are placed at the root of WT blades, offering several
aerodynamic, structural, and aeroelastic advantages. From an aerodynamics
point of view, FB airfoils can provide higher lift values due to the
reduced adverse pressure gradient over the aft part of the suction side
\cite{Baker2006}. Furthermore, their aerodynamic performance is less sensitive
to surface roughness than traditional
sharp TE airfoils \cite{Baker2006}. Early studies showed that blades
with FB airfoils can be up to 16\% lighter than those using traditional airfoils, without any performance penalty \cite{Griffith2014}. Additionally, due to the blunt TE and increased
blade cross-sectional area, FB blades have increased flapwise stiffness.
However, FB airfoils come with an increase in drag force and noise
\cite{Baker2006,Barone2009}. Consequently, several TE flow
control devices have been proposed to improve their aerodynamic
performance and decrease the associated drag penalty
\cite{Barone2009,Manolesos2016,Papadakis2020}.

The literature on FB airfoils has mostly focused on their
aerodynamic or aeroacoustic performance
\cite{Barone2009,Stone2009,Doosttalab2019,Fuchs2022,Jaffar2023} or
the wake characteristics of their flow
\cite{Wang2018,Papadakis2020,Papadakis2020a,Hongpeng2020} at low angles of attack (AoA). Recently,
studies were performed up to near-stall or post-stall AoA
\cite{Manolesos2021,Bangga2022} to provide new insights
regarding the three-dimensional wake behavior using high-fidelity tools.
Both experimental
\cite{Smith1950,Post2008,Baker2008,Barone2009,Manolesos2016} and numerical
\cite{Winnemoeller2007,Barone2009,Soerensen2011,Xu2014,Manolesos2021} studies have observed 
a low-drag ``pocket'' emerging in the region of AoA near stall, but the aerodynamics in this regime remain largely underexplored.

This study uses high-fidelity simulations and data-driven wake analysis to enhance the understanding of flow behavior around FB airfoils within the low-drag pocket. Gaining deeper insights into this flow regime can guide the development of passive and/or active flow control strategies to mitigate the increased drag associated with FB airfoils. The blunt TE in FB airfoil results in vortex shedding and complex interactions between its wake and the upstream boundary layers (BL). The following subsections offer a concise review of the vast literature on bluff-body vortex shedding (Section \ref{bluff-body-shedding}) and the influence of boundary layers on the drag in bluff bodies (Section \ref{sec-BLdrag}). Section \ref{sec:thiswork} presents an overview of the main contribution of this work and the article organization.

\subsection{Bluff body vortex shedding}\label{bluff-body-shedding}

A challenge in FB airfoil investigations is that the flow in the wake is
unsteady due to the blunt TE, leading to the formation of
counter-rotating vortices detaching periodically from the upper and
the lower edge of the base, i.e., the generation of the B\'enard-von K\'arm\'an
vortex street. Furthermore, at high Reynolds numbers, three-dimensional
secondary wake instabilities, namely streamwise vortices called braids,
are present \cite{Williamson1996}. Understanding these instabilities
is of great importance, as they have been utilized in the past as bases
for flow control strategies \cite{NaghibLahouti2012,Yang2018}. To
this end, several studies have investigated the wake of elongated bluff
bodies
\cite{Ryan2005,NaghibLahouti2012,NaghibLahouti2014,Gibeau2018,Gibeau2020}
up to
$\mathrm{Re}_{h_{TE}} = U_{\infty}h_{TE} / \nu=5\times10^{4}$,
where $h_{TE}$ is the TE height. A recent review can be found in
\citet{ForouziFeshalami2022}.

The characterization of the secondary instabilities found in bluff body
wakes is based on circular cylinder wakes \cite{Williamson1996}.
Direct Numerical Simulations and Floquet stability analysis by \citet{Ryan2005} for an generalized bluff body showed that Mode A of circular cylinder wakes is
probably the most unstable for $c/h_{TE}<7.5$, where $c$ is the body
chord. They also showed a new mode that dominates the flow for
$c/h_{TE}>7.5$, similar to Mode B of circular cylinders but with a
different predicted wavelength of $2.2h$ compared to Mode B's
$1.0h_{TE}$, thus characterized as B$^{\prime}$. Furthermore, for
larger $c/h_{TE}$, they predicted that another type of instability
described as Mode S$^{\prime}$, similar to Mode S for circular
cylinders, with a wavelength of $0.7-1.0h_{TE}$ dominates the flow.

Experimental investigations conducted by
\citet{NaghibLahouti2012,NaghibLahouti2014} suggested that Mode
B$^\prime$ is the dominant mode for an elongated bluff body with
$c/h_{TE}=12.5$ and $250\le\mathrm{Re}_{h_{TE}}\le 5\times10^{4}$,
with a spanwise wavelength ranging from $2.0h_{TE}$ to $2.5h_{TE}$. The
analysis of \citet{NaghibLahouti2012,NaghibLahouti2014} used the
spanwise velocity undulations in the wake and the Proper Orthogonal
Decomposition (POD) of the velocity fields to conclude that Mode
B$^\prime$ exists in the wake.
However, the effect of applying POD to the velocity data and the subsequent results was later questioned by \citet{Gibeau2018}, where the authors demonstrated that POD acted as a low-pass filter to the velocity undulations and directly influenced the spanwise wavelength estimation.
More recent studies, experimentally
investigated cases for $c/h_{TE}=46.5$ \cite{Gibeau2018} and
$c/h_{TE}=12.5$ \cite{Gibeau2020} up to
$Re_{h_{TE}}=2.5\times10^{4}$ using correlation-based techniques in
order to estimate the wake wavelengths.
In both \cite{Gibeau2018,Gibeau2020}, Mode B was identified as the dominant
mode, with a spanwise wavelength $0.7h_{TE}-0.9h_{TE}$.

Although there is a significant body of work in bluff body wakes, this
is not the case for FB airfoils. Recently, \citet{Manolesos2021}
observed these coherent structures in the wake of a FB airfoil, i.e., an
elongated bluff body with varying thickness, with
$c/h_{TE}\approx9.43$ for two different AoA,
$\alpha=0^{\circ}$ and $\alpha=12^{\circ}$, at
$\mathrm{Re}_{h_{TE}}=15.9\times10^{4}$. The spanwise velocity
undulations of instantaneous fields revealed a spanwise correlation
length of $1.0h_{TE}$ and $1.4h_{TE}$, but no attempt was made to
characterize the mode.

\subsection{Boundary layers and the drag on bluff bodies}\label{sec-BLdrag}

In the literature on the flow past bluff bodies, the upstream BL has
been linked to base pressure recovery and, consequently, to base
drag reduction for elongated bluff bodies. The first mention of this
relationship is due to \citet{Hoerner1950}, who proposed
a jet pump mechanism for base drag. According to this concept, the high-energy outer flow mixes with the low-energy fluid in the base, pumping low-energy fluid away from the base area and decreasing the static
pressure on the base, thereby increasing base drag. The separating
BL acts as an insulation sheet, preventing the high-energy outer flow
from mixing with low-energy fluid in the base. Therefore, increasing the
BL thickness reduces the effectiveness of the jet pump mechanism and
leads to lower base drag. Further studies by
\citet{Bearman1965,Bearman1967}, demonstrated that the base pressure
of a blunt TE airfoil is a function of the ratio of the
trailing edge height, $h_{TE}$, to the separating BL
momentum thickness $\theta$. This dependence is
also shown in \cite{Petrusma1994}, where it was found that base
pressure increases up to a critical value of $h_{TE}/\theta$,
beyond which it remains constant. This critical ratio differs between laminar and turbulent BLs, with 
$\left(h_{TE}/\theta\right)_{crit.}$ being approximately 25 for laminar and 35 turbulent BLS.

\citet{Rowe2001} demonstrated that, for incompressible and compressible flows and both laminar and turbulent boundary layers (BLs), increasing the BL thickness increases the base pressure and reduces the vortex shedding frequency. These authors suggest that this behavior is driven by the
BL's displacement thickness: as this increases, it takes
longer for the circulation fed by the upstream BL layer on one side to
be carried across the wake and to initiate the
formation of a vortex on the opposite side and, hence, the beginning of
vortex shedding. Therefore, for a given state of boundary layers at the
trailing edge, this effective diffusion length which depends solely on
the effective distance between the two boundary layers is the main
factor in controlling the vortex shedding frequency and base drag.
However, these authors acknowledge that this explanation does not fully account for how the state of the BL (whether laminar or turbulent) at the trailing edge
influences the vortex shedding frequency and the base drag. Instead,
they proposed that the shape factor of the BL is the key parameter, regardless of the BL state.
This claim was later challenged by \citet{Mariotti2013} for a turbulent BL, as they observed an increase in base pressure when increasing the BL thickness, while the shape factor remained constant.

Furthermore, using data-driven decompositions and low-order
reconstructions of the velocity field in the near wake, \citet{Durgesh2013} demonstrated that an increase in boundary-layer thickness leads to more organized vortex shedding in the near wake,
with reduced vortex strength, faster downstream advection, and reduced base drag.
From a control perspective, \citet{Durgesh2013} suggested that the modification of the BL at
separation represents an entirely different base-drag reduction
mechanism from other approaches that seek to disrupt vortex shedding or
limit vortex interactions, even though some of these methods appear to be
influenced by the upstream BL \cite{Park2006}. Similar
findings were reported in more recent studies \cite{Trip2014,Trip2017},
which investigated the BL state of a perforated axisymmetric elongated body with a smooth leading edge and a blunt trailing edge, utilizing BL suction for flow control.

It is important to note that all the works mentioned above refer to
cases without adverse pressure gradient along the streamwise direction.
In contrast, FB airfoils experience both adverse and favorable
pressure gradients, depending on the AoA. Additionally, \citet{ElGammal2007} studied the wake
topology of the flow past a blunt and divergent TE airfoil with
$h_{TE}=1.8\%c$ at $Re_{h_{TE}}=1.26\times10^{4}$ for 4 AoA. They found that while the BL thickness varied with AoA, the distribution of spanwise braids remained unchanged. Similarly,
\citet{NaghibLahouti2012,NaghibLahouti2014} argued that the upstream
BL plays a rather limited role in the wake dynamics, a thesis echoed by \citet{Gibeau2018}. 
However, \citet{Gibeau2020} suggested that the upstream BL may have some impact, though further investigation is required. Additionally, \citet{Ryan2005} reports that the influence of secondary instabilities on the upstream BL for elongated bluff bodies is implicitly observed via the variation
of the dominant modes with $c/h_{TE}$, because altering $c/h_{TE}$
leads to a variation in the distance between the separating shear
layers. To the authors' knowledge, no effort has been made to link the drag in FB
airfoils with the properties of the separating BL.

\subsection{Contribution of this work and manuscript organization}\label{sec:thiswork}

This work extends the contributions in \cite{Manolesos2016,Papadakis2020,Papadakis2020a,Manolesos2021} by focusing on the dynamics of the wake past a FB airfoil in the low drag regime, combining several advanced data-driven investigation methods. Specifically, the flow around the LI30-FB10 airfoil is simulated using Detached Eddy Simulations (DES) for a wide range of AoA, analyzing the link between the BL thickness and the base drag. This investigation was complemented with a statistical analysis of the wake formation length. Then, focusing on two AoA, namely $\alpha=0^{\circ}$
(outside the low-drag regime) and $\alpha=12^{\circ}$ (inside the low-drag
regime), the spanwise wavelength of the secondary instability of the
flow is estimated using statistical methods, while the leading three-dimensional
coherent structures were identified using the multiscale Proper Orthogonal Decomposition (mPOD\cite{Mendez2019,Mendez2023}). The mPOD combines multiresolution analysis and traditional Proper Orthogonal Decomposition (POD, \cite{Lumley1964}) to decompose a dataset into a linear combination of elementary contributions (modes) acting at different scales. This offers a compromise between the energy optimality of the Proper Orthogonal Decomposition (POD, \cite{Lumley1964}) and the spectral purity of the Dynamic Mode Decomposition (DMD,\cite{Schmid2010}), enabling scale separation. While several authors have explored the data-driven decomposition of the wake behind elongated bluff bodies (e.g., \cite{NaghibLahouti2012,NaghibLahouti2014}) and FB airfoils (e.g., \cite{Bangga2022}), to the authors' knowledge, no study has yet examined the modal decomposition of a 3D flow in the wake of an FB airfoil. Finally, the insights regarding the wake dynamics are combined with the observations of the upstream BL and base pressure variation with AoA, leading to a better understanding of the flow behavior inside the low-drag regime.

The rest of the article is organized as follows. Section \ref{sec-Methods} gives an overview of the numerical setup and the data processing methods used in this work. Section \ref{sec-Results} presents the results while \autoref{sec-Conclusions} closes the article with conclusions and outlooks.

\section{Methods}\label{sec-Methods}

This section is divided into three subsections. Section \ref{the-case-and-numerics} describes the selected test case and the numerical methods for the simulation. Section \ref{mPOD} reports on the computation of the mPOD, while section \ref{vortex_span_wise} describes the vortex identification and the analysis of spanwise coherency. 

\subsection{Selected test case and numerical methods}\label{the-case-and-numerics}

We consider the flow past an LI30-FB10 airfoil with TE thickness of $h_{TE}/c=10.6\%$ and maximum thickness of $30\%c$, at a Reynolds number of $\mathrm{Re}=1.5\times10^{6}$, corresponding to $Re_{h_{TE}}=15.9\times10^{4}$. This has been previously investigated experimentally \cite{Manolesos2016} and numerically \cite{Papadakis2020,Papadakis2020a,Manolesos2021}. The numerical and experimental investigations in \cite{Manolesos2016,Manolesos2021} showed good agreement in terms of time-averaged pressure distribution and wake frequencies. The experiments were carried out using both free and fixed transition, using a zig-zag tape located at $3\%$ of the chord of sides of the airfoil. Since no appreciable difference was observed in the performance of the airfoil in terms of pressure distribution, all the numerical investigations in this study are carried out without specific BL transition modeling, assuming fully turbulent flow in the entire numerical domain.

\begin{figure}[h]
\centering
\includegraphics[width=0.55\linewidth]{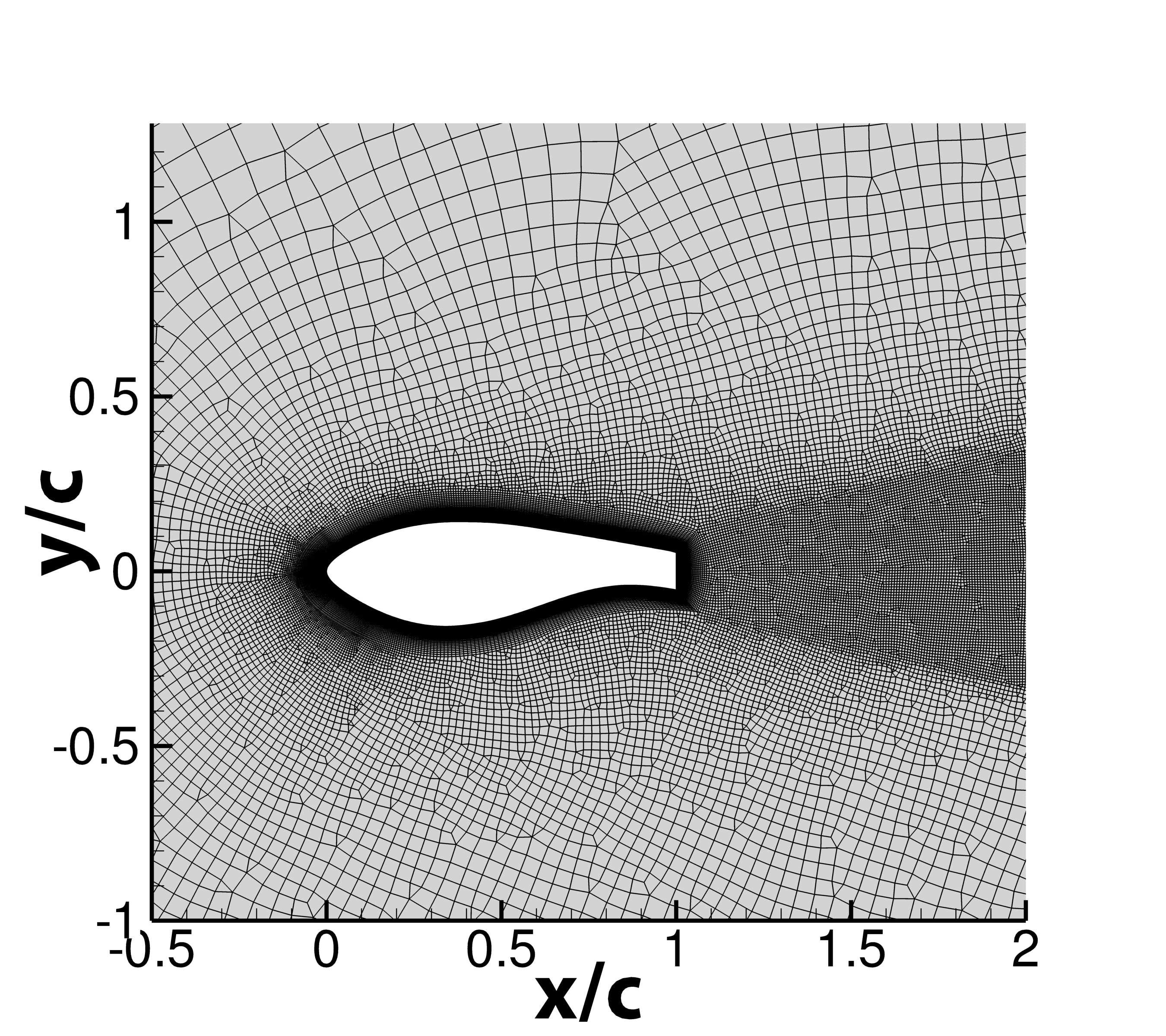}
\caption{The LI30-FB10 profile and details of the grid around it.}
\label{fig-LI30-FB10-grid}
\end{figure}%

The numerical simulations in this work are performed using a similar numerical setup as in \cite{Manolesos2021}, but with a more extensive data export to facilitate the post-processing discussed in this study. All simulations were conducted using the in-house MaPFlow solver developed at the National Technical University of Athens \cite{Papadakis2015}. MaPFlow is a cell-centered Computational Fluid Dynamics (CFD) solver that supports both structured and unstructured grids and is capable of handling compressible as well as incompressible flows using the artificial compressibility method \cite{Ntouras2020} and low Mach preconditioning.

The solver uses the approximate Rieman Solver by \citet{Roe1981} for the convective fluxes and a second-order piecewise linear interpolation
scheme to reconstruct the flow field \cite{Papadakis2019}.
The viscous fluxes are discretized using a central second-order scheme and the Green--Gauss approach is used to evaluate flow gradients at cell centers.

Time integration is achieved with an implicit manner, permitting large
Courant--Friedrichs--Lewy (CFL) numbers, using a
second-order time-accurate scheme combined with the dual time-stepping
technique to facilitate convergence \cite{Biedron2005}. Consequently, all computations are second-order accurate both in time and space.

Regarding turbulence closures, the simulations utilized the Improved Delayed Detached Eddy Simulation (IDDES) \cite{Diakakis2019,Shur2008}. This approach combines Large Eddy Simulation (LES) in the far field with Reynolds Averaged Navier Stokes (RANS) modeling near the solid boundaries. In this work, the RANS modeling was performed using the Spalart--Allmaras (SA)
\citet{Spalart1992} model. As shown in \cite{Papadakis2020a}, the IDDES allows for avoiding the prohibiting cost of performing LES inside the BL region while still capturing detailed flow characteristics.

In general, for hybrid models such as the IDDES, grid resolution is key
\cite{Manolesos2021}. In the near-wall region, the grid needs to be coarse enough to ensure proper activation of the RANS (Reynolds-Averaged Navier-Stokes) component of the model. Conversely, the grid must be suitably refined in the wake region, where the wall-modeled LES (Large Eddy Simulation) part of the model becomes relevant. Building upon previous investigations \cite{Papadakis2020,Papadakis2020a,Manolesos2021}, a grid with approximately 12 million cells ($12 \times 10^6$) is employed. A slice of the computational domain and the grid are shown in \autoref{fig-LI30-FB10-grid}. Near the airfoil, this consists
mainly of hexahedral cells while the rest of the domain is unstructured and contains tetrahedral cells. The hexahedral structured-like region extends $0.1$ chords
around the airfoil and consists of 100 points in the normal-to-the-wall
direction while the airfoil is discretized using 430 nodes.

In the near-airfoil region, the first cell center is located at
$1\times10^{-6}c$ from the wall, achieving non-dimensional wall
distance $y^{+}\leq0.1$. This allows for employing a RANS approach
inside this region without the use of a wall model. Additionally, the
cells are two orders of magnitude larger in the spanwise and chordwise
directions. This type of spacing ensures that the IDDES model in the BL
region is switched off; see
\cite{Diakakis2019,Papadakis2020,Manolesos2021}.

In the wake region and up to five chords downstream, a refinement region
of the unstructured grid is utilized. A portion of it shown in
\autoref{fig-LI30-FB10-grid}; beyond that point, the grid becomes coarser towards the domain boundaries. Based on the findings in
\citet{Papadakis2020a,Manolesos2021}, the domain is extruded in the spanwise direction by one chord, with a grid having constant spacing of
$\Delta z = 0.008c$ or $\Delta z \approx 0.075h_{TE}$ (125 spanwise
locations).

The simulations were carried out using the ARIS HPC cluster of GRNET, running on 1200 cores thanks to the MPI capabilities of MaPFlow. Finally, concerning the boundary conditions, symmetry boundary conditions were enforced on the sides boundaries of the domain in the spanwise direction. The far-field inlet was located 50 chord lengths away from the airfoil. Here, the flow is assumed fully turbulent, with a uniform $\Tilde{\nu}=50 \nu$, with $\Tilde{\nu}$ the viscosity-like variable in the SA model and $\nu$ the kinematic viscosity.

The non-dimensional timestep was taken as
$\Delta t=0.002 << {c}/{100 U_{\infty}}$, to ensure proper resolution of the vortex shedding behavior (approximately 250 timesteps per cycle) and a maximum $\mathrm{CFL} \approx 1$.

\subsection{Multiscale Proper Orthogonal Decomposition (mPOD)}\label{mPOD}

The Multiscale Proper Orthogonal Decomposition (mPOD) method enables the identification of optimal modes within specified frequency ranges. For a detailed theoretical background, the reader is referred to \cite{Mendez2019,Mendez2023}, while open-source implementations can be found in \cite{Poletti_MODULO_2023,Ninni2020}. Given a 3D dataset $\mathbf{d}(\mathbf{x},t)$, with $\mathbf{x}=(x,y,z)\subseteq\Omega$ and $t\in [0,T]$, modal decompositions can be written as

\begin{equation}
\mathbf{d}(\mathbf{x},t) = \sum^{r_c}_{r=1} \sigma_r \bm{\phi}_r(\mathbf{x}) \psi_r(t)\,,
\end{equation} where $\bm{\phi}_r(\mathbf{x})$ and $\psi_r(t)$ are the spatial and temporal structure of each mode and $\sigma_r$ the associated amplitudes. The number of modes $r_c$ defines the rank of the dataset $\mathrm{rk}\left(\mathbf{d}\right)$ in the case of the POD. For other decompositions, $r_c\ge\mathrm{rk}\left(\mathbf{d}\right)$.

The mPOD computes orthogonal modes $\psi_r(t)$ by diagonalizing the multi-resolution decomposition of the temporal correlation matrix $\mathbf{K}$. Given a uniform time discretization $t_k=(k-1)\Delta t$, with $k=1\dots n_t$, this matrix is $\mathbf{K}\in \mathbb{R}^{n_t\times n_t}$, defined as

\begin{equation}
\label{K}
\mathbf{K}_{i,j} = \langle \mathbf{d}(\mathbf{x},t_i)\,,\mathbf{d}(\mathbf{x},t_j) \rangle _x=\frac{1}{||\Omega||} \int_{\Omega} \mathbf{d}(\mathbf{x},t_i) \mathbf{d}(\mathbf{x},t_j) d\Omega\,,
\end{equation} having introduced a measure $||\Omega||$ of the domain (areas in 2D or volumes in 3D) and the inner product in space $\langle \,,\, \rangle _x$. The multiresolution decomposition of the temporal correlation matrix reads 
 
\begin{equation}
\mathbf{K} = \sum_{m=1}^{M}\mathbf{K}_{m} = \sum_{m=1}^{M} \mathbf{\Psi}_{F} \left[\hat{\mathbf{K}} \odot \mathbf{H}_{m} \right] \mathbf{\Psi}_{F}
\end{equation} where $\mathbf{\Psi}_{F}$ is the Fourier matrix, $\hat{\mathbf{K}}=\overline{\mathbf{\Psi}}_{F}\mathbf{K}\overline{\mathbf{\Psi}}_{F}$ is
the 2D Fourier transform of $\mathbf{K}$, $\mathbf{H}_{m}$ is the transfer function matrix associated to the scale $m$ and $\odot$ is the shur (entry by entry) product. The transfer functions are designed to have $\sum^M_m \mathbf{H}_{m}= \mathbf{I}$, with $\mathbf{I}$ the identity matrix. The scale contributions $\mathbf{K}_m$ do not share frequencies if the matrices $\mathbf{H}_{m}\cap\mathbf{H}_{n}=0$ for $m\neq n$. 

The temporal basis of the multiscale POD is constructed by concatenating the eigenvectors $\psi^{\{m\}}_n$ of all the scale contributions $\mathbf{K}_{m}$, with $n=1,\dots n_m$ the non-zero eigenvalues at scale $m$, i.e:

\begin{equation}
\bm{\Psi}_{\mathcal{M}}=\left[\psi^{\{1\}}_{1}\dots \psi^{\{1\}}_{n_1},  \psi^{\{2\}}_{1}\dots \psi^{\{2\}}_{n_2},\dots\psi^{\{M\}}_{1}\dots, \psi^{\{M\}}_{n_M}\right ] \quad \mbox{with} \quad \mathbf{K}_m \psi_n^{\{m\}} = \sigma^{\{m\}}_n \psi_j^{\{m\}}\,\, ,\,\,n\in[1,n_m]\,.
\end{equation} 

Similarly to the snapshot POD \cite{Sirovich1987}, this basis is then used to compute the spatial structures via projection in time, i.e. $\bm{\phi}_r(\mathbf{x}) = \langle \mathbf{d}(\mathbf{x},t), \psi_r(t) \rangle _t/\sigma_r$ with

\begin{equation}
 \langle \mathbf{d}(\mathbf{x},t), \psi_r(t) \rangle _t=\frac{1}{T} \int_T \mathbf{d}(\mathbf{x},t), \psi_r(t) dt \approx \frac{1}{n_t} \sum^{n_t}_k\mathbf{d}(\mathbf{x},t_k), \psi_r(t_k)\quad \mbox{and}\quad \sigma_r=||\langle \mathbf{d}(\mathbf{x},t), \psi_r(t) \rangle _t||_2\,.
\end{equation}

\begin{figure}[h]
\centering
\includegraphics[width=0.55\linewidth]{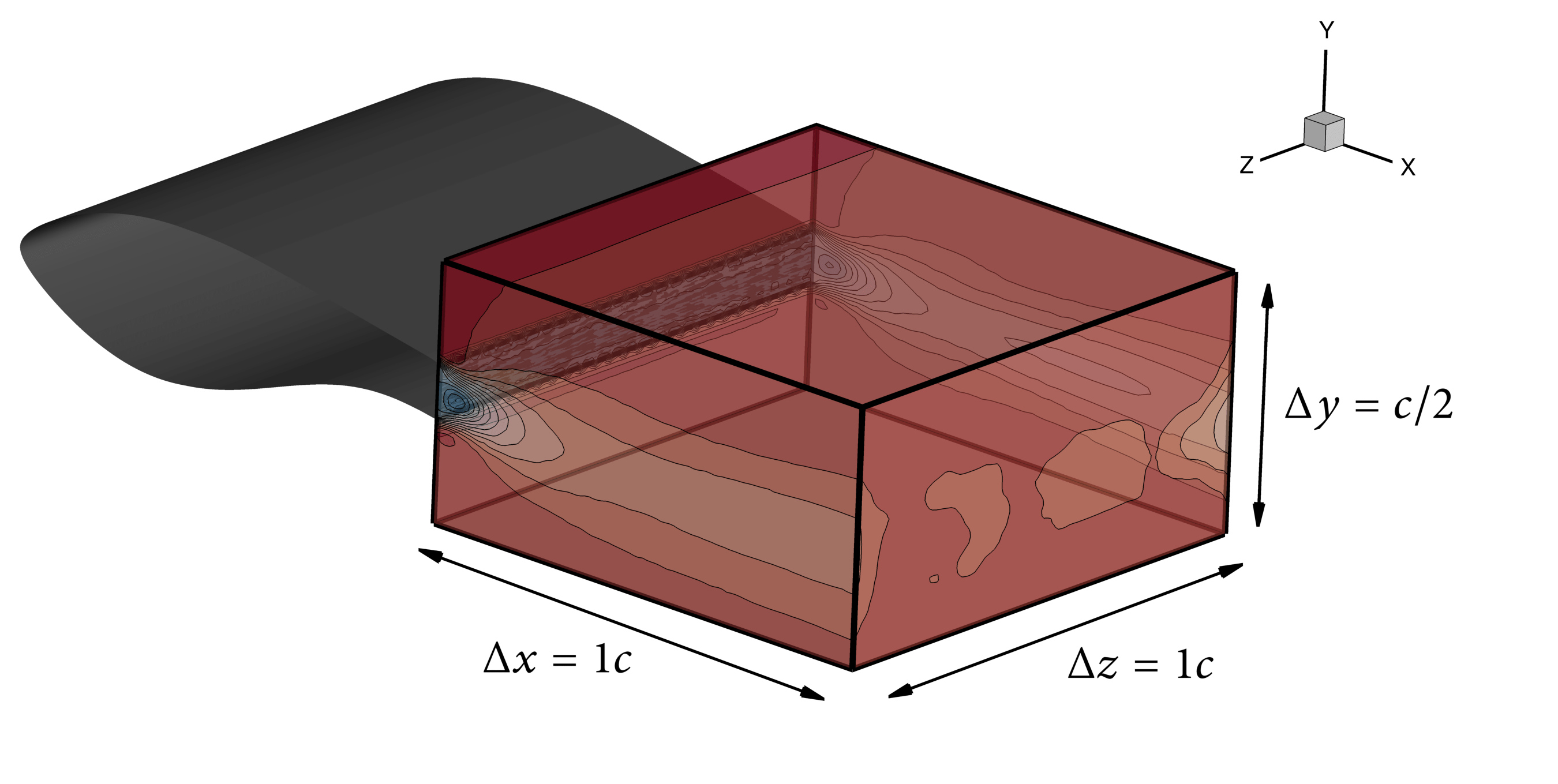}
\caption{Sampling box definition, along with its relevant dimensions.}
\label{fig-samplingbox}
\end{figure}

In this work, the inner product in space in \eqref{K} was computed via interpolation onto a uniform grid and Euler inner product. The analysis was carried out in a subdomain $\left(\Delta_{x},\Delta_{y},\Delta_{z}\right)=\left(1c,0.5c,1c\right)$ shown in \autoref{fig-samplingbox} discretized with $\left(n_{x},n_{y},n_{z}\right)=\left(100,50,100\right)$ points, resulting in a resolution $dx=dy=dz=0.01c$. The interpolation of the cell-centered values onto the uniform grid is
performed using an inverse distance weighting technique coupled with a
k-d tree search algorithm \cite{Kennel2004}, using
only cell centers within a sphere of radius $r=0.01c$
around each box point.

The sampling interval is equal to the time step used in the simulations,
i.e., $\Delta t = 0.002$ non-dimensional time units $c/U_{\infty}$, with a duration
of 65000 timesteps, representing approximately 260 vortex shedding
cycles and thus providing adequate time resolution for each examined
AoA. The resolution (granularity) in the frequency domain is thus 
\begin{equation}
\Delta f = \frac{f_{sampling}}{N_{FFT}} = \frac{1/\Delta t}{2^{16}} = \frac{500}{65536} \approx 0.008 
\label{eq-granularity_def}
\end{equation}

The computations were carried out using the MODULO
\cite{Ninni2020,Poletti_MODULO_2023} software package, with mean removal and using the block-partitioned memory saving option. 

\subsection{Vortex identification and spanwise wake correlation}\label{vortex_span_wise}

\begin{figure*}
\centering
    \begin{subfigure}{.47\textwidth}
      \centering
      \includegraphics[width=0.7\linewidth]{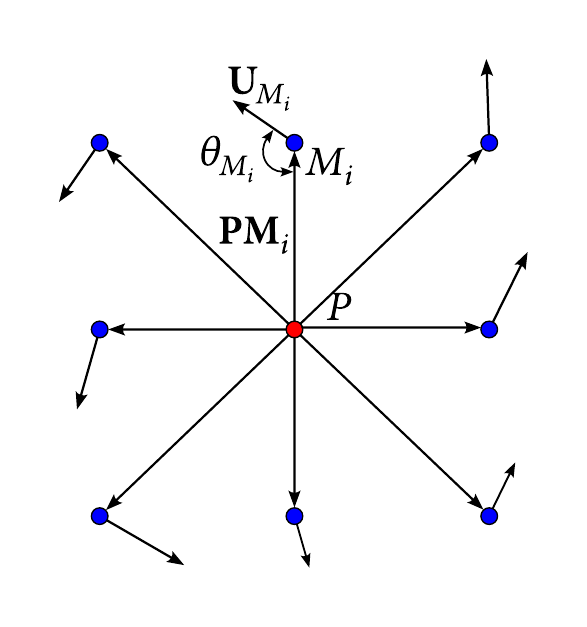}
      \caption{}
\label{fig-Lzdefs}
\end{subfigure}
    \begin{subfigure}{0.47\textwidth}
      \centering
      \includegraphics[width=1\linewidth]{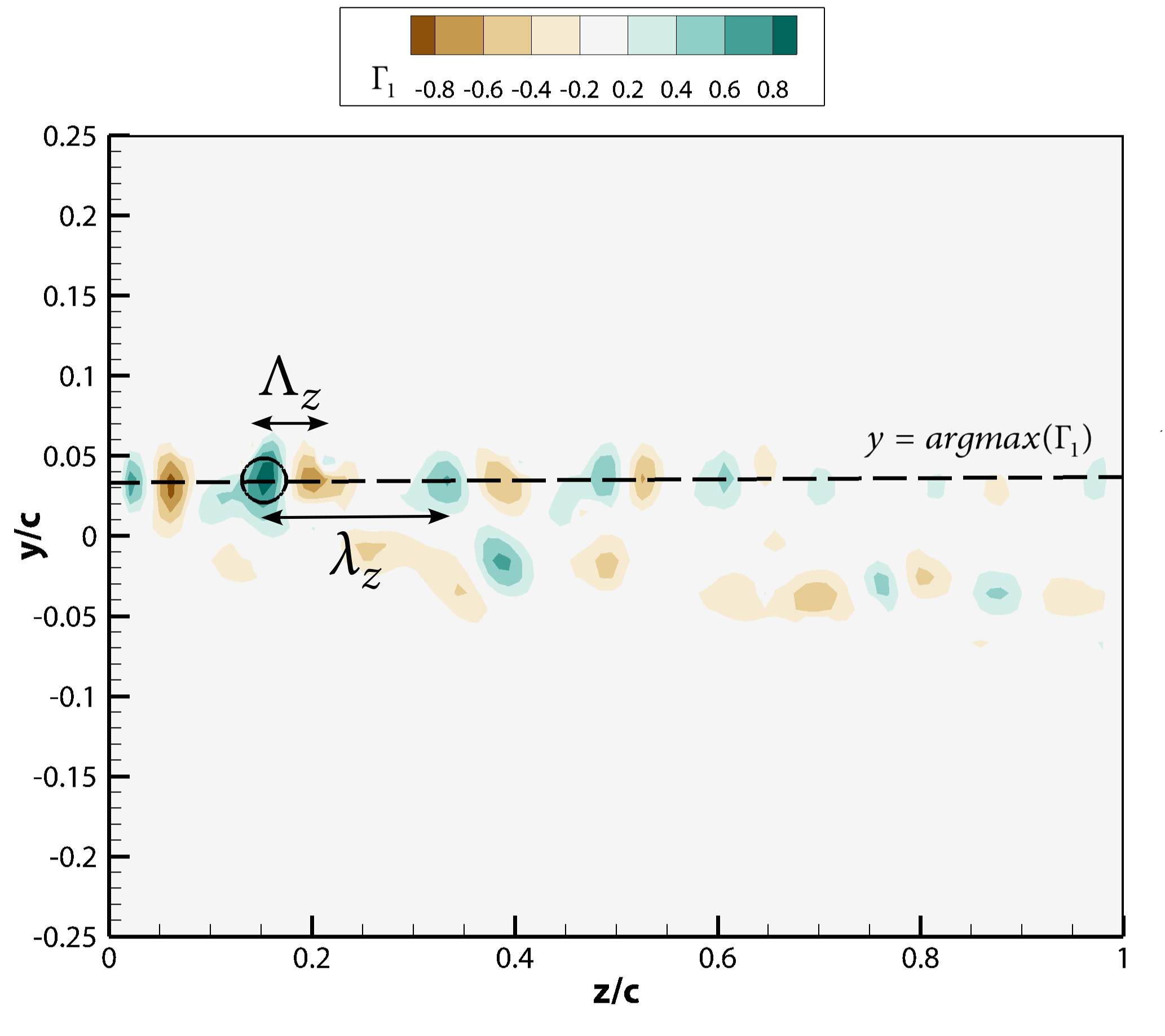}
      \caption{}
\label{snapshot_T}
    \end{subfigure}
    \caption{(a) Definition of the 8 point stencil to compute $\Gamma_{1}$ at a given point. b) Example of a $\Gamma_{1}$ contour for a
snapshot at $x/c=1.1$ for the $\alpha=12^{\circ}$ (low-drag) case. The
black dotted line denotes $y=\mathrm{argmax}(\Gamma_{1})$ and the
black circle indicates the location of $\max(\Gamma_{1})$.
Additionally, the physical meaning of the distance between two vortices
of the same vortex pair, $\Lambda_z$, and the distance between
adjacent pairs, $\lambda_{z}$, are shown. }
    \label{figures_merged}
\end{figure*}

The analysis and identification of coherent structures in 2D and 3D relied on three quantities and criteria.
The first criterion is that of defining vortex cores as regions of high vorticity $\mathbf{\omega} = \nabla \times \mathbf{U}$ \cite{Spalart1988}. The second is to identify vortices as regions of positive values of the second invariant of the velocity field gradient. This is the Q function \cite{Hunt1988} defined as 

\begin{equation}
Q = \frac{1}{2}\left[||\bm{\Omega}||^2_F - ||\bm{S}||^2_F \right]
\end{equation} where $\bm{\Omega}=(\nabla \bm{U}-\nabla \bm{U}^T)/2$ and $\bm{S}=(\nabla \bm{U}+\nabla \bm{U}^T)/2$ are antisymmetric and the symmetric portions of the velocity gradient tensor $\nabla \bm{U}$.
These quantities require the computations of spatial derivatives, which were carried out using a least square scheme \cite{Raffel2018} to minimize the truncation error. Finally, the third criterion is Graftieux's method \cite{Graftieaux2001}, which is a 2D gradient-free approach. In this method, the centers of vortices are identified through a scalar function $\Gamma_{1}|S\left(\mathbf{x}_p\right)$ that evaluates the flow topology around a point $\mathbf{x}_p$ and with respect to a plane $S$. This is defined as 

\begin{equation}
\Gamma_{1}|S\left(\mathbf{x}_p\right) = \frac{1}{N}\sum^{N}_{i=1} \frac{\left(\mathbf{PM}_{i} \times \mathbf{U}_{M_{i}} \right) }{\| \mathbf{PM}_{i}\| \cdot \|\mathbf{U}_{M_{i}}\|}\,
\label{eq-Gamma1def}
\end{equation} where $\mathbf{PM}{i}$ is the vector connecting the point $\mathbf{x}_{p}$ to a neighboring point $M_i$, and $\mathbf{U}_{M{i}}$ is the velocity projection onto the plane $S$ evaluated at $M_i$. All neighboring points lie on the plane $S$. In this work, the evaluation was performed using a stencil of $N=8$ points, as shown in \autoref{fig-Lzdefs}. Larger stencils were found not to significantly improve vortex detection accuracy \cite{SotoValle2022}. In this work, this function is evaluated on the plane $S=(x,z)$, hence the notation is simplified to $\Gamma_{1}|S\left(\mathbf{x}_p\right)=\Gamma_{1}\left(\mathbf{x}_p\right)$. 

\autoref{snapshot_T} shows the contour-plot of $\Gamma_1$ at the plane $x/c=1.1$ for a snapshot at $\alpha=12^{\circ}$ (low-drag) case. Vortices are observed to form in counter-rotating pairs. To analyze the secondary instabilities in the flow, we estimate the distance between two vortices within the same pair, defined as $\Lambda_z$, and the distance between two adjacent pairs, defined as $\lambda_{z}$. These are also illustrated in \autoref{snapshot_T}.

\begin{figure}[h]
\centering
\includegraphics[width=0.55\linewidth]{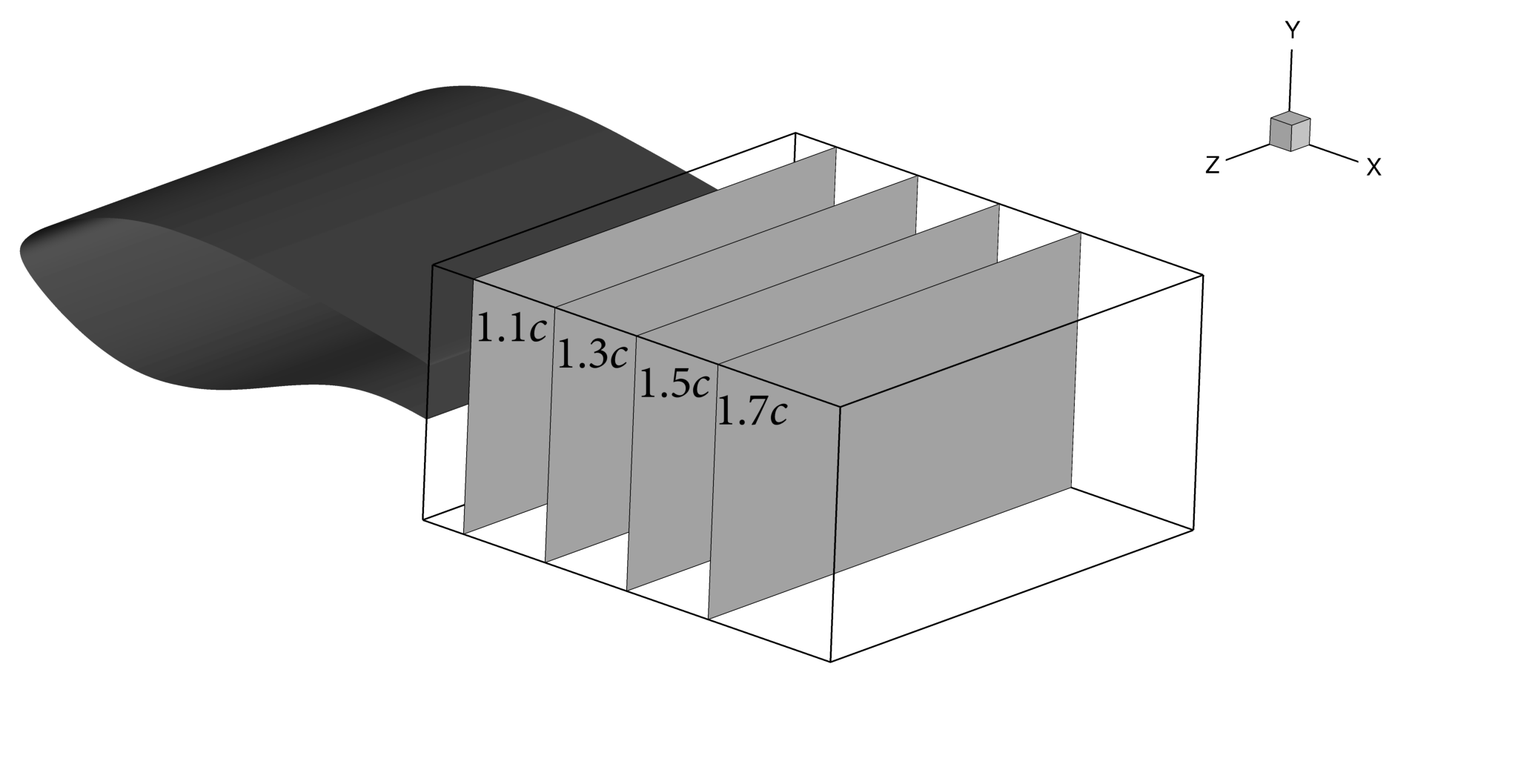}
\caption{Sampling planes used for the
spanwise correlation length estimation.}
\label{fig-sampling-box-planes}
\end{figure}%

These two quantities were computed in each time step along isolines $y/c=const.$ for which $y=\mathrm{argmax}(\Gamma_{1}^{*})$ on the planes in \autoref{fig-sampling-box-planes}. The average of the distances between a local maxima and the nearest local minima was used to estimate $\Lambda_{z}$ in each time step, following the algorithm detailed in Algorithm~\ref{alg:Gamma1Lambda_z}.

\begin{algorithm}[h]
\caption{Intra-pair spanwise distance $\Lambda_{z}$ calculation}\label{alg:Gamma1Lambda_z}
% \SetAlgoLined
\DontPrintSemicolon
\KwData{$N_{\mathrm{snapshots}}$ of velocity data at a $x/c=const.$ plane}
\KwResult{$\Lambda_{z}$ distribution}
initialization\;
\For{$i\leftarrow 1$ \KwTo $N_{\mathrm{snapshots}}$}{
Obtain interpolated flowfield at the $x/c=const.$ plane\;
Calculate $\Gamma_{1}$ from \eqref{eq-Gamma1def}\;
Find the $(y_m,z_m)$ coordinates of $\max\left(\Gamma_{1}\right)$\;
$\mathbf{sl} \gets$ $\Gamma_{1}$ along the $y=y_{m}$ line\;
Estimate the positions of maxima (peaks) and minima (valleys) of $\mathbf{sl}$\;
$\mathbf{cmin} \gets$ position of the valley closest to each peak\;
$\mathbf{\Delta_{z}} \gets \left\| z[\mathbf{peaks}] - z[\mathbf{cmin}]  \right\|$\;
$\mathbf{\Lambda_{z}}[i] \gets \overline{\mathbf{\Delta_{z}}}$\;
}
Obtain mean and mode values of $\Lambda_{z}$\;
\end{algorithm}

Along the same $y/c=const.$ line on $y-z$ planes at each time step, the average vortex pair distance $\lambda_{z}$ was computed from the spatial autocorrelation of $\Gamma_{1}$, which at each time step is defined as

\begin{equation}
\rho_{\Gamma_{1}}\left(\Delta z, t_{i}\right) = \frac{\int_z  \Gamma^{\prime}_1 (x_0,y_0,z,t_i)\Gamma^{\prime}_1 (x_0,y_0,z+\Delta z,t_i)  dz }{\int_z  {\Gamma^{\prime}_{1}}^{2} (x_0,y_0,z,t_i)  dz }
\label{eq-acorr_def}
\end{equation} with $\Gamma'_1=\Gamma_1-\mu_\Gamma(t_i)$ and $\mu_\Gamma$ the spatial average of $\Gamma_1(x_0,y_0,z,t_i)$ along $z$, $x_{0}$ denotes the $x/c=const.$ sampling plane, $y_{0}$
refers to the $y/c=const.$ isoline and $\Delta z$ is the spanwise
shift.
The procedure is summarized in Algorithm~\ref{alg:Acorrlambda_z}

The statistical distributions both $\Lambda_z$ and $\lambda_z$ are identified from the 65000 time steps available for each case.

\begin{algorithm}[h]
\caption{Spanwise distance between vortex pairs $\lambda_{z}$ calculation}\label{alg:Acorrlambda_z}
\DontPrintSemicolon
\KwData{$N_{\mathrm{snapshots}}$ of velocity data at a $x/c=const.$ plane}
\KwResult{$\lambda_{z}$ distribution}
initialization\;
\For{$i\leftarrow 1$ \KwTo $N_{\mathrm{snapshots}}$}{
Obtain interpolated flowfield at the $x/c=const.$ plane\;
Calculate $\Gamma_{1}$ \eqref{eq-Gamma1def}\;
Find the $(y_m,z_m)$ coordinates of $\max\left(\Gamma_{1}\right)$\;
$\mathbf{sl} \gets$ $\Gamma_{1}$ along the $y=y_{m}$ line\;
Calculate the spatial autocorrelation of $\mathbf{sl}$, $\rho_{\Gamma_{1}}$ \eqref{eq-acorr_def}\;
Estimate the positions of maxima (peaks) of $\rho_{\Gamma_{1}}$\;
$\mathbf{\Delta_{z}} \gets \left\| z[\mathbf{peaks}] - z[\mathbf{cmin}]  \right\|$\;
$\Delta_{z} \gets 0$\;
\For{$j\leftarrow 1$ \KwTo $N_{peaks}$-1}{
$\Delta_{z}\gets\Delta_{z}+\frac{1}{N_{peaks}-1}\left\|
\mathbf{z}\{\mathbf{peaks}[j+1]\}-
\mathbf{z}\{\mathbf{peaks}[j]\}\right\|$
}
$\mathbf{\lambda_{z}}[i] \gets\Delta_{z}$\;
}
Obtain mean and mode values of $\lambda_{z}$\;
\end{algorithm}

\section{Results}\label{sec-Results}

This section is divided into four parts. Section~\ref{main-flow-characteristics} reports on the main flow characteristics in terms of force coefficients, wake flow statistics, and based pressure. Section~\ref{separating-boundary-layer} reports on the separating BL, while section~\ref{wake-statistical-analysis} presents the primary wake statistics and analyses the spanwise wakec orrelation. Finally, section~\ref{three-dimensional-coherent-structures} analyses the three-dimensional coherent structures for different AoA outside and inside the low drag region.

\subsection{Main flow characteristics}\label{main-flow-characteristics}

\begin{figure}[h]
    \centering
   \begin{subfigure}{.45\textwidth}
      \centering
      \includegraphics[width=0.98\linewidth]{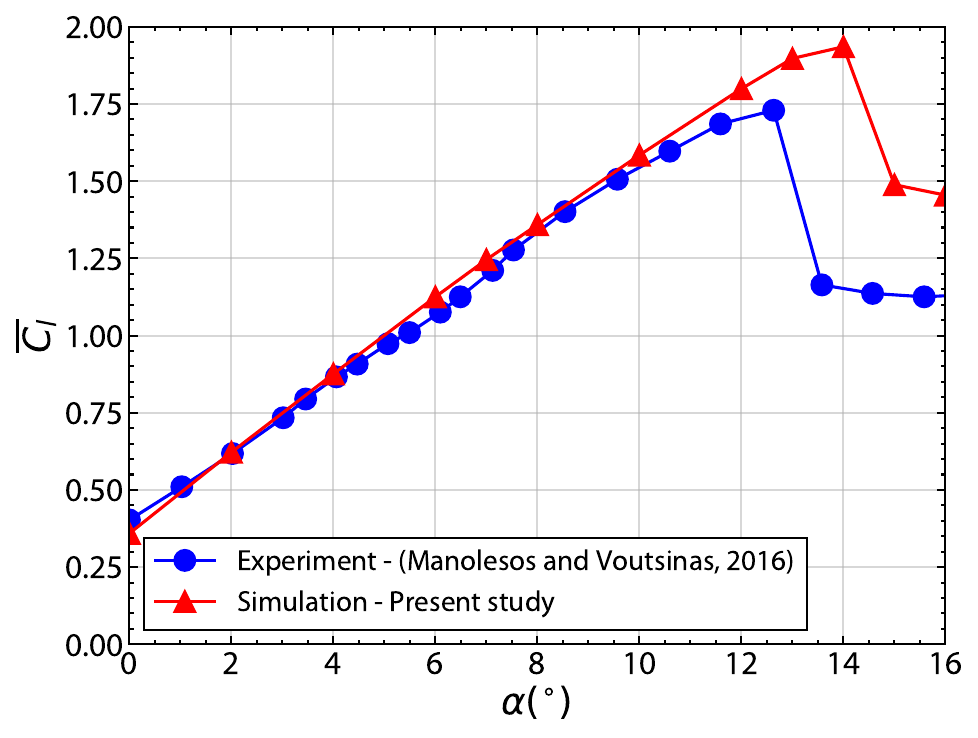}
      \caption{$C_{l}$}
      \label{fig-Cl-polar}
    \end{subfigure}
    \begin{subfigure}{.45\textwidth}
      \centering
      \includegraphics[width=0.98\linewidth]{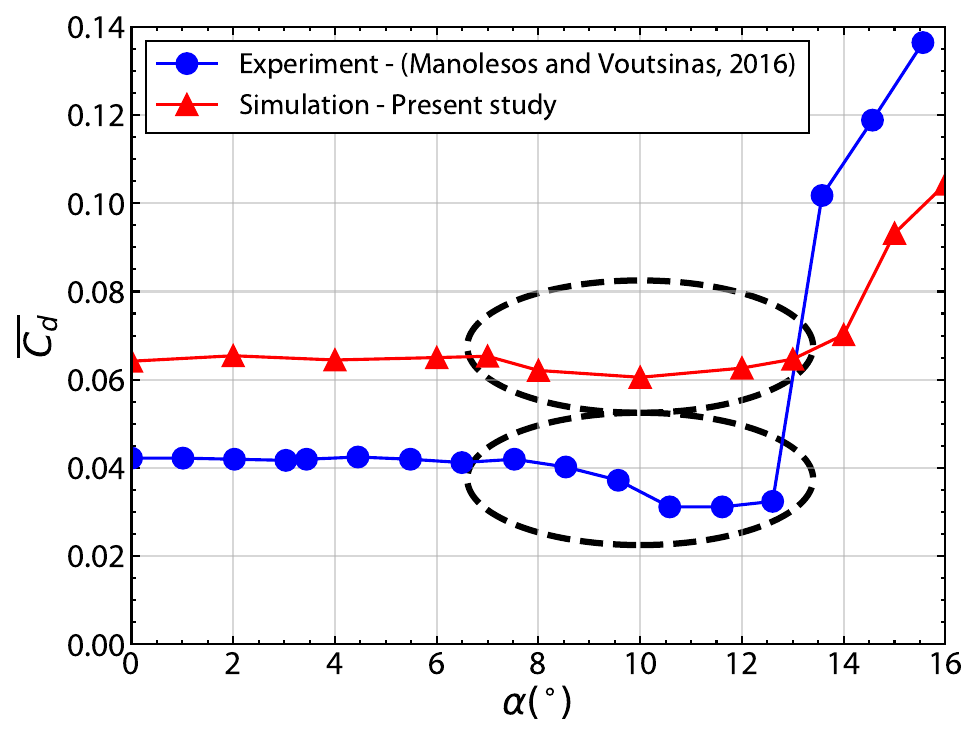}
      \caption{$C_{d}$}
      \label{fig-Cd-polar}
    \end{subfigure}
    \caption{Lift (a) and drag (b) force coefficients for the LI30-FB10 airfoil. The low-drag regime is highlighted using dashed ellipses.}
    \label{fig-Polars}
\end{figure}

\autoref{fig-Polars} shows the variation of the mean values of the
lift and drag coefficients of the airfoil as the AoA increases. The
averaging is performed for the last 65000 timesteps of the simulation.
The results are compared with a previous experimental investigation in
\cite{Manolesos2016}. The results from the simulations and the
experiment are in good agreement on the linear region of the lift coefficient
variation, but the simulation overpredicts the $Cl_{max}$ and the
stall AoA, as shown in \autoref{fig-Cl-polar}. The low-drag ``pocket'' (highlighted with dashed ellipses) is evident for both the simulation and the experiment, as shown in
\autoref{fig-Cd-polar}, and is in agreement with the literature on
cases where both experiments and simulations were performed of FB
airfoils \cite{Stone2009,Xu2014}. The numerical simulations
performed here show a drag decrease up to $5.7\%$ inside the low-drag
regime compared to $\alpha=0^{\circ}$.

\begin{figure}[h]
    \centering
   \begin{subfigure}{.45\textwidth}
      \centering
      \includegraphics[width=0.98\linewidth]{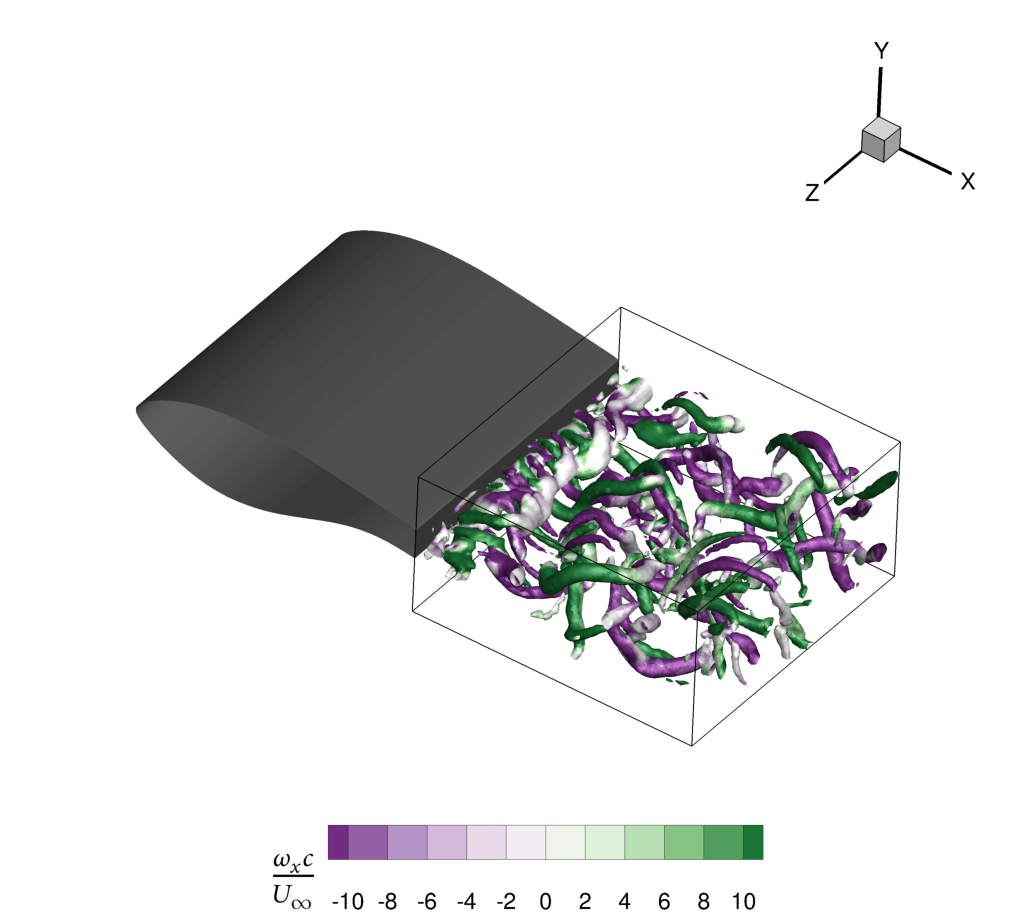}
      \caption{$\alpha=0^{\circ}$ --
$\omega_{x}c/U_{\infty}$}
      \label{fig-2WAKESNAP_X_0}
    \end{subfigure}
   \begin{subfigure}{.45\textwidth}
      \centering
      \includegraphics[width=0.98\linewidth]{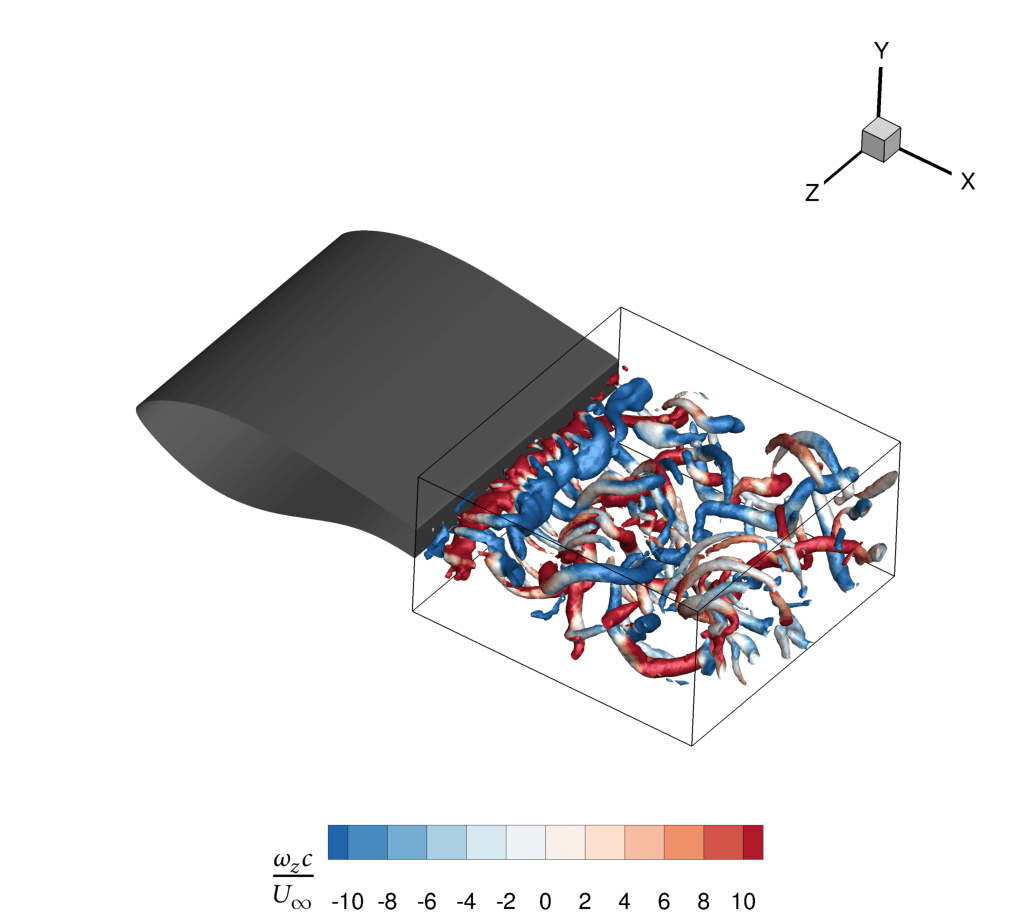}
      \caption{$\alpha=0^{\circ}$ --
$\omega_{z}c/U_{\infty}$}
      \label{fig-2WAKESNAP_Z_0}
    \end{subfigure}
   \begin{subfigure}{.45\textwidth}
      \centering
      \includegraphics[width=0.98\linewidth]{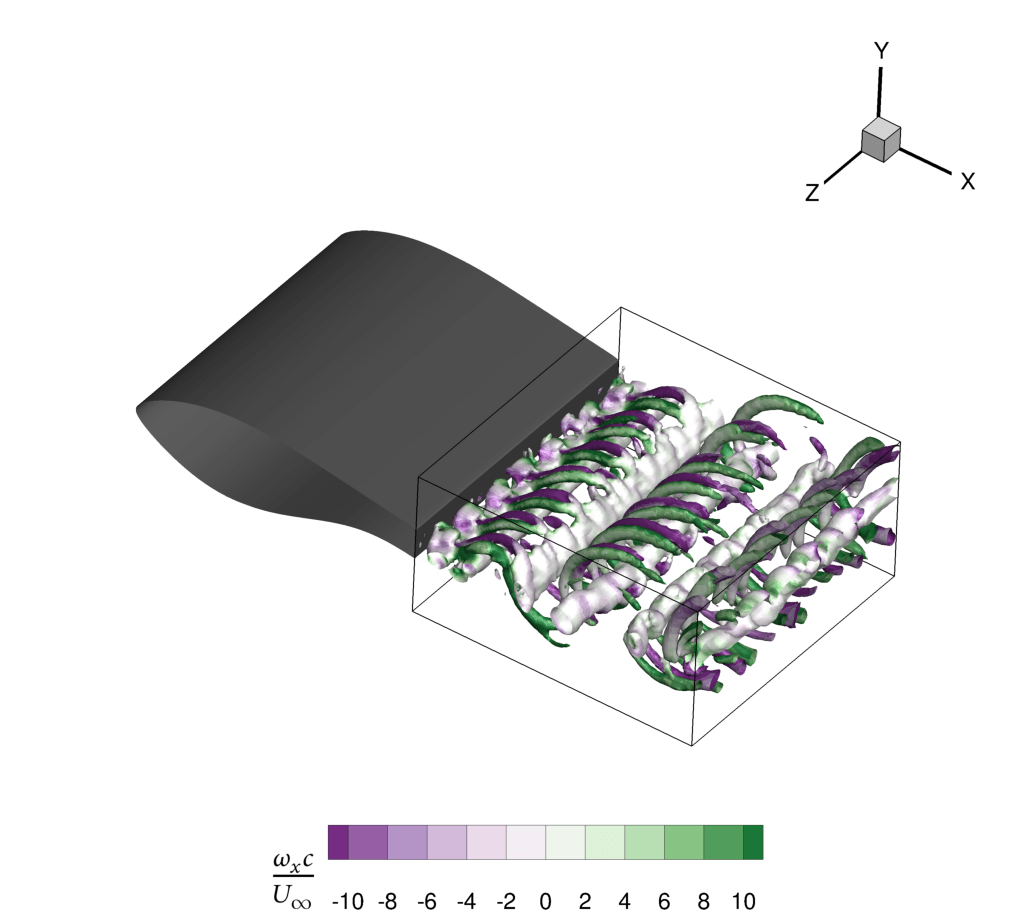}
      \caption{$\alpha=12^{\circ}$ --
$\omega_{x}c/U_{\infty}$\label{fig-2WAKESNAP_X_12}}
    \end{subfigure}
   \begin{subfigure}{.45\textwidth}
      \centering
      \includegraphics[width=0.98\linewidth]{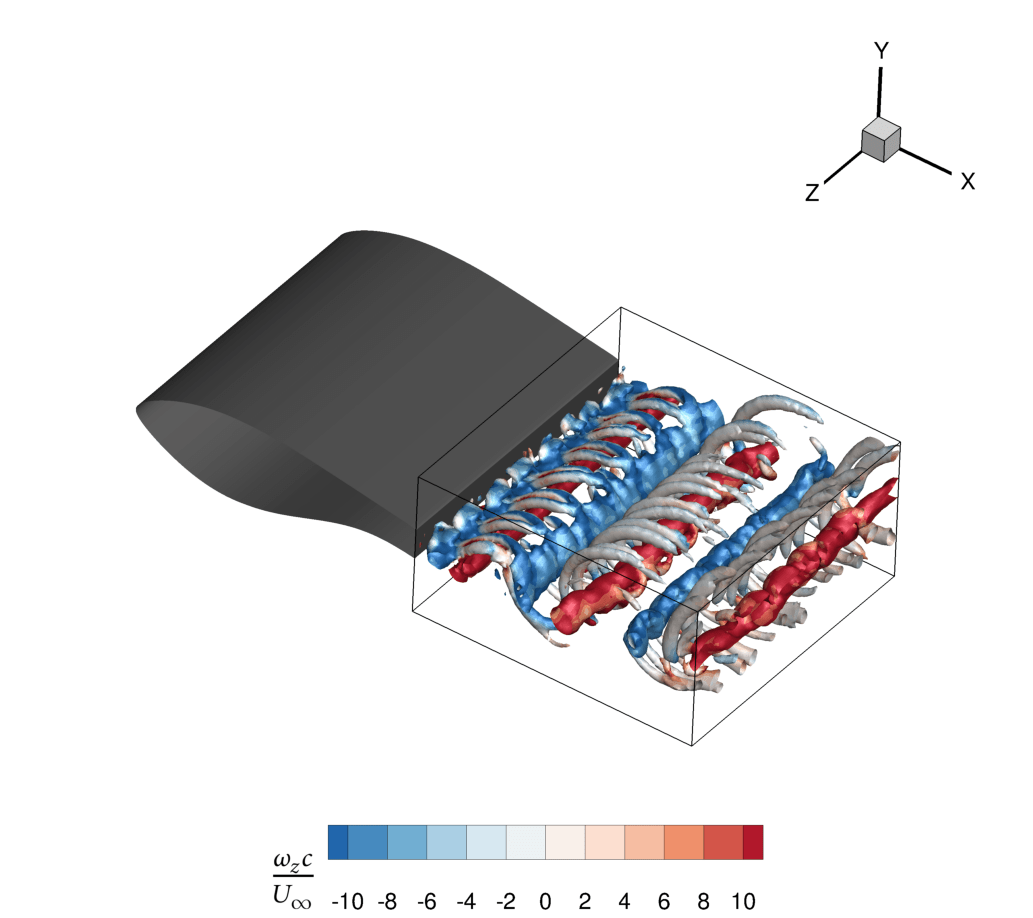}
      \caption{$\alpha=12^{\circ}$ --
$\omega_{z}c/U_{\infty}$}
      \label{fig-2WAKESNAP_Z_12}
    \end{subfigure}
    \caption{Instantaneous Q-criterion isosurfaces
colored with normalized streamwise $\omega_{x}c/U_{\infty}$ and
spanwise $\omega_{z}c/U_{\infty}$ vorticities for the same timestep.
Each row corresponds to a different AoA, and each column corresponds to
a different vorticity vector component. In the top row, results are
shown for high-drag $\left(\alpha=0^{\circ}\right)$ colored with (a)
$\omega_{x}c/U_{\infty}$ and (b) $\omega_{z}c/U_{\infty}$. In the
bottom row, results are shown for low-drag $\left(\alpha=12^{\circ}\right)$
colored with c) $\omega_{x}c/U_{\infty}$ and d)
$\omega_{z}c/U_{\infty}$. Flow is from left to right.\label{fig-2WAKESNAPSHOTS}}
\end{figure}

\autoref{fig-2WAKESNAPSHOTS} illustrates the Q isosurfaces for two AoA, namely $\alpha=0^{\circ}$ (outside the low-drag regime) and $\alpha=12^{\circ}$ (within the low-drag regime). The isosurfaces are colored according to the normalized streamwise vorticity $\omega_{x}c/U_{\infty}$ and spanwise vorticity $\omega_{z}c/U_{\infty}$.

In both cases, the main B\'{e}nard-von K\'{a}rm\'{a}n vortex street is clearly visible, with large-scale distortions being more pronounced at $\alpha=0^{\circ}$, resembling the oblique shedding behavior described in \cite{Williamson1989}. Additionally, the main B\'{e}nard-von K\'{a}rm\'{a}n vortices appear to break up closer to the trailing edge (TE) in the $\alpha=0^{\circ}$ case. Streamwise vortex pairs, corresponding to the secondary instability, connect the primary vortices in both cases. Each vortex pair links two consecutive counter-rotating vortices, and the rotational direction of the braids shed from the same side of the TE alternates. As a result, two complete cycles of the primary instability are required for one full cycle of the secondary instability, i.e., the shedding of braids from the same side of the TE in alternating directions.

The low-drag case $\left(\alpha=12^{\circ}\right)$ exhibits
clearer bidimensinoal and regular pattern allowing for estimating
$\lambda_{z}\approx1.4h_{TE}$ as in \cite{Manolesos2021} from the istantaneous field. The high-drag case $\left(\alpha=0^{\circ}\right)$ is more three-dimensional and does not allow for a simple estimation of the distance between vortices from the instantaneous fields.

\begin{figure}[h]
    \centering
   \begin{subfigure}{.45\textwidth}
      \centering
      \includegraphics[width=0.98\linewidth]{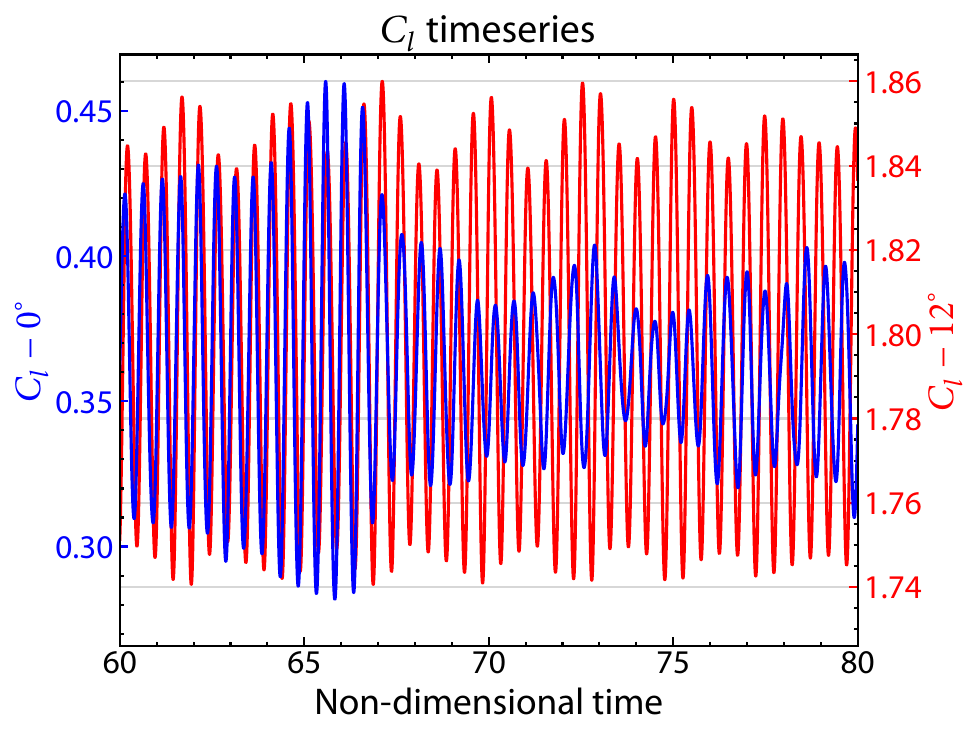}
      \caption{$C_{l}$}
      \label{fig-Cltimeseries}
    \end{subfigure}
    \begin{subfigure}{.45\textwidth}
      \centering
      \includegraphics[width=0.98\linewidth]{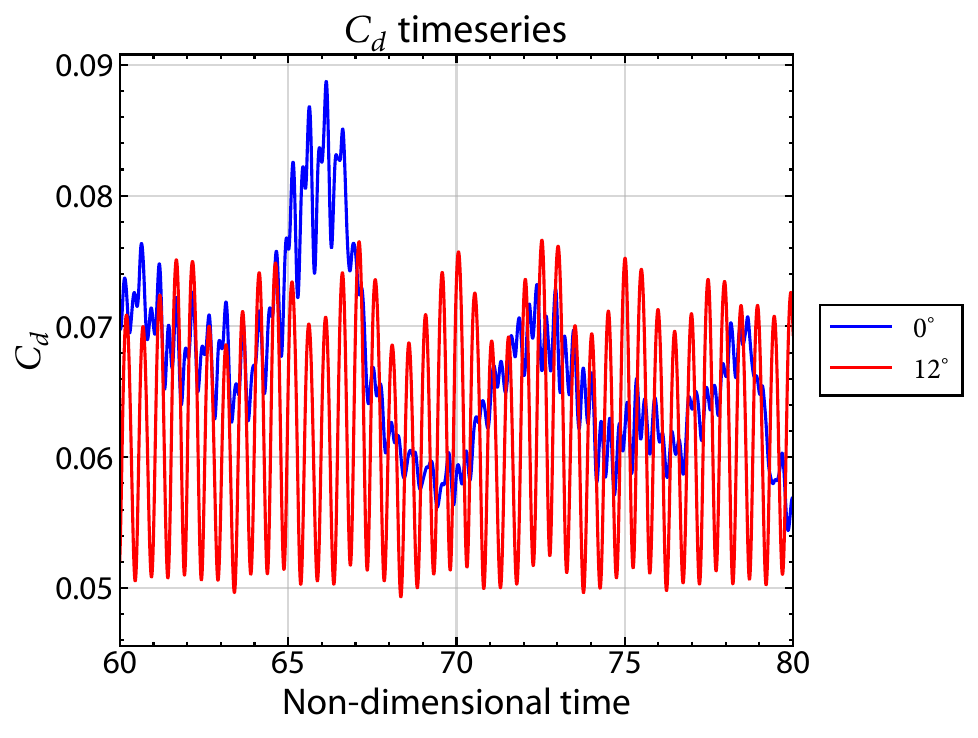}
      \caption{$C_{d}$}
      \label{fig-Cdtimeseries}
    \end{subfigure}
    \caption{Timeseries for (a) $C_{l}$, (b)
$C_{d}$ for $\alpha=0^{\circ}$ (blue) and $\alpha=12^{\circ}$ (red).}
    \label{fig-CLCDtimeseries}
\end{figure}

\begin{figure}[h]
\centering
\includegraphics[width=0.45\linewidth]{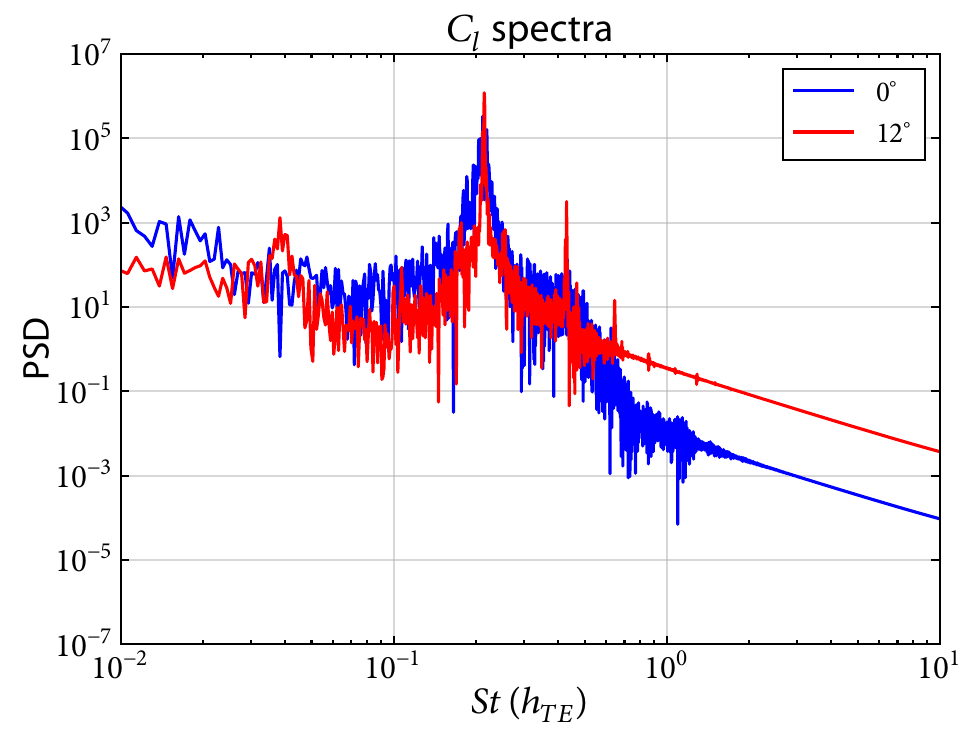}
\caption{Frequency spectra for (a) $C_{l}$.}
\label{fig-Clspectra}
\end{figure}%

\autoref{fig-CLCDtimeseries} presents the $C_{l}$ and $C_{d}$ time series for both cases. While both exhibit clear periodic behavior with the similar leading frequency, the case in the high-drag regime ($\alpha=0^{\circ}$) shows stronger modulation over time and a broader spectrum. \autoref{fig-Clspectra} displays the Power Spectral Density (PSD) of the $C_{l}$ coefficient from \autoref{fig-CLCDtimeseries}, in terms of the Strouhal number $St(h_{TE}) = f_{s}h_{TE}/U_{\infty}$. The PSD was computed via Fast Fourier Transform (FFT) on the full time series of 65000 samples (equivalent to $\hat{t} = 130 c/U_{\infty} \approx 1226h_{TE}/U_{\infty}$ dimensionless time units), giving a resolution of $\Delta St = 0.0008$. The primary shedding frequency is $St \approx 0.214$ for the low-drag case, with a pronounced second harmonic at $St\approx0.413$. For the high-drag case at $\alpha = 0^{\circ}$, the primary frequency is similar at   $St\approx0.212$ and no higher harmonics are distinguishable. This behavior aligns with previous high-fidelity studies on the flow past FB airfoils \cite{Wang2018}.

\begin{figure}[h]
\centering
\includegraphics[width=0.45\linewidth]{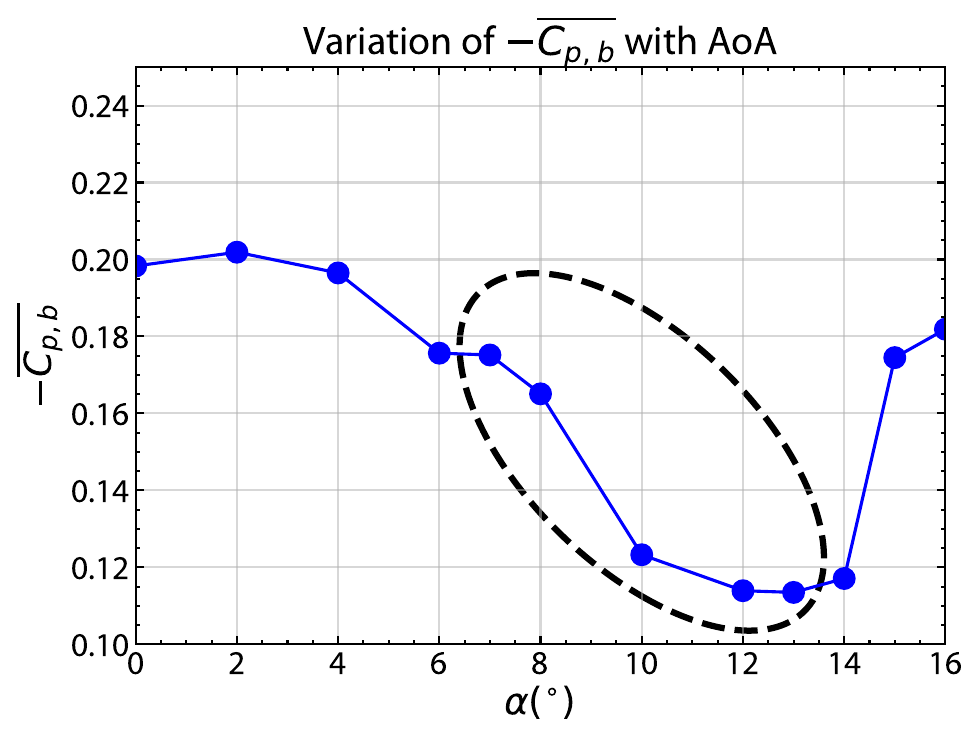}
\caption{Variation of $-C_{p,b}$ with AoA. The
low-drag regime is denoted with a dashed ellipse.}
\label{fig-Cpb-polar}
\end{figure}

To better locate the region of low drag behavior versus AoA, \autoref{fig-Cpb-polar} illustrates the base pressure coefficient, defined as 

\begin{equation}
C_{p,b} = \frac{2\left(p_{b}-p_{\infty}\right)}{\rho U_{\infty}^{2}}
\label{eq-Cpb}
\end{equation}
where $p_{b}$ is the base pressure obtained by averaging the
contribution of all points located at the midspan location of the TE,
i.e., $z/c=0.5$ and $x/c=1$, $p_{\infty}$ and
$U_{\infty}$ are the free-stream pressure and velocity, respectively.This quantity represents the pressure drag contribution and is a significant component in bluff bodies \cite{Durgesh2013}. \autoref{fig-Cpb-polar} illustrates an increase of up to approximately 44\% (relative to $\alpha=0^{\circ}$) in the base pressure coefficient $C_{p,b}$ as the AoA varies. Specifically, $C_{p,b}$ remains relatively constant up to $\alpha=4^{\circ}$, after which it increases with a changing gradient up to $\alpha=14^{\circ}$. Beyond the stall angle, a reduction in base pressure is observed, correlating well with the increase in $C_d$ (see \autoref{fig-Cd-polar}). This reduction in base drag, along with the decreased vortex shedding intensity noted earlier (\autoref{fig-Clspectra}), was also reported in \cite{ElGammal2007, Durgesh2013} for elongated bluff bodies.

\subsection{Separating boundary layer}\label{separating-boundary-layer}

As discussed in \autoref{sec-BLdrag}, a link between the base drag
and the separating BL exists for bodies with blunt TE
\cite{Hoerner1950,Petrusma1994}. To link the decrease in base
drag reported in \autoref{fig-Cpb-polar} with the BL characteristics, we extracted the relevant BL properties slightly upstream of the TE for
both the pressure and suction sides, specifically at $x/c=0.97$. These were computed from the mean velocity profiles along surface-normal lines from the resulting time-averaged flow field. 

\begin{figure}[h]
\centering
\includegraphics[width=0.45\linewidth]{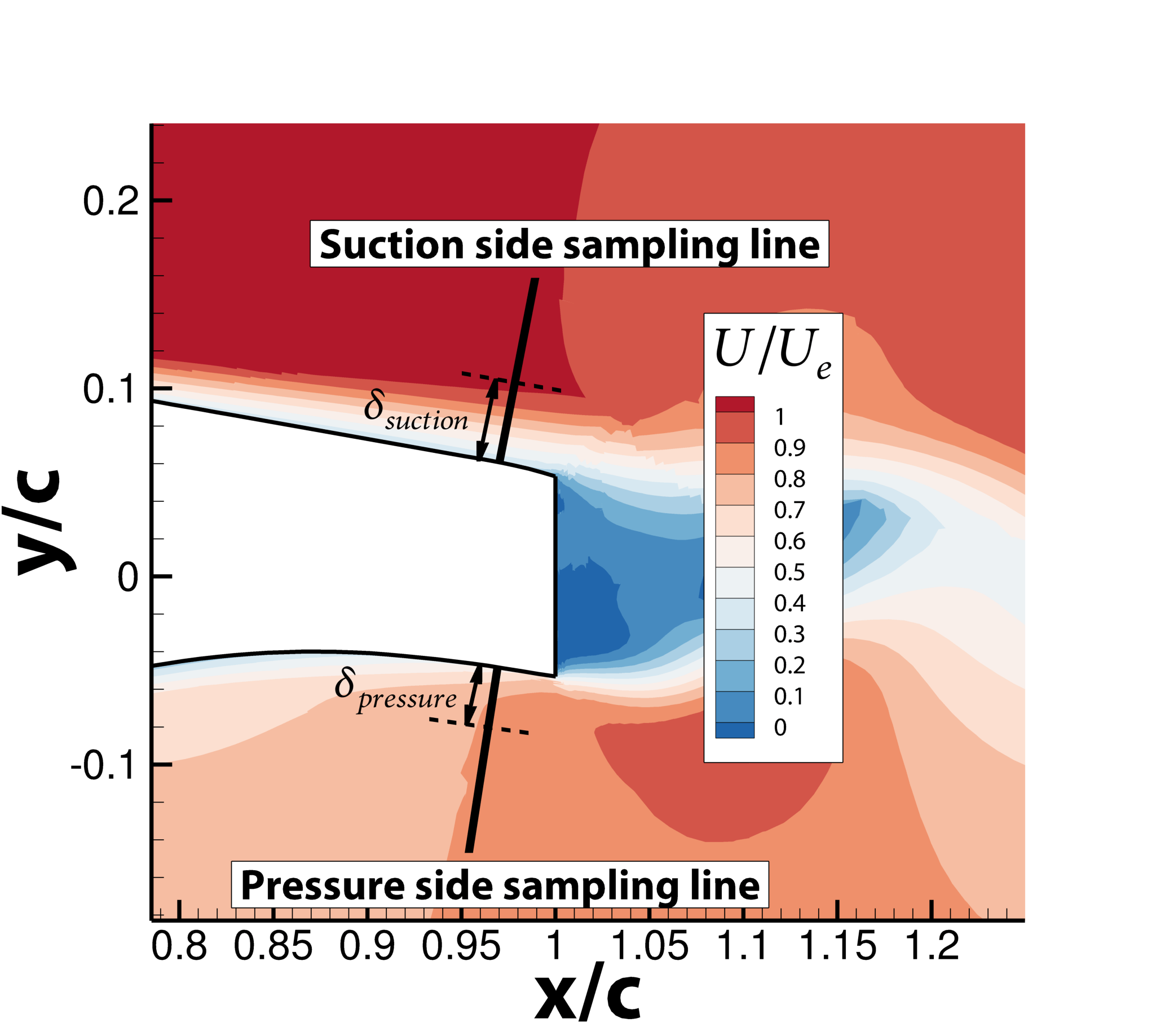}
\caption{Boundary layer definition for the suction and
pressure sides of the airfoil.}
\label{fig-BL_def}
\end{figure}

The total length of each sampling line, as shown in \autoref{fig-BL_def}, is
$\Delta n = h_{TE}$, with 500 points used for the sampling. The BL
thickness $\delta$ is estimated as the distance at which ${U}/{U_{e}}\bigl|_{d=\delta} = 0.99$, with  $d$ the wall distance and $U_{e}$ is the edge velocity, approximated as $U_{e} = \max_{\mathrm{line}}(U)$. The effective TE height is then defined as $h_{TE}^{\prime}=h_{TE}+\delta_{\mathrm{pressure}}+\delta_{\mathrm{suction}}$,
where $\delta_{\mathrm{pressure}}$ and $\delta_{\mathrm{suction}}$
are the BL thicknesses on the pressure and suction sides respectively.

The velocity values are then integrated along the surface-normal line up to the BL height $\delta$ to calculate the displacement thickness $\delta^{*}$, momentum thickness $\theta$, and the shape factor $H$ for each side of the airfoil, defined as

\begin{align}
\delta^{*} =\int_{0}^{\delta}\left(1-\frac{U}{U_{e}}\right)\mathrm{d}n \quad \quad
\theta =\int_{0}^{\delta}\frac{U}{U_{e}}\left(1-\frac{U}{U_{e}}\right)\mathrm{d}n \quad \quad
\mbox{and}\quad H = \frac{\delta^{*}}{\theta}
\end{align} respectively.

In order to allow for comparisons with the literature available for
elongated bluff bodies, where the BL on the pressure and suction sides
is the same, the suction and pressure sides' values are averaged, to define an ``effective'' BL height: 

\begin{equation}
\delta_{\mathrm{effective}}=\frac{\delta_{\mathrm{pressure}}+\delta_{\mathrm{suction}}}{2}\,.
\label{eq-Delta_eff_def}
\end{equation}

Similarly, the effective displacement
$\delta^{*}_{\mathrm{effective}}$ and momentum
$\theta_{\mathrm{effective}}$ thicknesses are also computed via averaging.

\begin{figure}[h]
    \centering
   \begin{subfigure}{.45\textwidth}
      \centering
      \includegraphics[width=0.98\linewidth]{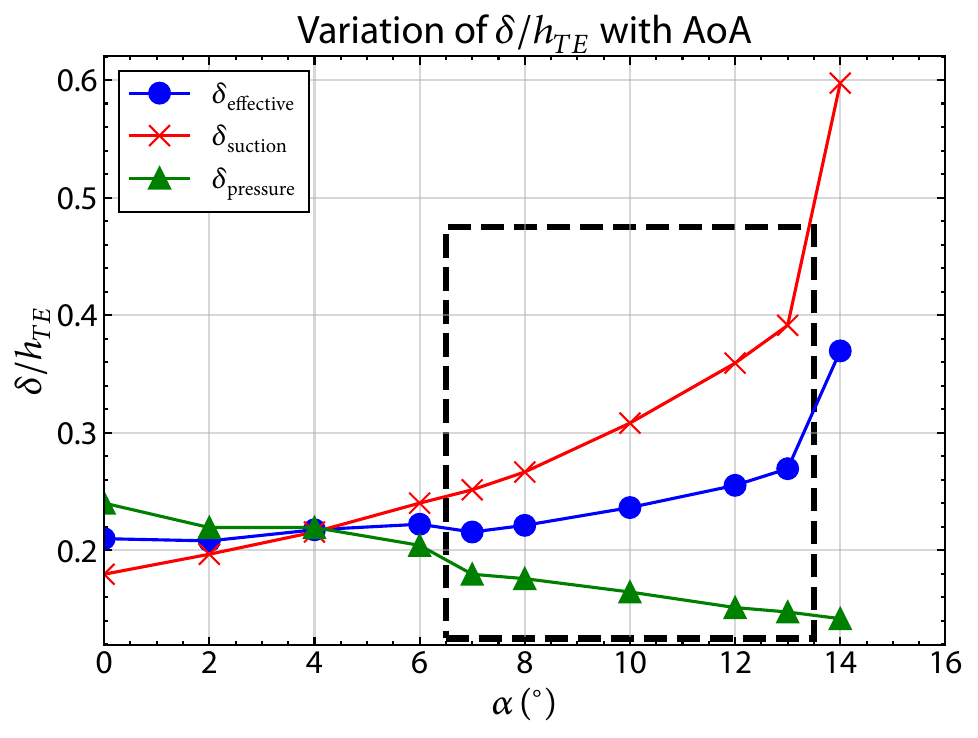}
      \caption{$\delta/h_{TE}$}
      \label{fig-DvsAoA}
    \end{subfigure}
    \begin{subfigure}{.45\textwidth}
      \centering
      \includegraphics[width=0.98\linewidth]{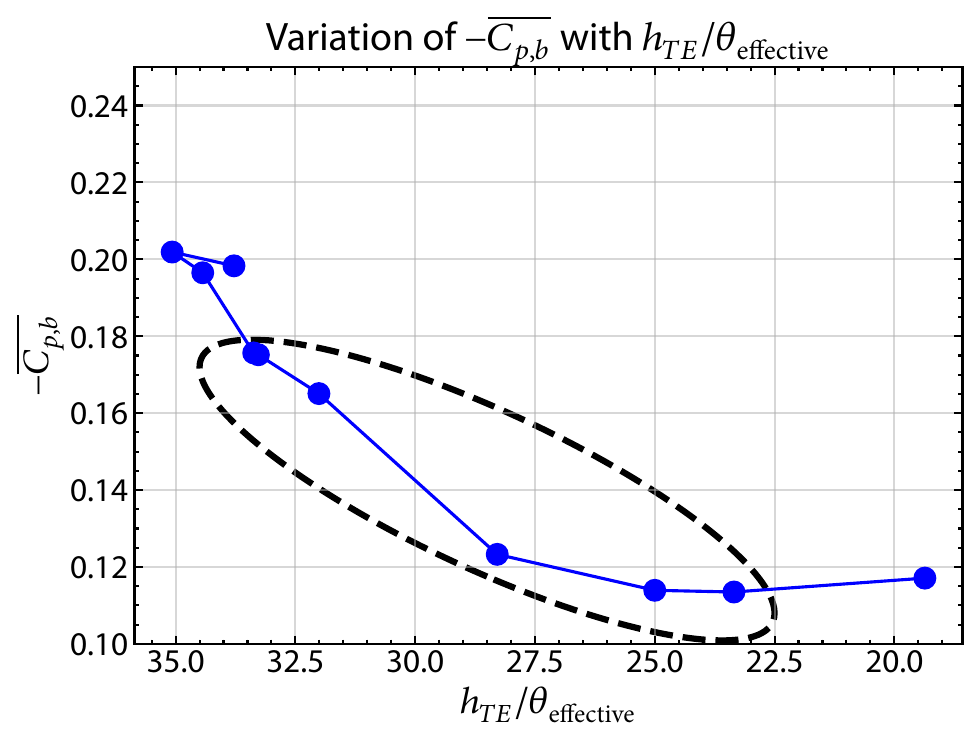}
      \caption{$h_{TE}/\theta_{\mathrm{effective}}$}
      \label{fig-THETvsCP}
    \end{subfigure}
    \caption{Boundary layer behavior for the
LI30-FB10 airfoil: (a) boundary layer height variation with AoA and (b)
$-C_{p,b}$ variation with $h_{TE}/\theta_{\mathrm{effective}}$. The
low-drag regime is denoted with a dashed square (a) and a dashed ellipse
(b).}
\label{fig-BLcomparisons}
\end{figure}

As the AoA increases, the BL thickness on the suction side increases, while it decreases on the pressure side, as shown in \autoref{fig-DvsAoA}. Furthermore, the effective thickness increases with the AoA, as also depicted in \autoref{fig-DvsAoA}. It is noted that values corresponding to post-stall AoA have been omitted due to flow separation.

\autoref{fig-THETvsCP} shows the variation of $C_{p,b}$ as a function of the
ratio $h_{TE}/\theta_{\mathrm{effective}}$. Firstly, it is observed that
for small AoA, $h_{TE}/\theta_{\mathrm{effective}}\approx35$ -- this is
the critical value after which the base pressure is constant
\cite{Petrusma1994}. As $h_{TE}/\theta_{\mathrm{effective}}$
decreases (for AoA inside the low-drag regime), $C_{p,b}$ increases
leading to lower drag, a result that is in agreement with the literature
\cite{Petrusma1994,Rowe2001,Mariotti2013,Durgesh2013}.

The increase in $\delta_{\mathrm{effective}}$ results in a greater distance between the two shear layers, i.e., the boundary layers (BLs) after separation from the trailing edge (TE). This behavior has been linked to a reduction in vortex shedding intensity \cite{ElGammal2007}, which is also observed in this case.

\subsection{Wake statistical analysis}\label{wake-statistical-analysis}

\subsubsection{Primary wake statistics}\label{primary-wake-statistics}

It is well known that vortex formation length $L_{f}$ is linked with base pressure and base drag for bodies with blunt TE for 2D
\cite{Bearman1967} and 3D wake vortex shedding behavior
\cite{Petrusma1994}. In line with \cite{NaghibLahouti2012}, in this study the vortex formation length is defined as the position of the downstream edge of the recirculation zone along the wake centerline, determined from the sectional streamlines of the mean flow field.

\begin{figure}[h]
    \centering
   \begin{subfigure}{.45\textwidth}
      \centering
      \includegraphics[width=0.8\linewidth]{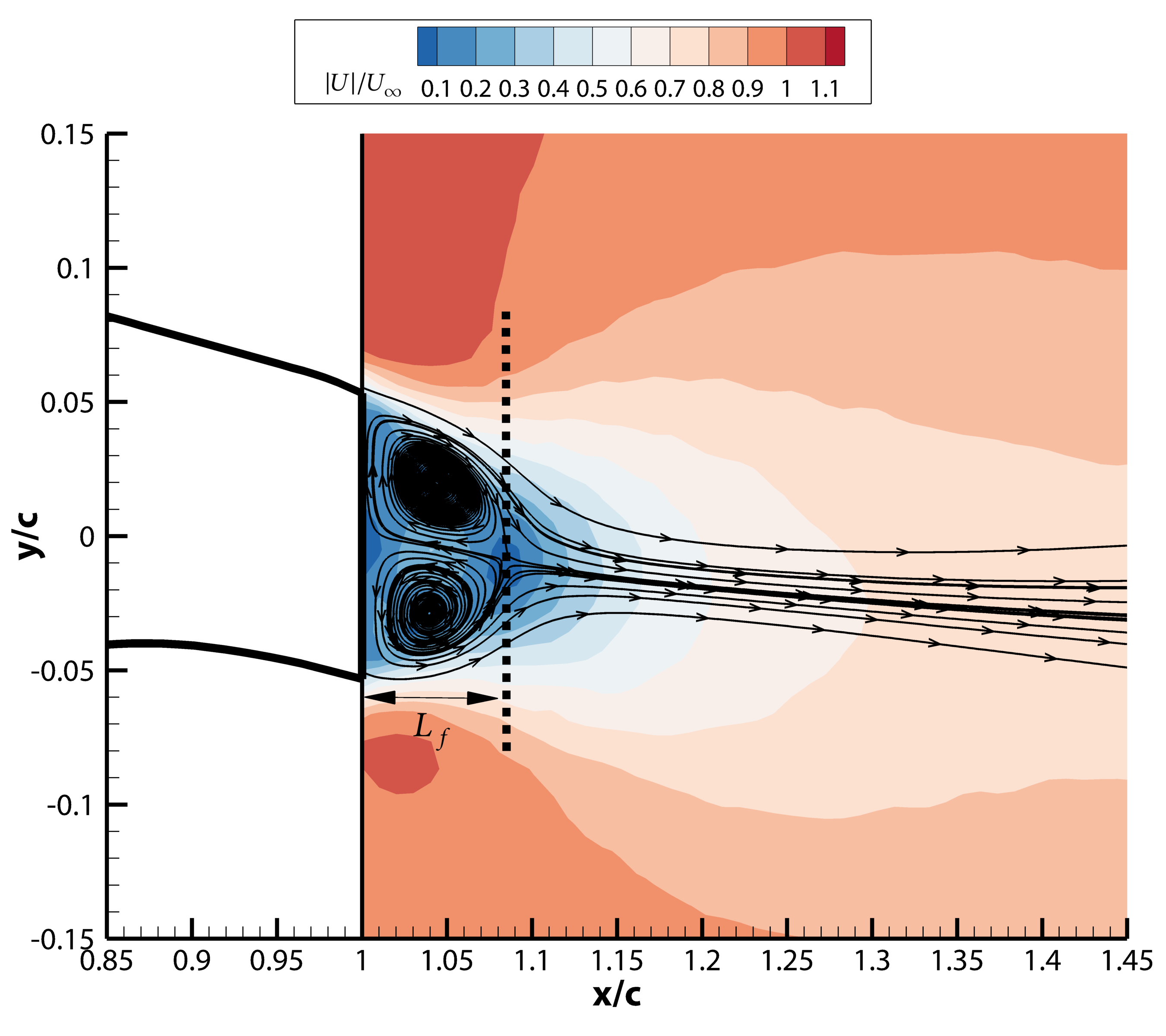}
      \caption{$\alpha=0^{\circ}$}
      \label{fig-LF0}
    \end{subfigure}
    \begin{subfigure}{.45\textwidth}
      \centering
      \includegraphics[width=0.8\linewidth]{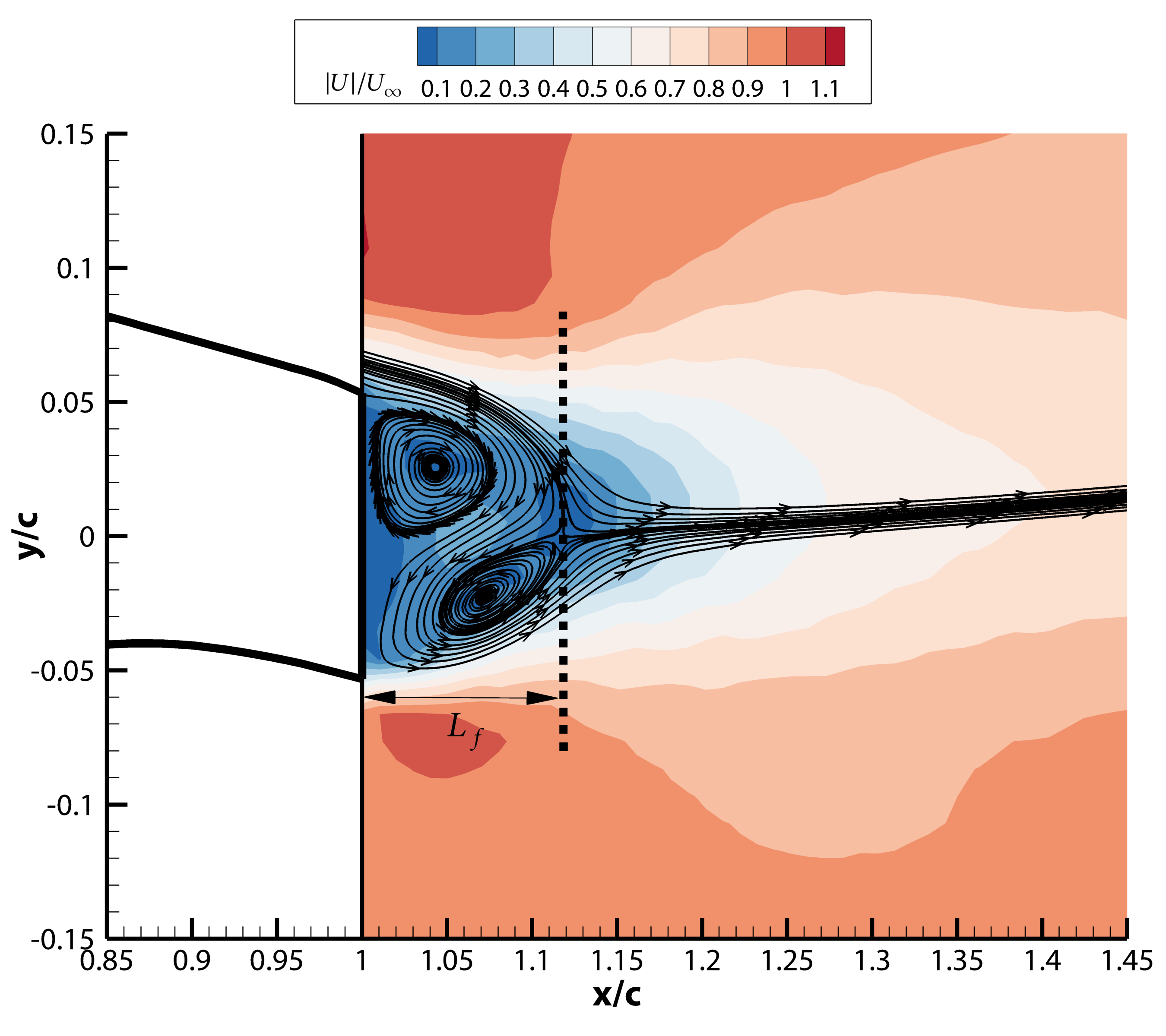}
      \caption{$\alpha=12^{\circ}$}
      \label{fig-LF12}
    \end{subfigure}
    \caption{Vortex formation lengths for (a) high-drag
$\left(\alpha=0^{\circ}\right)$ and (b) low-drag
$\left(\alpha=12^{\circ}\right)$.}
\label{fig-LF}
\end{figure}

The estimation of the vortex formation length is shown in
\autoref{fig-LF} for $\alpha=0^{\circ}$ and $\alpha=12^{\circ}$. The
estimated value of $L_{f}\approx0.77h_{TE}$ at $\alpha=0^{\circ}$ is in
good agreement with the results of previous studies for bluff bodies
\cite{NaghibLahouti2014,Gibeau2018,Gibeau2020}. At the low-drag case
$\left(\alpha=12^{\circ}\right)$, the formation length is increased,
$L_{f}\approx1.11h_{TE}$; this result is in line with the base drag
reduction noted in previous investigations
\cite{Manolesos2016,Bearman1965} that aimed to increase the
formation length for flow control (drag reduction) purposes.

\begin{figure}[h]
    \centering
    \begin{subfigure}{.45\textwidth}
      \centering
      \includegraphics[width=0.98\linewidth]{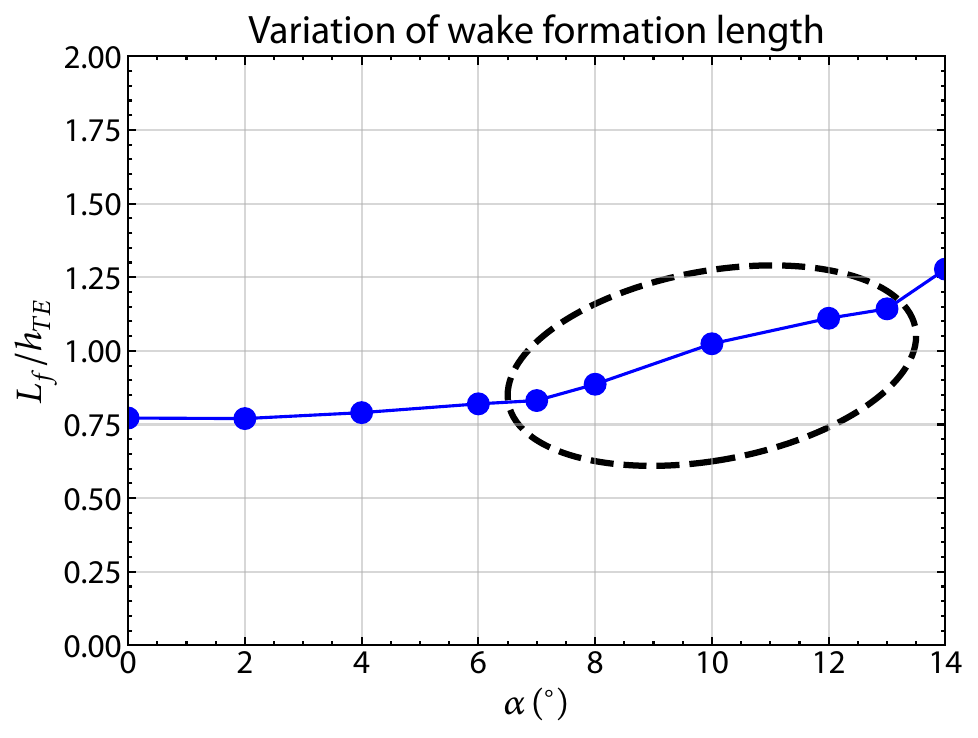}
      \caption{$L_{f}$}
      \label{fig-LfAoA}
    \end{subfigure}
   \begin{subfigure}{.45\textwidth}
      \centering
      \includegraphics[width=0.98\linewidth]{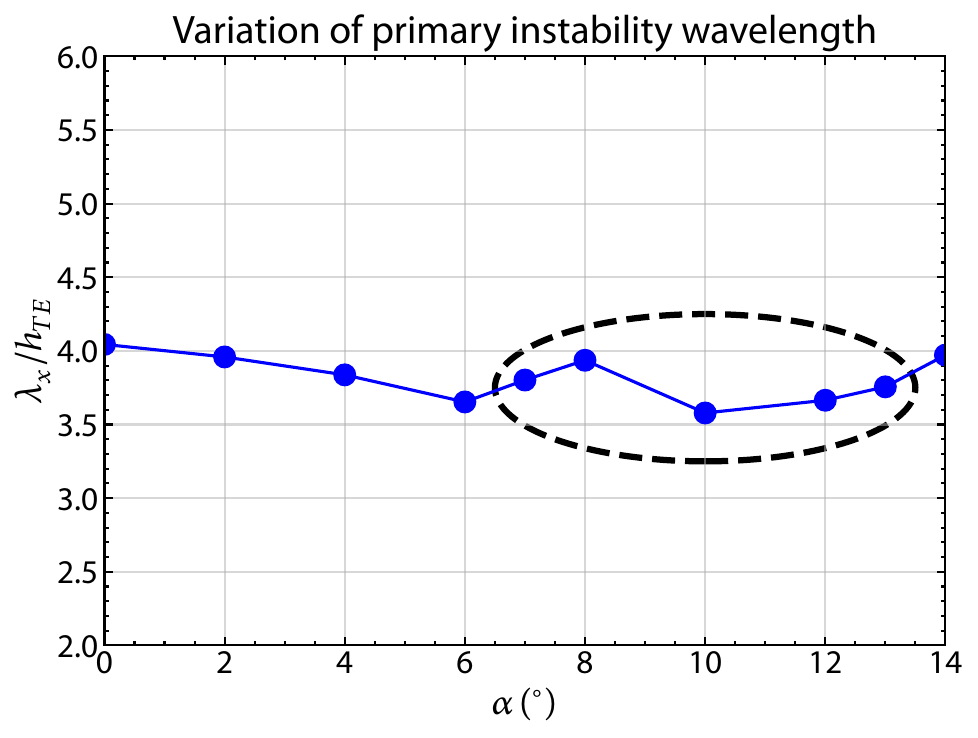}
      \caption{$\lambda_{x}$}
      \label{fig-lambdax}
    \end{subfigure}
    \caption{Primary wake statistics variation with
AoA. (a) Vortex formation length $L_{f}$ and (b) Primary instability wavelength $\lambda_{x}$. The low-drag regime is denoted with a dashed
ellipse.}
\label{fig-primWakeStats}
\end{figure}

Additionally, \autoref{fig-LfAoA} shows the variation of $L_{f}$
with the AoA. The formation length does not seem to change outside the low-drag regime, while it increases inside the low-drag regime (up to $65\%$ compared to $\alpha=0^{\circ}$). This behavior is consistent with previous studies \cite{Rowe2001,ElGammal2007,Durgesh2013}, which suggest that the increase in vortex formation length, resulting from the greater separation between the boundary layers (BLs), corresponds to a reduction in vortex-shedding strength, an increase in base pressure, and a decrease in base drag.

Next, the streamwise correlation length of the primary instability
$\lambda_{x}$ is considered, as it
quantiﬁes the spatial structure of the B\'enard-von K\'arm\'an vortex street
downstream of the formation region. This is defined as the streamwise distance between two consecutive vortices shed from the same edge of the
blunt TE. For consistency with previous works
\cite{NaghibLahouti2012,Manolesos2021}, $\lambda_{x}$ is calculated as: $\lambda_{x} = U_{c}/f_{s}$, where $U_{c}$ is the convective velocity, defined as the mean velocity on the wake centerline at $4h_{TE}$ away from the wing TE. As shown in \autoref{fig-lambdax}, $\lambda_{x}$ decreases as the AoA increases, with the minimum values occurring inside the low-drag regime
(reduction up to $12.5\%$ compared to $\alpha=0^{\circ}$). The plot contains only AoA before stall, as the separated flow post stall significantly alters the B\'enard-von K\'arm\'an vortex street.

For the two AoA considered, $\alpha=0^{\circ}$ and $\alpha=12^{\circ}$, we have 
$\lambda_{x} \approx 4.04h_{TE}$ and $\lambda_{x} \approx 3.66h_{TE}$ respectively. These values correspond to a relative decrease of $\approx9.5\%$ and is in good agreement with the decrease of $\approx 8.3\%$ reported in \cite{Manolesos2021}.

\subsubsection{Intra-pair spanwise
distance}\label{intra-pair-spanwise-distance-1}

\begin{figure}[h]
\centering
\includegraphics[width=0.5\linewidth]{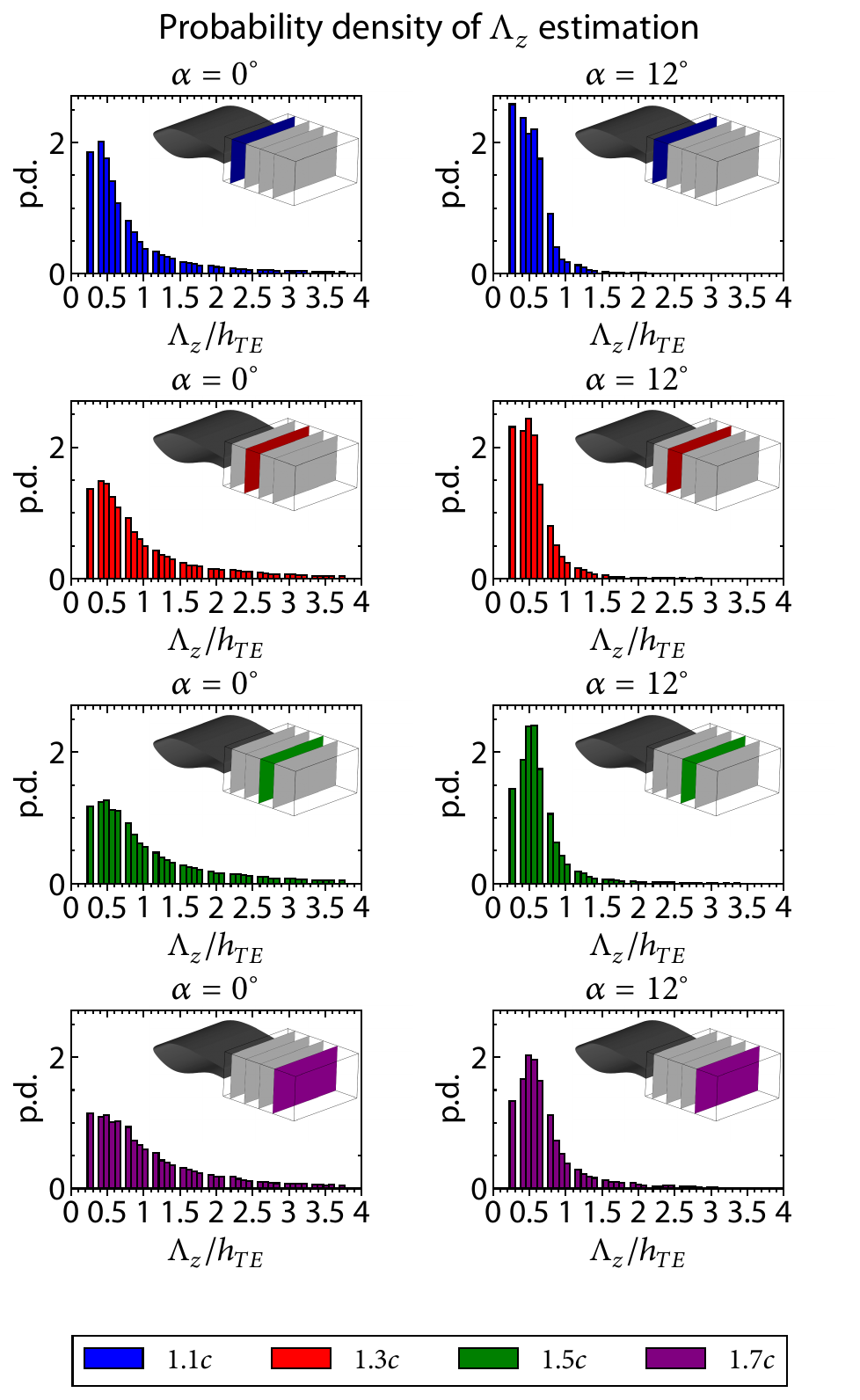}
\caption{Histograms and kernel
density estimations for $\Lambda_{z}$ using the spanwise
$\Gamma_{1}$ variations for varying
$y(t)/c=\mathrm{argmax}{\Gamma_{1}(t)}$.}
\label{fig-LAMBDA_Z_HIST_GAMMA1_moving}
\end{figure}

Next, the distance between consecutive streamwise vortices of the same pair $\Lambda_{z}$ is obtained for each timestep on four $x/c=const.$ planes, as described in Algorithm~\ref{alg:Gamma1Lambda_z}, for the $\alpha=0^{\circ}$ (high-drag) and $\alpha=12^{\circ}$ (low-drag) cases.
The results are binned and displayed via histograms in
\autoref{fig-LAMBDA_Z_HIST_GAMMA1_moving}.
The distributions for both AoA are unimodal, but it is important to note that the distributions for $\alpha=0^{\circ}$ are more skewed than those for $\alpha=12^{\circ}$.

The histograms shown in \autoref{fig-LAMBDA_Z_HIST_GAMMA1_moving}
also allow for the estimation of the distributions' most probable values
(modes). For the $\alpha=0^{\circ}$ case, the distribution gets more skewed
as one moves away from the TE, and thus, the most probable value is more
difficult to estimate after $x/c=1.7$. Conversely, at
$\alpha=12^{\circ}$, the unimodality of the distribution becomes clearer as
one moves away from the TE as the more organized wake structures are
convected downstream. This behavior suggests that, streamwise vortical pairs break up
closer to the TE at $\alpha=0^{\circ}$.

\subsubsection{Spanwise distance between vortex
pairs}\label{spanwise-distance-between-vortex-pairs-1}

\begin{figure}[h]
\centering
\includegraphics[width=0.5\linewidth]{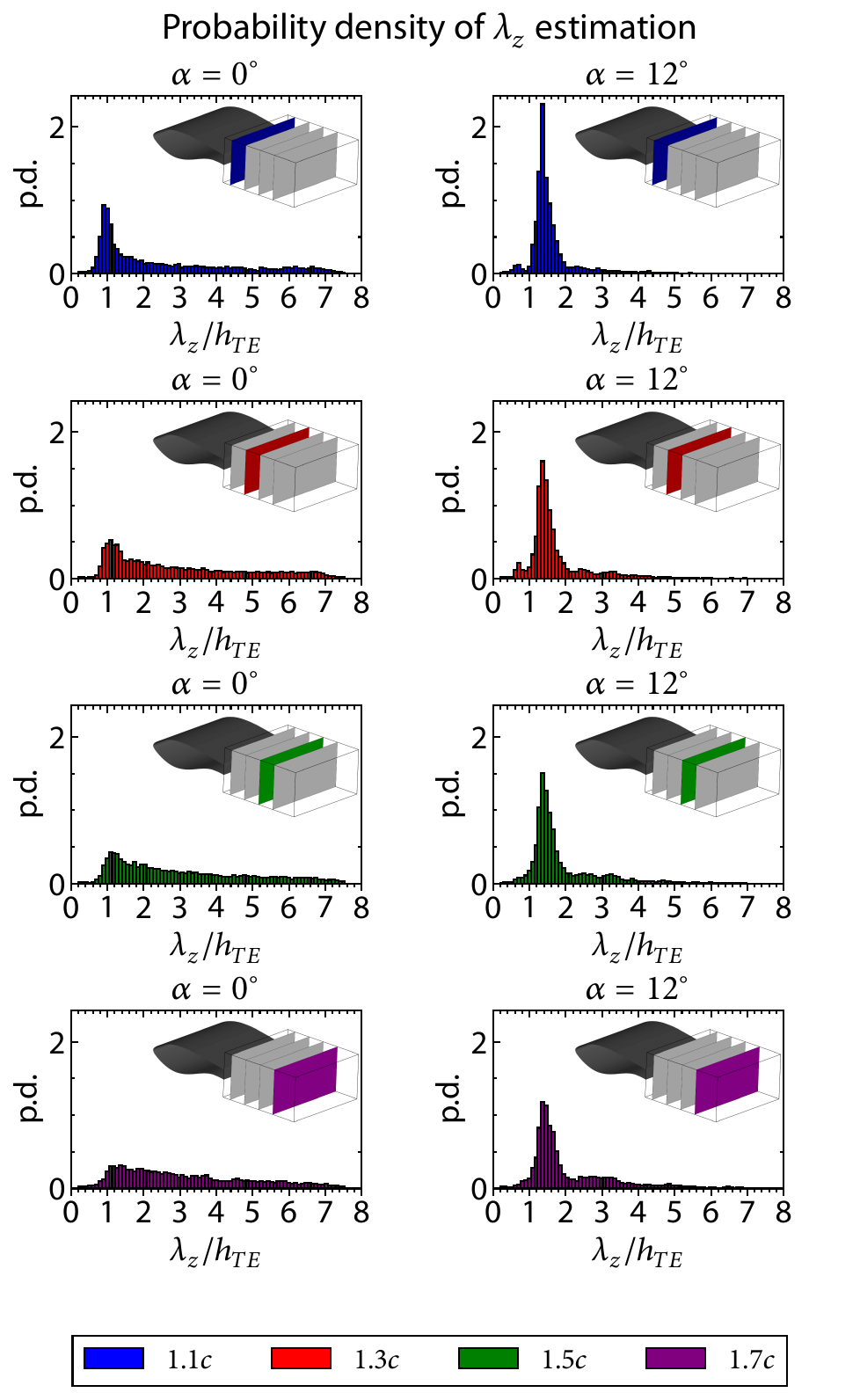}
\caption{\label{fig-LAMBDA_Z_HIST_ACORR_moving}Histograms and kernel
density estimations for $\lambda_{z}$ using the spanwise
$\Gamma_{1}$ autocorrelation for varying
$y(t)/c=\mathrm{argmax}{\Gamma_{1}(t)}$.}
\end{figure}

\begin{figure}[h]
\centering
\includegraphics[width=0.45\linewidth]{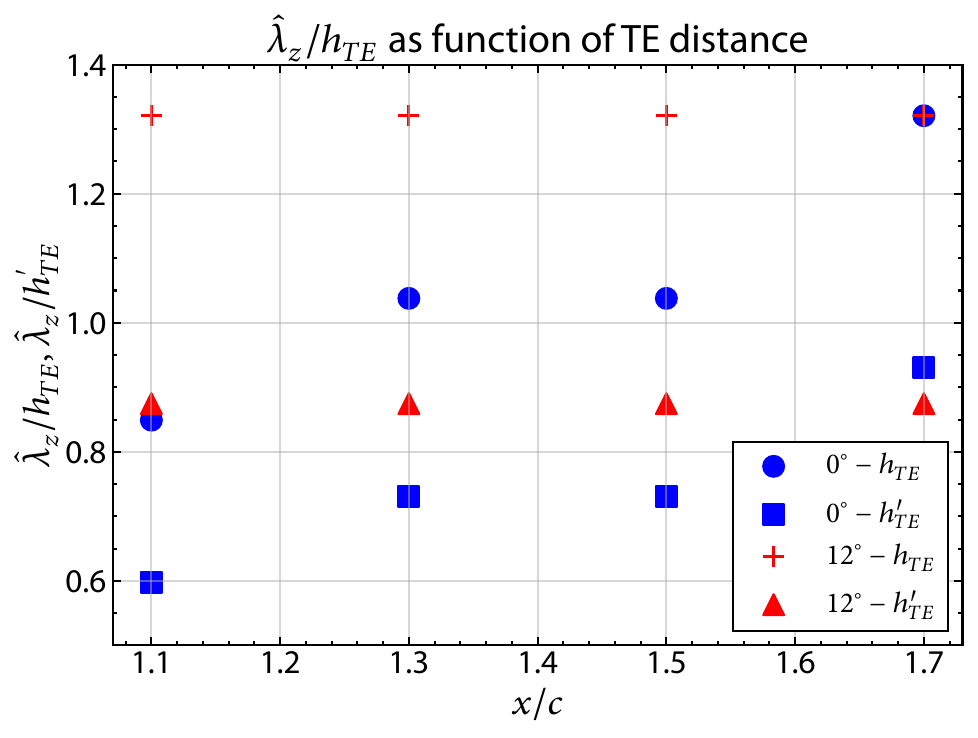}
\caption{\label{fig-autocorrelation-modes-scatter}Most probable values
(modes) of $\lambda_{z}$ normalized with TE height (circles and
squares) and the effective TE height (crosses and triangles) using the
spanwise correlation of $\Gamma_{1}$ for different downstream
sampling planes. The high-drag case $\left(0^{\circ}\right)$ is shown
in blue and the low-drag case $\left(12^{\circ}\right)$ is shown in red.}
\end{figure}

Also. the distance between adjacent pairs of streamwise vortices is
estimated for each timestep as described in Algorithm~\ref{alg:Acorrlambda_z} for the $\alpha=0^{\circ}$ (high-drag) and $\alpha=12^{\circ}$ (low-drag) cases.
The results are shown in
\autoref{fig-LAMBDA_Z_HIST_ACORR_moving}. The distributions for $\alpha=0^{\circ}$ are skewed, but peaks are observed at approximately
$\lambda_{z}/h_{TE}=1.0$ for the near-wake sampling planes. Similarly,
for $\alpha=12^{\circ}$, a unimodal distribution is evident near TE, and it
gradually reduces as the sampling plane moves downstream. \autoref{fig-autocorrelation-modes-scatter} shows the variation of
the most probable $\lambda_{z}$ normalized with $h_{TE}$ and $h_{TE}^{\prime}$ for $\alpha=0^{\circ}$ and $\alpha=12^{\circ}$. A dependency
on the streamwise distance from the TE is observed for the $\alpha=0^{\circ}$ case, while no variation is observed for the $\alpha=12^{\circ}$ case. It is noteworthy that the results are identical for both cases at $x/c=1.7$. This is observed because there are 100
points along each $y/c=const.$ line and a finite amount of possible values exist.

The predicted $\hat{\lambda}_{z}$ increases with the distance from the TE for the $\alpha=0^{\circ}$ cases, while it remains constant for the $\alpha=12^{\circ}$ case. This behavior is consistent with the previous
discussion as the increase of $\lambda_{z}$ means that the wake becomes less correlated, i.e., breaks up, as the sampling plane moves
away from the TE. Conclusions can be drawn only for the $12^{\circ}$ case. The resulting $\lambda_{z}/h_{TE}=0.9-1.4$ and $\lambda_{z}/h_{TE}^{\prime}=0.6-0.9$ agree with the
prediction of previous studies for elongated bluff bodies at lower Reynolds numbers (up to $Re_{h_{TE}}=2.5\times10^{4}$ compared with
the present $Re_{h_{TE}}=15.9\times10^{4}$) with similar $c/h_{TE}$
ratio \cite{Gibeau2020}, especially when using the
$h_{TE}^{\prime}$ normalization which showcases the dependency on the upstream BL. 

\subsubsection{Secondary instability
characterization}\label{secondary-instability-characterization}

\begin{figure}[h]
    \centering
   \begin{subfigure}{.4\textwidth}
      \centering
      \includegraphics[width=\linewidth]{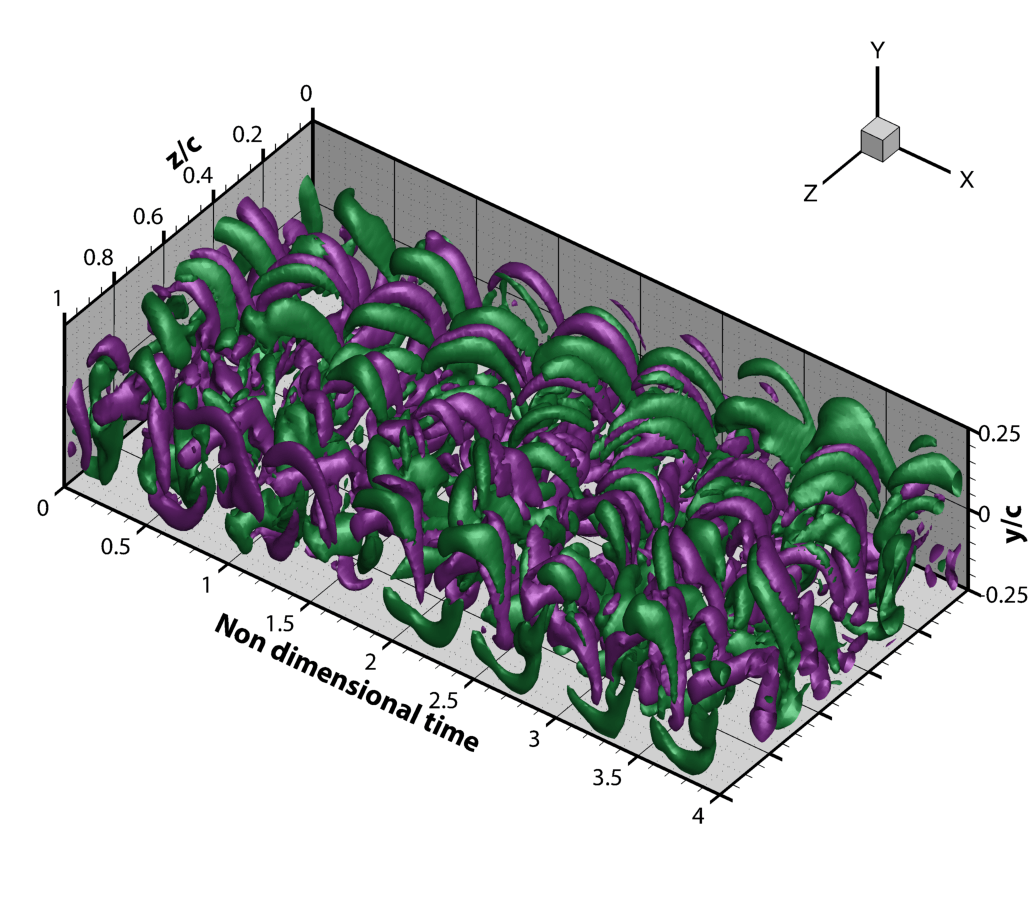}
      \caption{$\alpha=0^{\circ}$ -- iso}
\label{fig-OMEGAX_ISO_0}
    \end{subfigure}
   \begin{subfigure}{.4\textwidth}
      \centering
      \includegraphics[width=\linewidth]{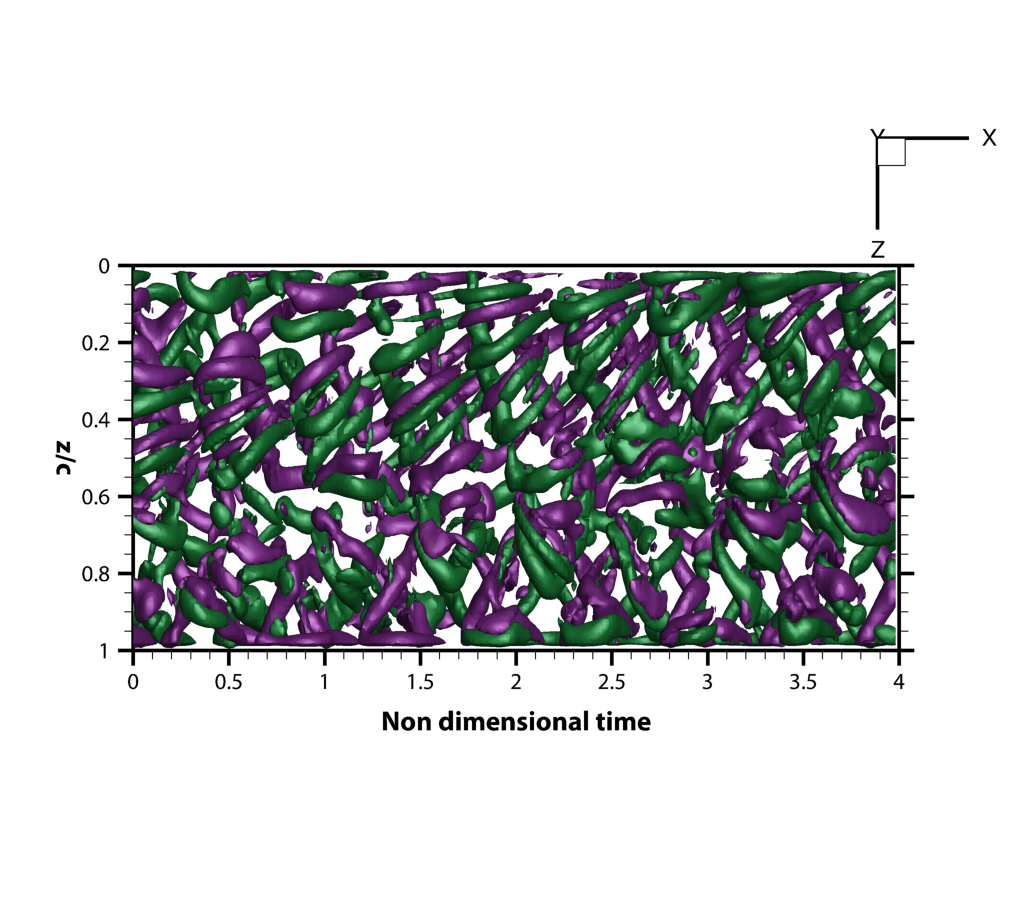}
      \caption{$\alpha=0^{\circ}$ -- top}
\label{fig-OMEGAX_TOP_0}
    \end{subfigure}
   \begin{subfigure}{.4\textwidth}
      \centering
      \includegraphics[width=\linewidth]{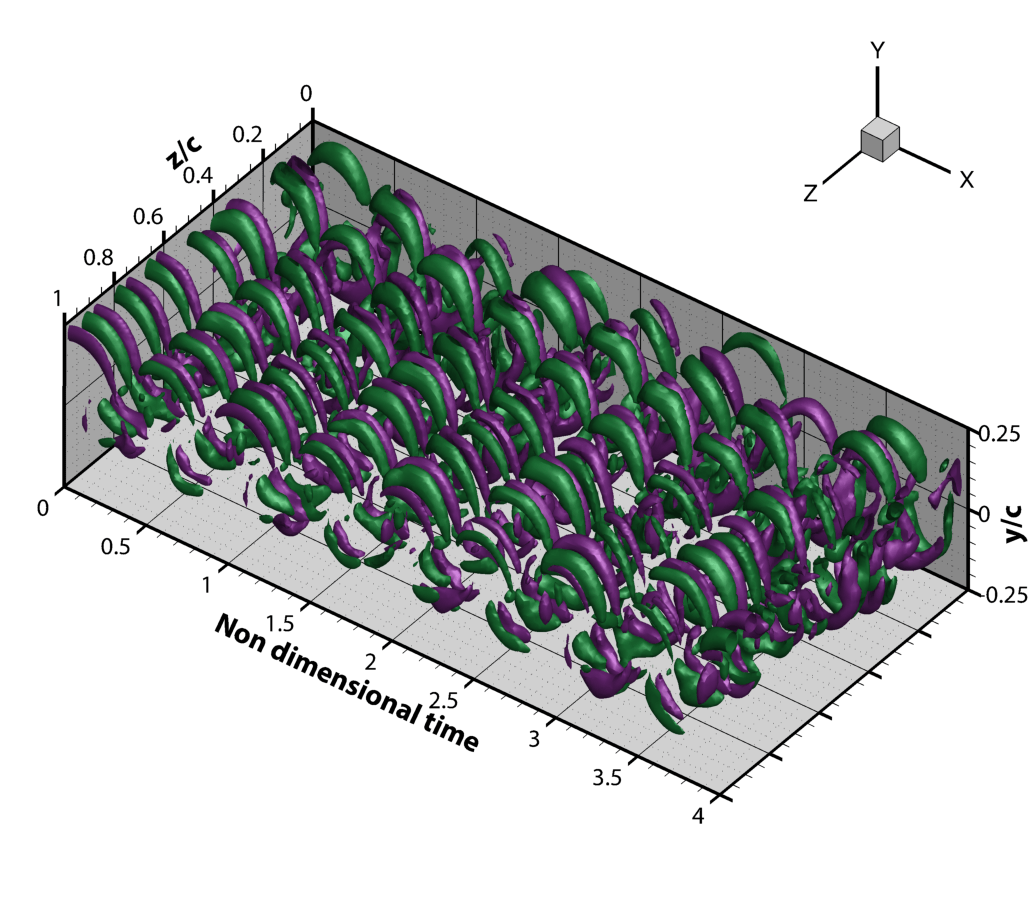}
      \caption{$\alpha=12^{\circ}$ -- iso}
\label{fig-OMEGAX_ISO_12}
    \end{subfigure}
   \begin{subfigure}{.4\textwidth}
      \centering
      \includegraphics[width=\linewidth]{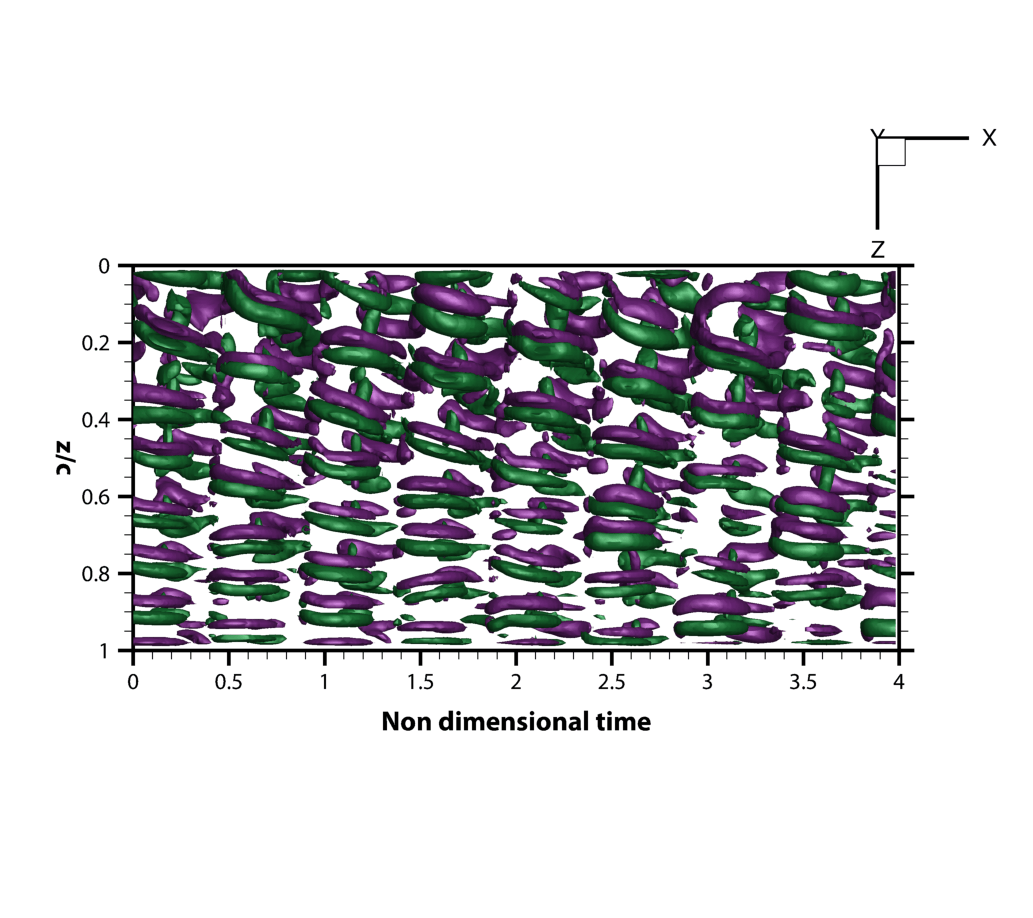}
      \caption{$\alpha=12^{\circ}$ -- top}
\label{fig-OMEGAX_TOP_12}
    \end{subfigure}
\caption{Normalized streamwise vorticity
${\omega_{x}}/{U_{\infty}}=\pm5$ isosurfaces for approximately 8
vortex shedding cycles at $x/c=1.7$. Green and purple represent
positive and negative rotational directions, respectively. An isometric
(iso) view and a top view are shown for the $\alpha=0^{\circ}$ (top row)
and $\alpha=12^{\circ}$ (bottom row) cases.}
\label{fig-OMEGAX_ISO}
\end{figure}

The $\lambda_{z}/h_{TE}\approx1-1.4$ values may correspond to the mode
B structure reported for the wakes of circular
\cite{Williamson1996b} and square \cite{Luo2007} cylinders and elongated bluff bodies \cite{Gibeau2018,Gibeau2020}. Another
possible framework that allows for the characterization of the mode is the one discussed in \cite{Ryan2005}. Both Floquet multiplier
analysis and Direct Numerical Simulations, at low Reynolds numbers $\left(\mathrm{Re}_{h_{TE}}\le600\right)$ revealed different structures that
correspond to the secondary instabilities for the flow past an elongated bluff body for different chord-to-thickness ratios. Namely, mode
S$^{\prime}$ and mode B$^{\prime}$ with
$\lambda_{z}/h_{TE}\approx1$ and $\lambda_{z}/h_{TE}\approx2$ were
predicted by \citet{Ryan2005}, and the mode B$^{\prime}$ structure
has been identified in previous studies for flows past elongated bluff
bodies \cite{NaghibLahouti2012,NaghibLahouti2014}.

In order to distinguish between the possible modes, planes normal to the streamwise direction are obtained from the first 2000 samples, corresponding to 8 vortex shedding cycles, at $x/c=1.7$ and merged
along the time dimension together with the reconstruction of $\omega_{x}c/U_{\infty}$. These isosurfaces are presented for both AoA
in \autoref{fig-OMEGAX_ISO}. Previous studies have used similar visualizations for elongated bluff bodies
\cite{Gibeau2018,Gibeau2020} at lower Reynolds numbers.

For $\alpha=0^{\circ}$, large-scale bi-directional wake distortions are evident, and no safe conclusion can be drawn. Then, for $\alpha=12^{\circ}$, although a distortion that seems to correspond to oblique shedding \cite{Williamson1989} appears, the streamwise vortex
pairs are clear. In addition, the rotational direction of the streamwise vortex pairs is altered after each vortex-shedding cycle of the primary vortex pairs, corresponding to approximately $0.5$ non-dimensional
time units.

This indicates that mode S$^{\prime}$ is present in the wake for the low drag case $\left(\alpha=12^{\circ}\right)$, as modes B and B$^{\prime}$
require that the rotational direction of the streamwise pairs not alternate throughout the primary vortex shedding cycle \cite{Ryan2005,Gibeau2018,Gibeau2020}. Additionally, by visual inspection, approximately seven streamwise vortex pairs can be identified, reinforcing the prediction of $\lambda_{z}/h_{TE}\approx1.4$ at $12^{\circ}$.

\begin{figure}[h]
\begin{subfigure}{.45\textwidth}
\centering
\includegraphics[width=\linewidth]{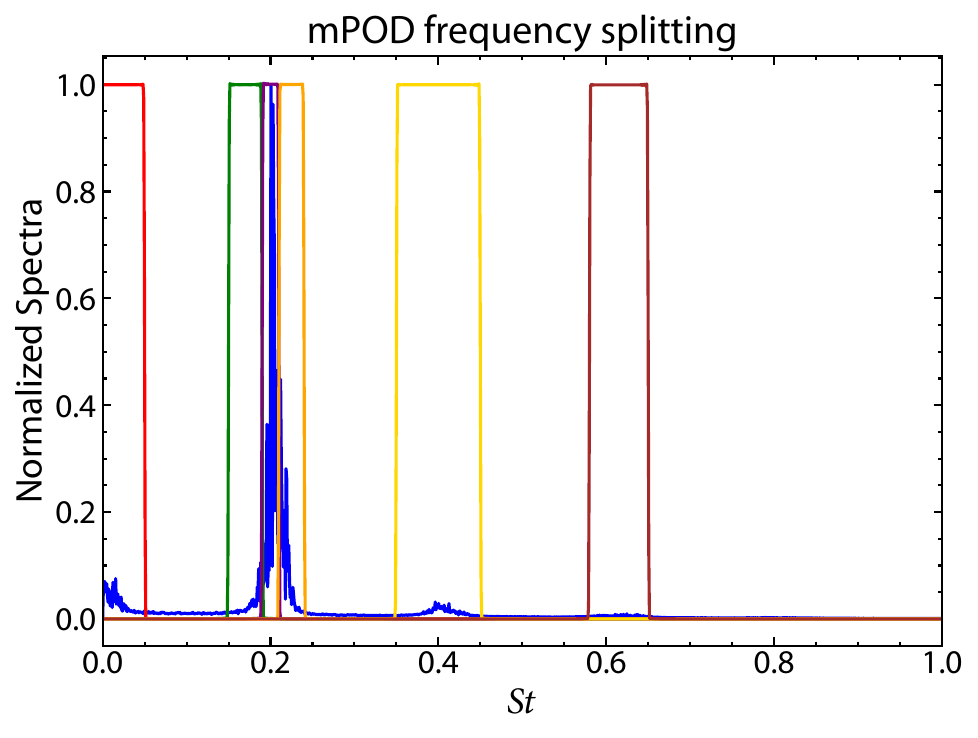}
\caption{$\alpha=0^{\circ}$}
\label{fig-freqsplit0}
\end{subfigure}
   \begin{subfigure}{.45\textwidth}
\centering
\includegraphics[width=\linewidth]{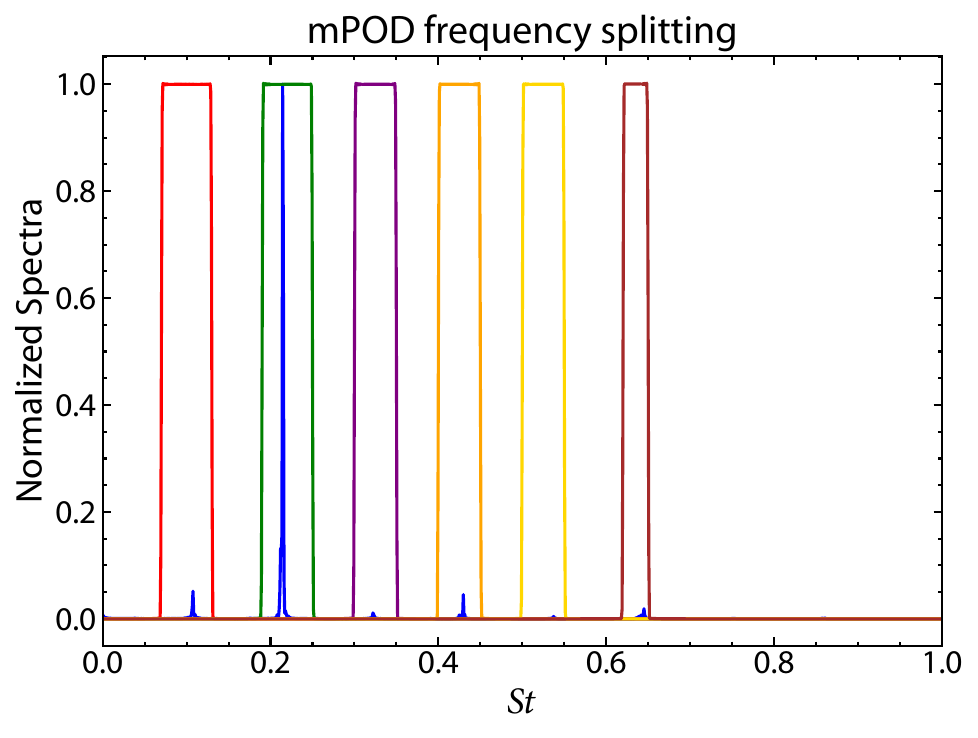}
\caption{$\alpha=12^{\circ}$}
\label{fig-freqsplit12}
\end{subfigure}
\caption{Frequency splitting filter banks used
in mPOD for (a) $\alpha=0^{\circ}$ and (b) $\alpha=12^{\circ}$.}
\label{fig-freqsplitting}
\end{figure}

\subsection{Three-dimensional coherent
structures}\label{three-dimensional-coherent-structures}

Building on the analysis of the previous section, this section presents the investigation of three-dimensional coherent structures via data-driven modal analysis as described in \autoref{mPOD}.

\subsubsection{Frequency splitting \& temporal coefficients}\label{sec-freqSplit}

The spectrum of the temporal correlation matrix $\mathbf{K}$ of the wake dataset was examined to determine the band-pass filter banks necessary for mPOD. The spectrum and the used filter banks are shown in \autoref{fig-freqsplitting} for both AoA. For $\alpha=0^{\circ}$, a
slow disturbance is noted near $St\approx0.01$, and two additional
peaks appear on both sides around the primary frequency at
$St\approx0.2$. Isolating these frequencies requires filtering with a sharp transfer function, hence very high order finite impulse response filters. Furthermore, two weak harmonics are shown at $St\approx0.4$ and $St\approx0.6$ and are taken into consideration.

\begin{figure}[h]
\centering
\includegraphics[width=0.5\linewidth]{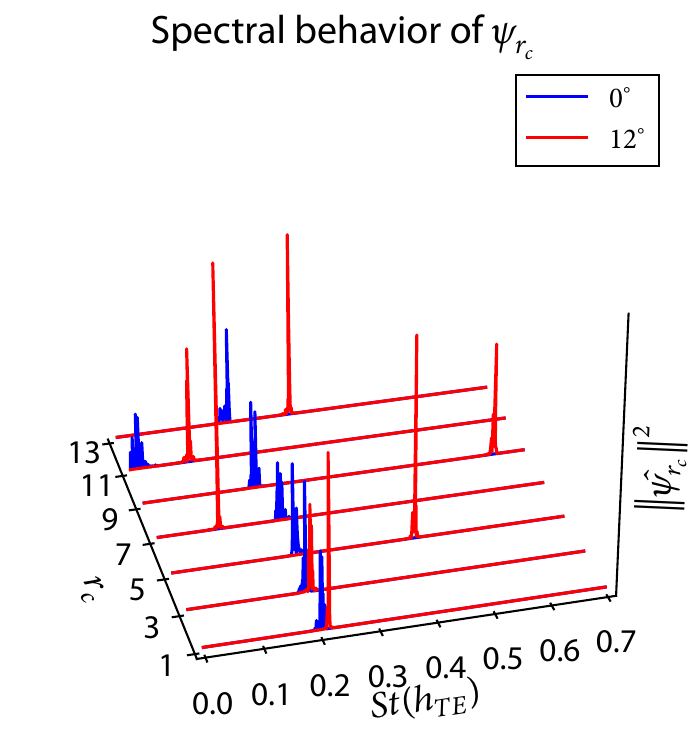}
\caption{Spectral behavior of the first 13 mPOD modes' temporal coefficients for $\alpha=0^{\circ}$ (blue) and $\alpha=12^{\circ}$ (red).}
\label{fig-mPOD_psi_fft}
\end{figure}

For $\alpha=12^{\circ}$, a subharmonic appears at $St\approx0.1$, and the
primary frequency at $St\approx0.21$ is significantly clearer. In
addition to the two harmonics of the primary frequency appearing at
$St\approx0.43$ and $St\approx0.64$, a trace of an additional
frequency is noted at $St\approx0.32$. It is noted that this
additional frequency cannot be correlated with the fundamental one and is further investigated in the sections that follow. The effect of the filter banks is evident in the spectral behavior of the temporal coefficients of the mPOD modes.
Namely, \autoref{fig-mPOD_psi_fft} shows the results of the Discrete Fourier Transform (DFT) of the temporal coefficients for the first 13 mPOD modes. 

All modes identified in the mean-shifted dataset are linked to traveling wave patterns, i.e. modes come in pairs with very similar eigenvalues (amplitudes) and spectra and $90^{\circ}$ phase shift in space and time\cite{Mendez2020}. Therefore, only the odd-numbered modes are presented in \autoref{fig-mPOD_psi_fft}. The peaks are sharper and more distinct in the low-drag case.

\subsubsection{Convergence \& spectral content}\label{convergence}

\begin{figure}[h]
\centering
\includegraphics[width=0.45\linewidth]{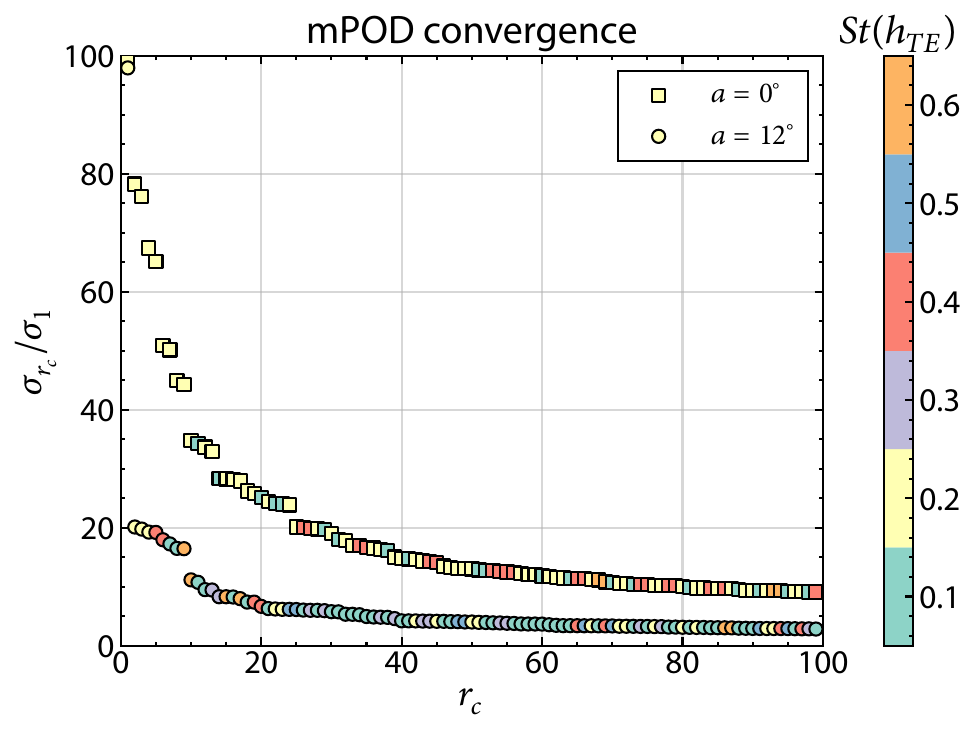}
\caption{mPOD convergence for $\alpha=0^{\circ}$
(filled squares) and $\alpha=12^{\circ}$ (filled circles) colored by dominant $St$ number.}
\label{fig-mPODconvergenceSt}
\end{figure}

\autoref{fig-mPODconvergenceSt} illustrates the ratio $\sigma_{r_{c}}/ \sigma_{1}$ of mPOD mode amplitude for both AoA.
We use filled markers colored by the dominant $St$ number of each mode (estimated as the frequency where the maximum of the PSD of the temporal coefficients $\psi_{r}$ occurs).
The ratio $\sigma_{r_{c}}/ \sigma_{1}$ can be considered as a measure of the convergence of the mPOD and consequently of the energy content contributed to the dataset from each mode. 

The mPOD convergences faster at the low-drag case $\left(\alpha=12^{\circ}\right)$, meaning that the dataset can be
reconstructed with fewer modes of similar importance. Additionally, the faster convergence for $\alpha=12^{\circ}$ indicates that the structures are more organized than the high-drag case $\left(\alpha=0^{\circ}\right)$.

As for the spectral content, it is unsurprising that the modes corresponding to the primary instability--specifically the primary shedding frequency and its harmonics--dominate the flow from an energy perspective. Nevertheless, none of these modes is perfectly harmonic, justifying the interest in the mPOD over purely harmonic formulations such as the Spectral POD \cite{Towne2018} or the traditional POD. 

The high-drag case $\left(\alpha=0^{\circ}\right)$ exhibits a variety of modes with $St\approx0.2$, corresponding to 3D undulations of the main vortices. Conversely, the low-drag case $\left(\alpha=12^{\circ}\right)$ has a very limited number of high energy modes at $St\approx0.21$. The first $\left(St\approx0.41\right)$ and second $\left(\approx0.62\right)$ harmonics of the primary instability appear in both cases.

For the high-drag case, several modes display very low frequencies $\left(St\approx0.01\right)$, while the low-drag case features several modes with $St\approx0.11$, as well as their harmonics with $St\approx0.32$ and $St\approx0.54$. 
This group of modes corresponds to the secondary instabilities present in the wake of the airfoil and are further discussed in \autoref{spatial-bases}.

\begin{figure}[h]
\begin{subfigure}{.45\textwidth}
\centering
\includegraphics[width=\linewidth]{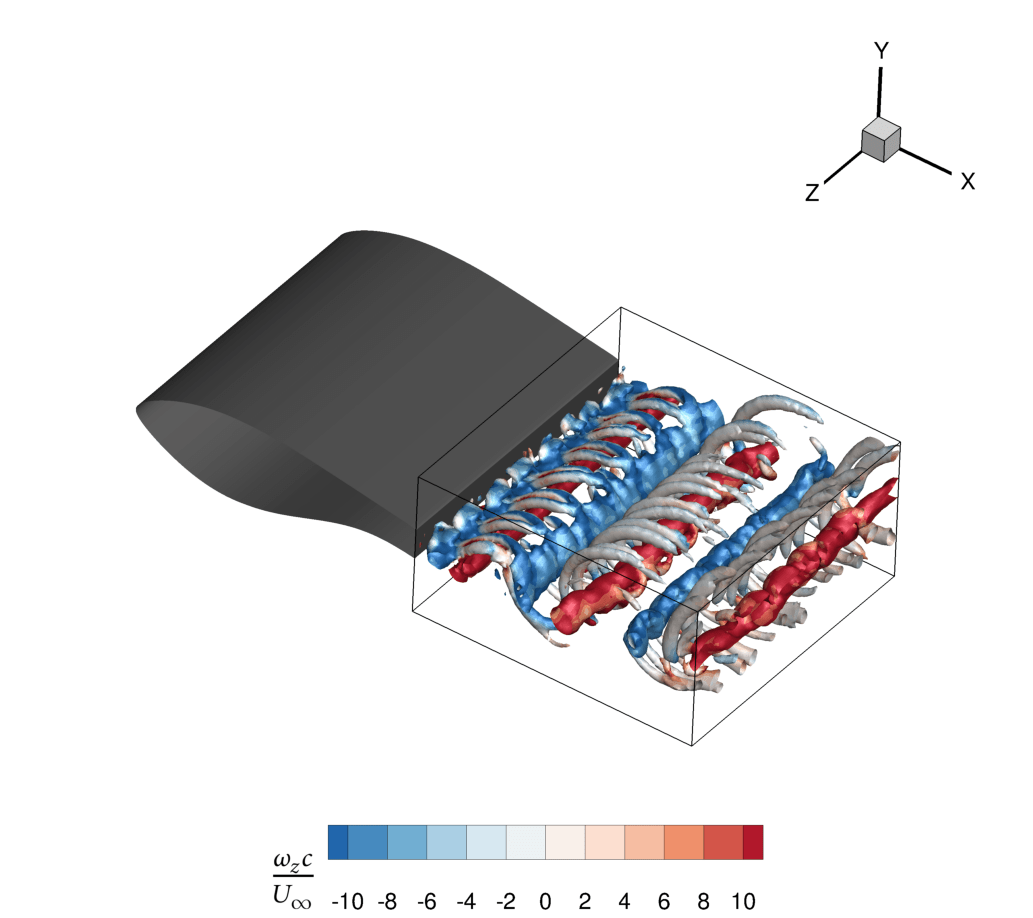}
\caption{Original timestep}
\label{fig-orig_12}
\end{subfigure}%
\begin{subfigure}{.45\textwidth}
\centering
\includegraphics[width=\linewidth]{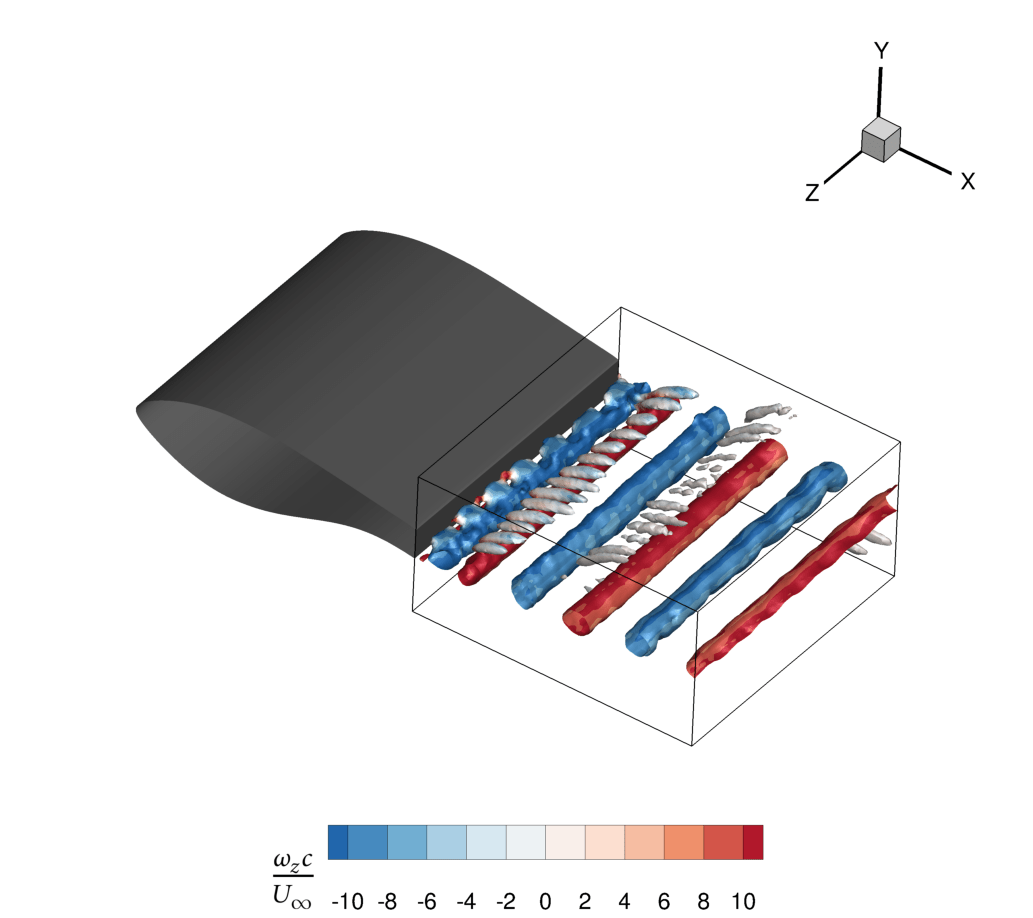}
\caption{Reconstructed timestep}
\label{fig-rec_12}
\end{subfigure}%
\caption{Q-criterion isosurfaces
colored by $\omega_{z}$ for (a) an original timestep and (b) the
reconstructed timestep using the first 20 mPOD modes at
$\alpha=12^{\circ}$.}
\label{fig-reconstructed-snapshot}
\end{figure}

A reconstruction of the flow field using the first 20 modes at $\alpha=12^{\circ}$ is shown in \autoref{fig-rec_12}.
It is evident that both the primary and secondary instabilities are captured and can be adequately represented by the first 20 modes of the dataset.

\begin{figure}[h]
\begin{subfigure}{0.47\textwidth}
\centering
\includegraphics[width=\linewidth]{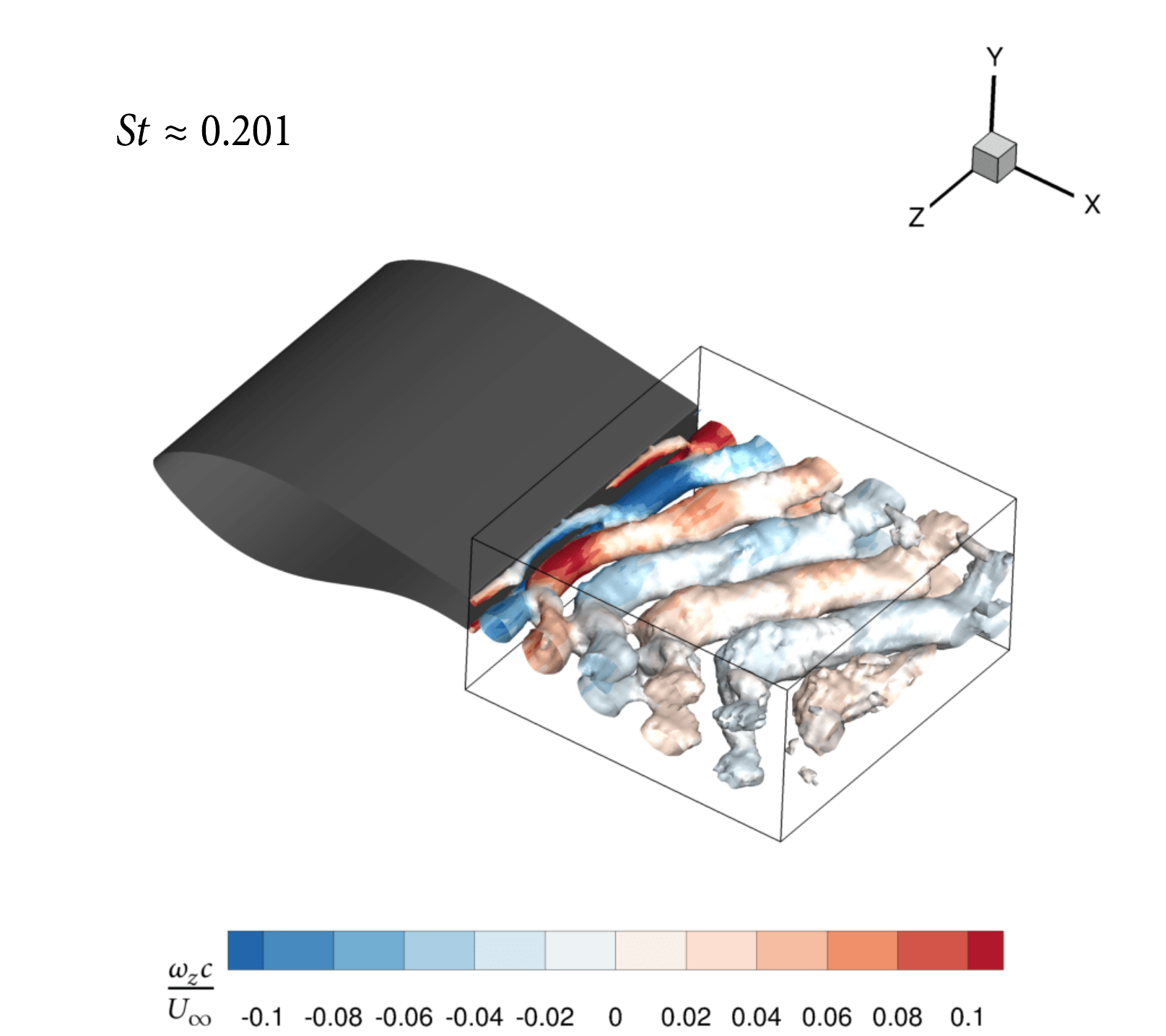}
\caption{$\alpha=0^{\circ}$ -- 1st mode}
\label{fig-mode1_0}
\end{subfigure}%
\begin{subfigure}{0.47\textwidth}
\centering
\includegraphics[width=\linewidth]{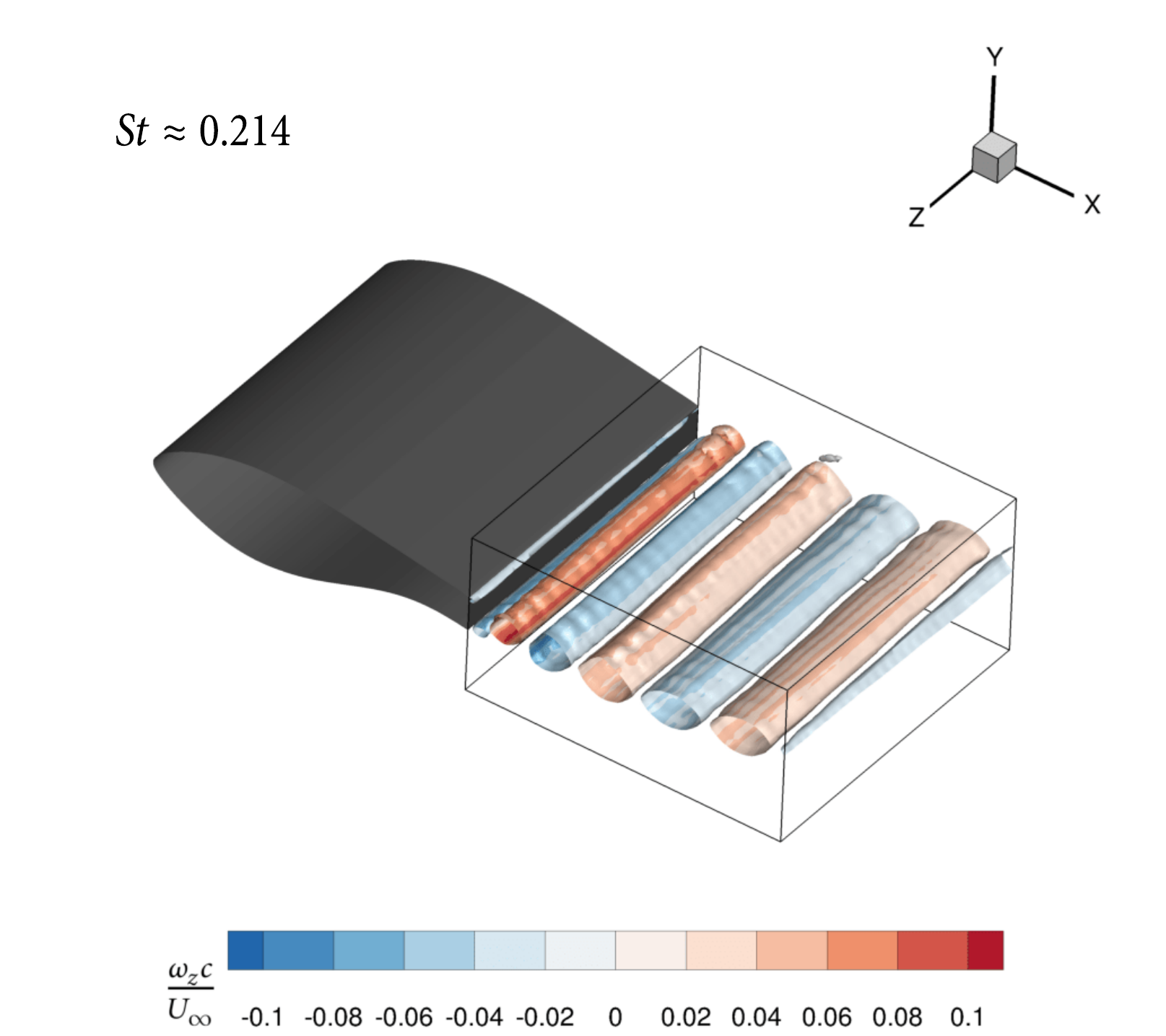}
\caption{$\alpha=12^{\circ}$ -- 1st mode}
\label{fig-mode1_12}
\end{subfigure}%
\\
\begin{subfigure}{0.47\textwidth}
\centering
\includegraphics[width=\linewidth]{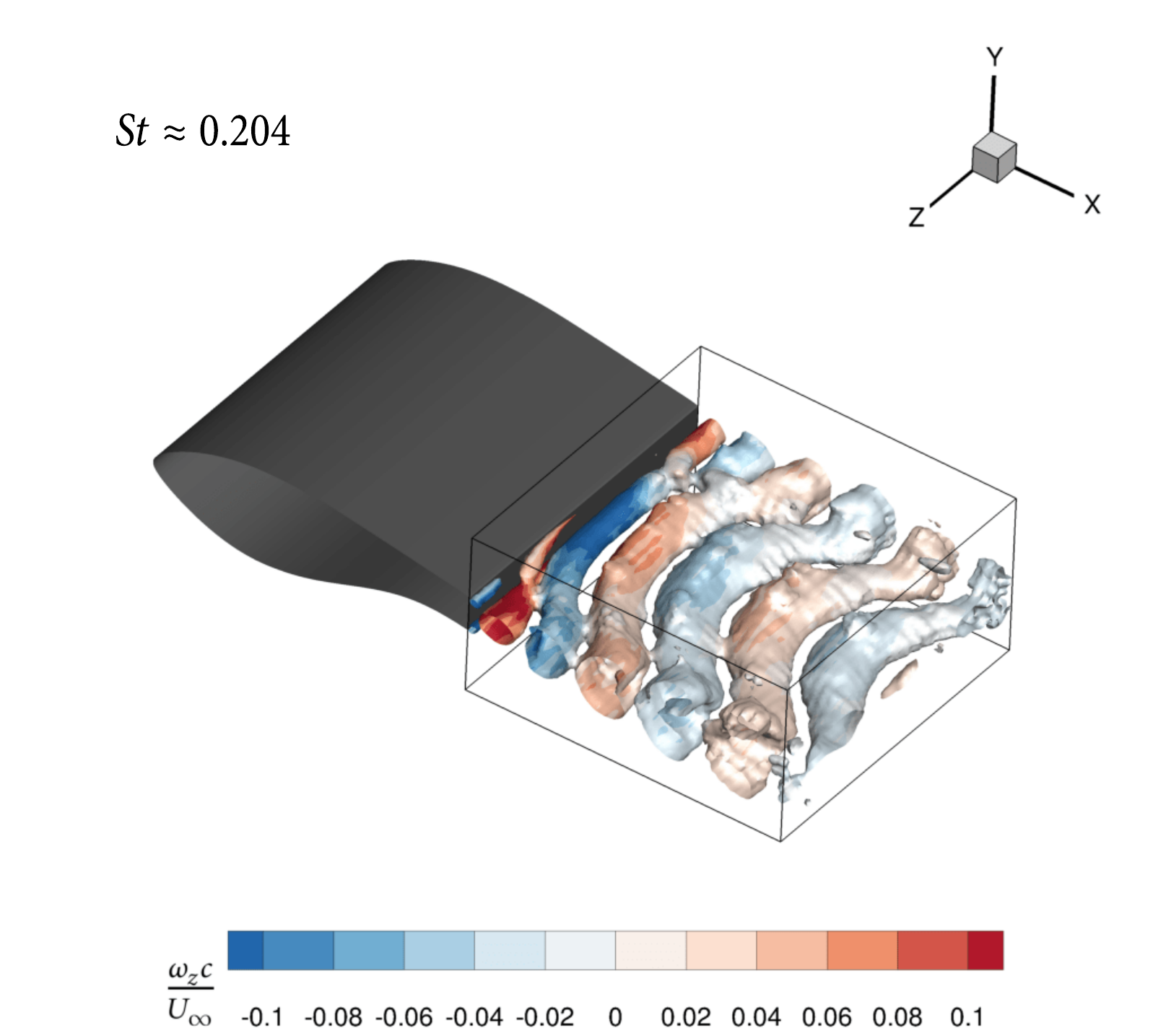}
\caption{$\alpha=0^{\circ}$ -- 3rd mode}
\label{fig-mode3_0}
\end{subfigure}%
\begin{subfigure}{0.47\textwidth}
\centering
\includegraphics[width=\linewidth]{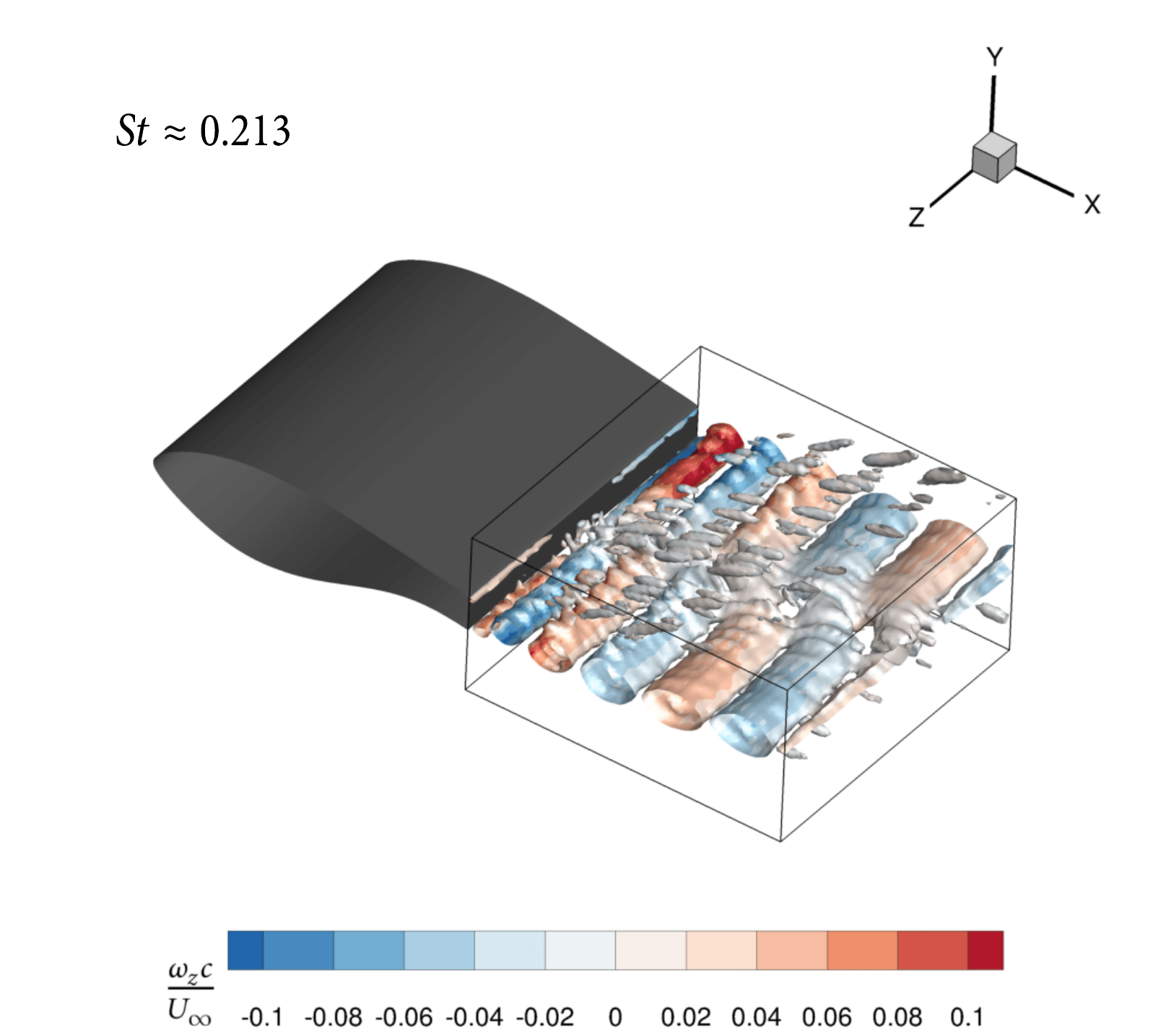}
\caption{$\alpha=12^{\circ}$ -- 3rd mode}
\label{fig-mode3_12}
\end{subfigure}%
\caption{Q-criterion isosurfaces colored by
$\omega_{z}c/U_{\infty}$ for the first group of mPOD modes, corresponding to the primary instability.
Panels (a) and (c) show the first two modes belonging to this group for $\alpha=0^{\circ}$, and panels (b) and (d) show the same for the $\alpha=12^{\circ}$ case.}
\label{fig-modes1}
\end{figure}%

\subsubsection{Spatial bases}\label{spatial-bases}

The three-dimensional spatial structures $\mathbf{\phi}_{r}$ of the leading mPOD modes are analyzed in terms of Q isosurfaces, colored by spanwise $\omega_{z}c/U_{\infty}$ or streamwise $\omega_{x}c/U_{\infty}$ normalized vorticity. 

The spatial bases, in conjunction with the spectral behavior of the modes discussed in the previous subsections,
allow for categorizing the modes into 4 families/groups displaying similar behavior. This is summarized in \autoref{tbl-modesSum}. 

\begin{table}[h]
\caption{Summary of the four mPOD modes' groups for the
high-drag $\left(\alpha=0^{\circ}\right)$ and low-drag
$\left(\alpha=12^{\circ}\right)$ cases.}
\label{tbl-modesSum}
\centering
\begin{tabular}{ccccccc} 
\toprule
  Group & $St$ & Characterization & Figure & Case observed\\ \midrule 
 1  & $\approx 0.20-0.21$ & Primary instability    &  \ref{fig-modes1} & Both \\
 2  & $\approx 0.4$ \& $\approx 0.6$ & Harmonics of primary instability    & \ref{fig-modes2} & Both \\
 3  & $\approx 0.01$ (high-drag) \& $\approx 0.11$ (low-drag)  & Secondary instability & \ref{fig-modes3} & Both \\
 4 & $\approx 0.32$ \& $\approx 0.54$ &  Harmonics of secondary instability & \ref{fig-modes4} & Low-drag only \\
 \bottomrule
\end{tabular}
\end{table}%

The first group of modes is shown for both AoA in \autoref{fig-modes1}. For brevity, only the two first modes for each case are included. This group contains modes that correspond to the primary B\'enard-von K\'arm\'an vortex street. 
The $\alpha=0^{\circ}$ case demonstrates large-scale distortions in the wake that can be attributed to oblique shedding
\cite{Williamson1989} and vortex dislocation
\cite{Williamson1992}.  On the contrary, the wake is much more
organized for the low-drag case $\left(\alpha=12^{\circ}\right)$ with the primary instability standing out, although distortions can be seen in \autoref{fig-mode3_12}. Additionally, the dominant frequency of the temporal coefficient $\mathbf{\psi}_{r}$ spectrum, obtained via DFT, shows an increase of $5\%$ at low-drag case, namely $St\approx0.201$ at $\alpha=0^{\circ}$ and $St\approx0.214$ at $\alpha=12^{\circ}$.

\begin{figure}[h]
\begin{subfigure}{0.47\textwidth}
\centering
\includegraphics[width=\linewidth]{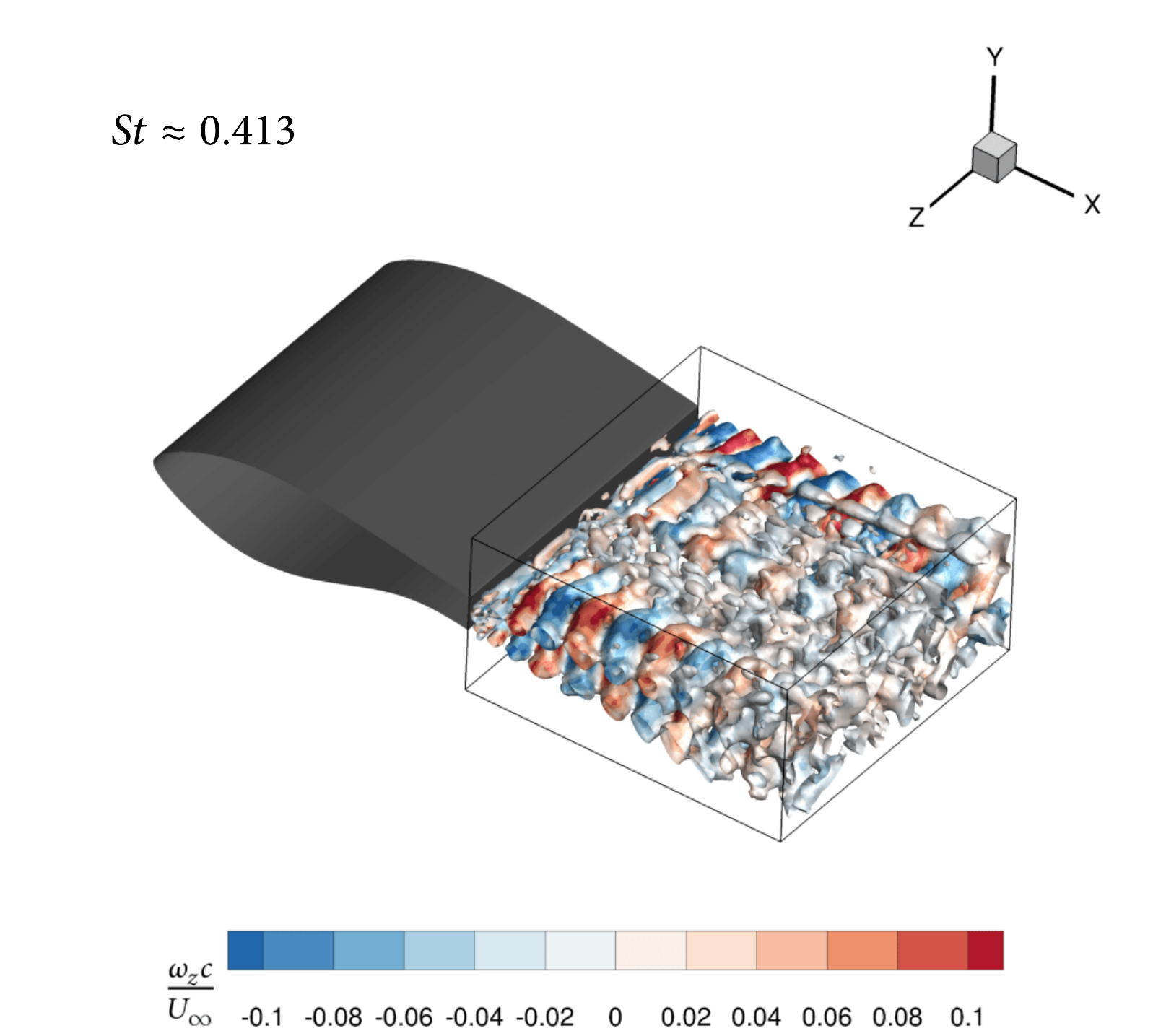}
\caption{$\alpha=0^{\circ}$ -- 27th mode}
\label{fig-mode27_0}
\end{subfigure}%
\begin{subfigure}{0.47\textwidth}
\centering
\includegraphics[width=\linewidth]{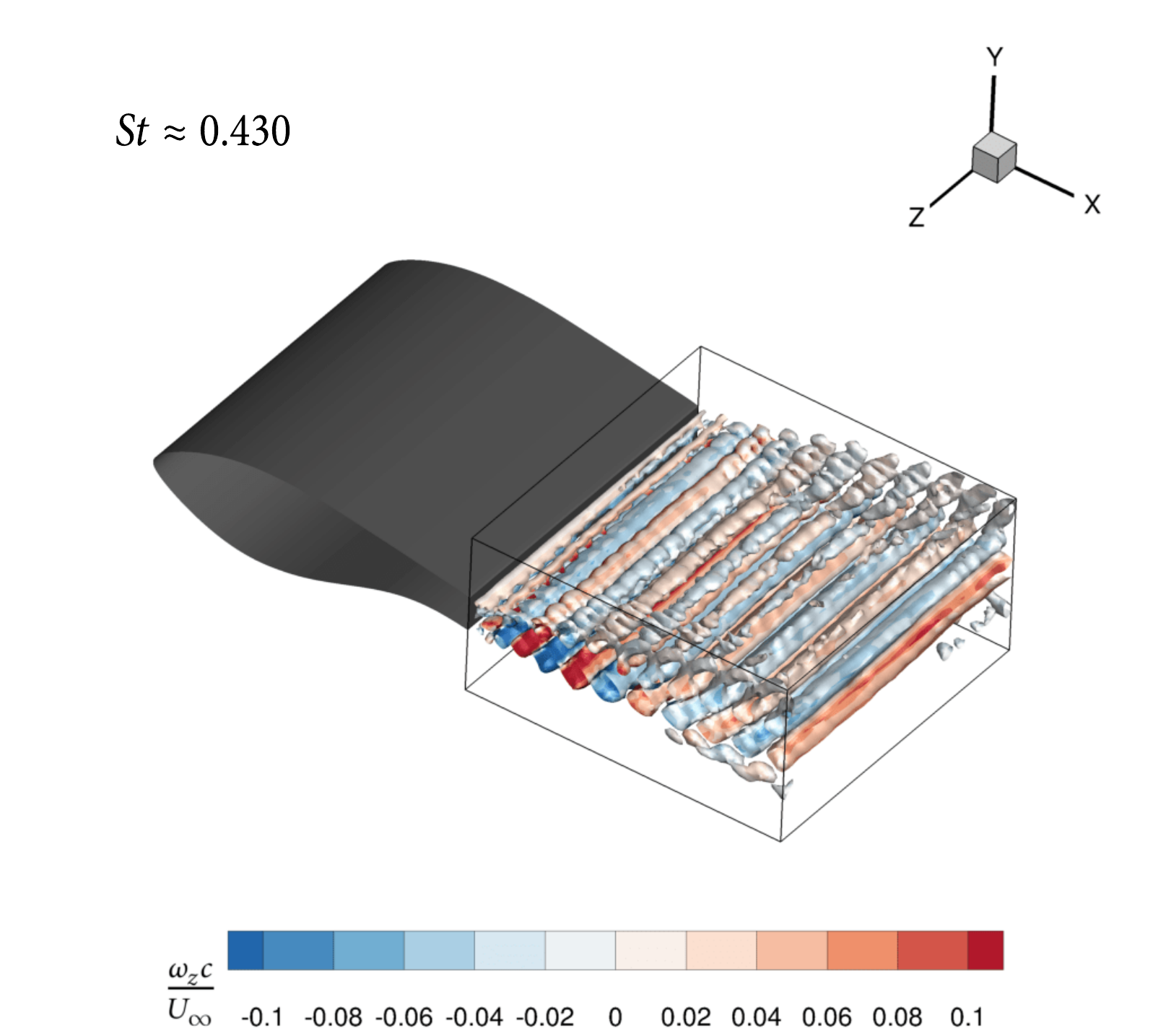}
\caption{$\alpha=12^{\circ}$ -- 5th mode}
\label{fig-mode5_12}
\end{subfigure}%
\\
\begin{subfigure}{0.47\textwidth}
\centering
\includegraphics[width=\linewidth]{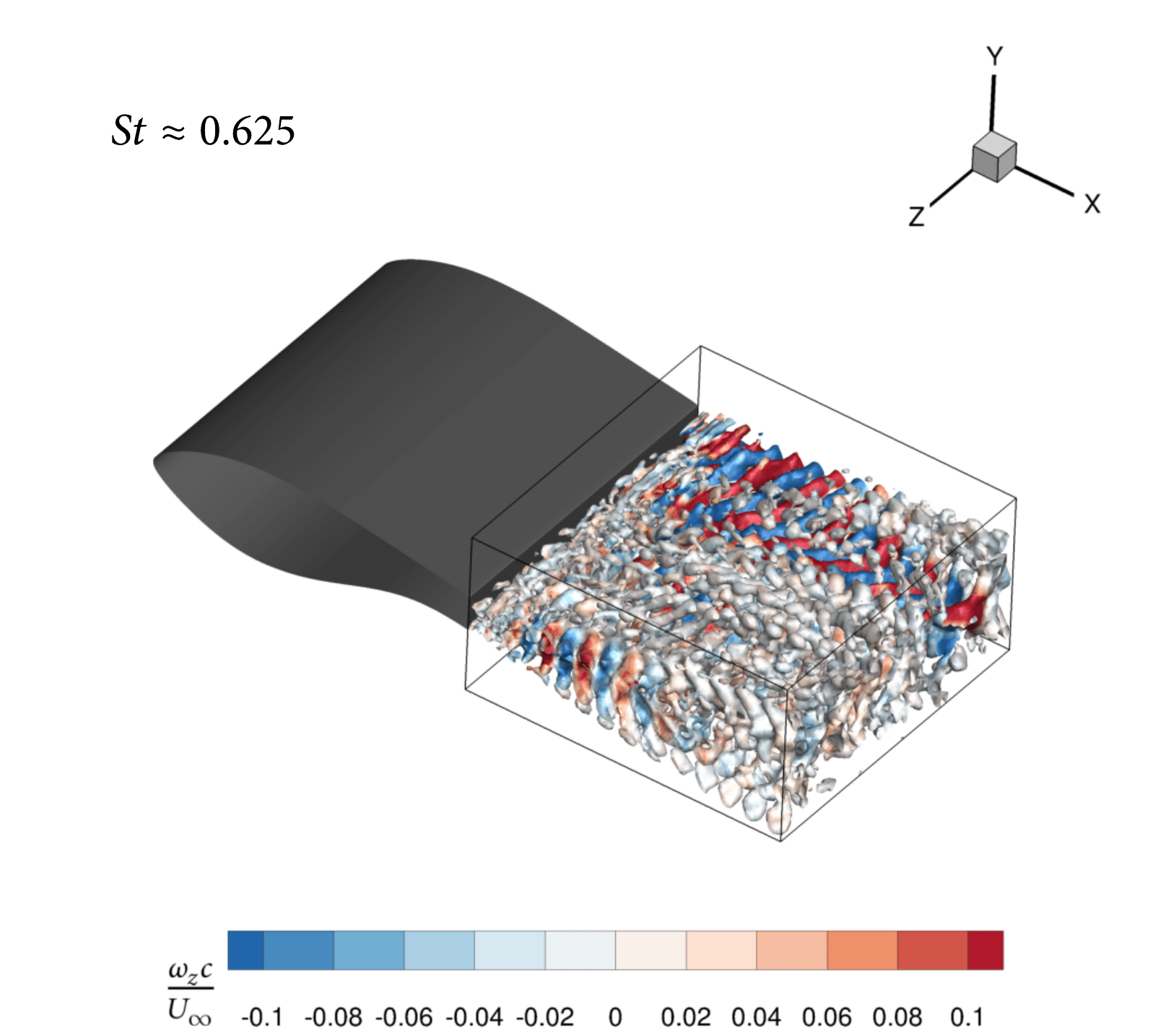}
\caption{$\alpha=0^{\circ}$ -- 69th mode}
\label{fig-mode69_0}
\end{subfigure}%
\begin{subfigure}{0.47\textwidth}
\centering
\includegraphics[width=\linewidth]{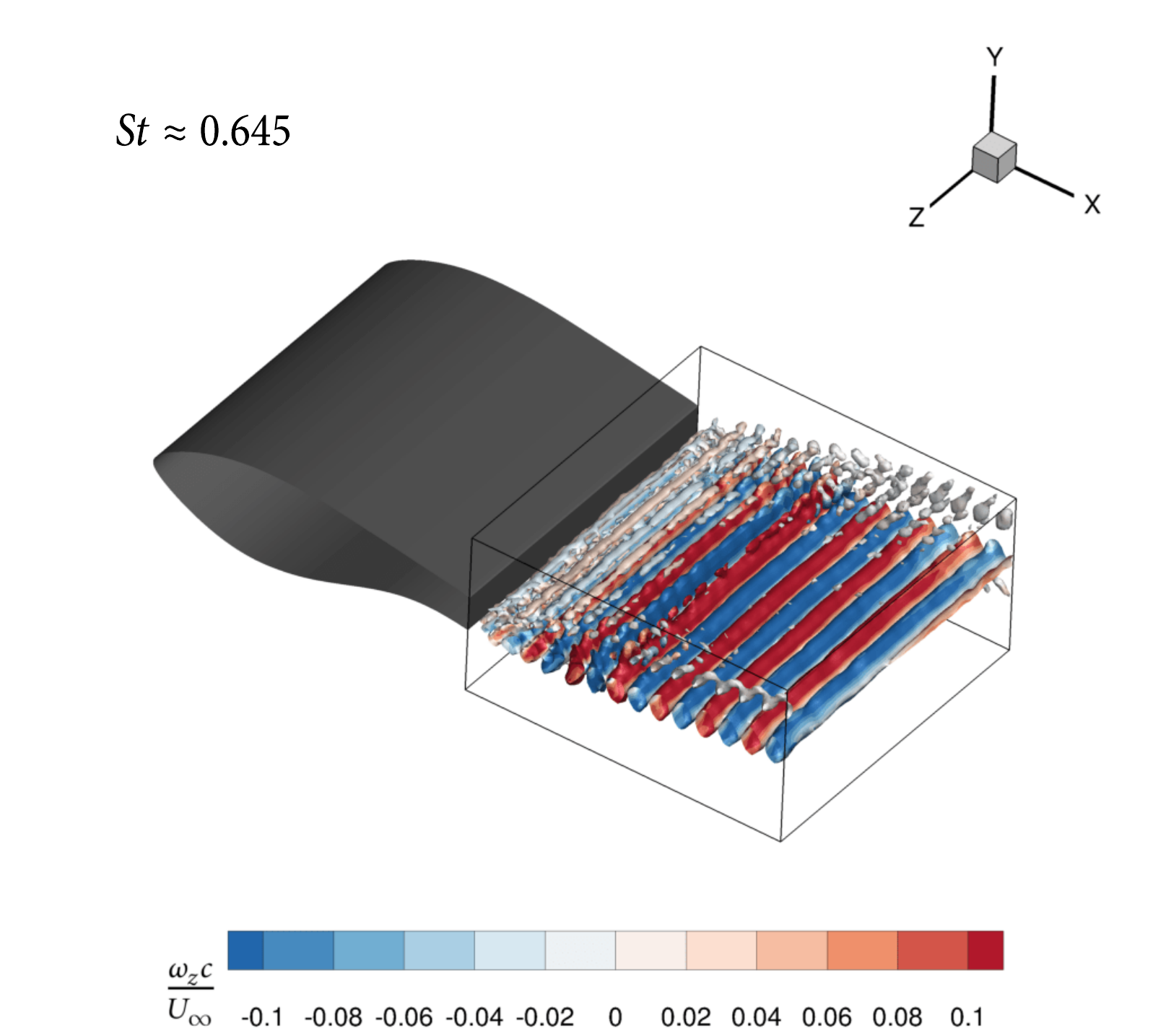}
\caption{$\alpha=12^{\circ}$ -- 9th mode}
\label{fig-mode9_12}
\end{subfigure}%
\caption{Q-criterion isosurfaces colored by
$\omega_{z}c/U_{\infty}$ for the second group of mPOD modes, corresponding to the first and second harmonics of the primary instability.
Panels (a) and (c) show the first two modes belonging to this group for $\alpha=0^{\circ}$, and panels (b) and (d) show the same for the $\alpha=12^{\circ}$ case.}
\label{fig-modes2}
\end{figure}

\autoref{fig-modes2} illustrates the group of mPOD modes corresponding to the primary instability's first and second harmonics. The difference in the leading $St$ numbers ($St\approx0.413$ versus $St\approx0.430$ for the high-drag and low-drag cases respectively) between the two cases is minimal. The same is true for the second harmonics ($St\approx0.625$ and $St\approx0.645$ respectively).

\begin{figure}[h]
\begin{subfigure}{0.47\textwidth}
\centering
\includegraphics[width=\linewidth]{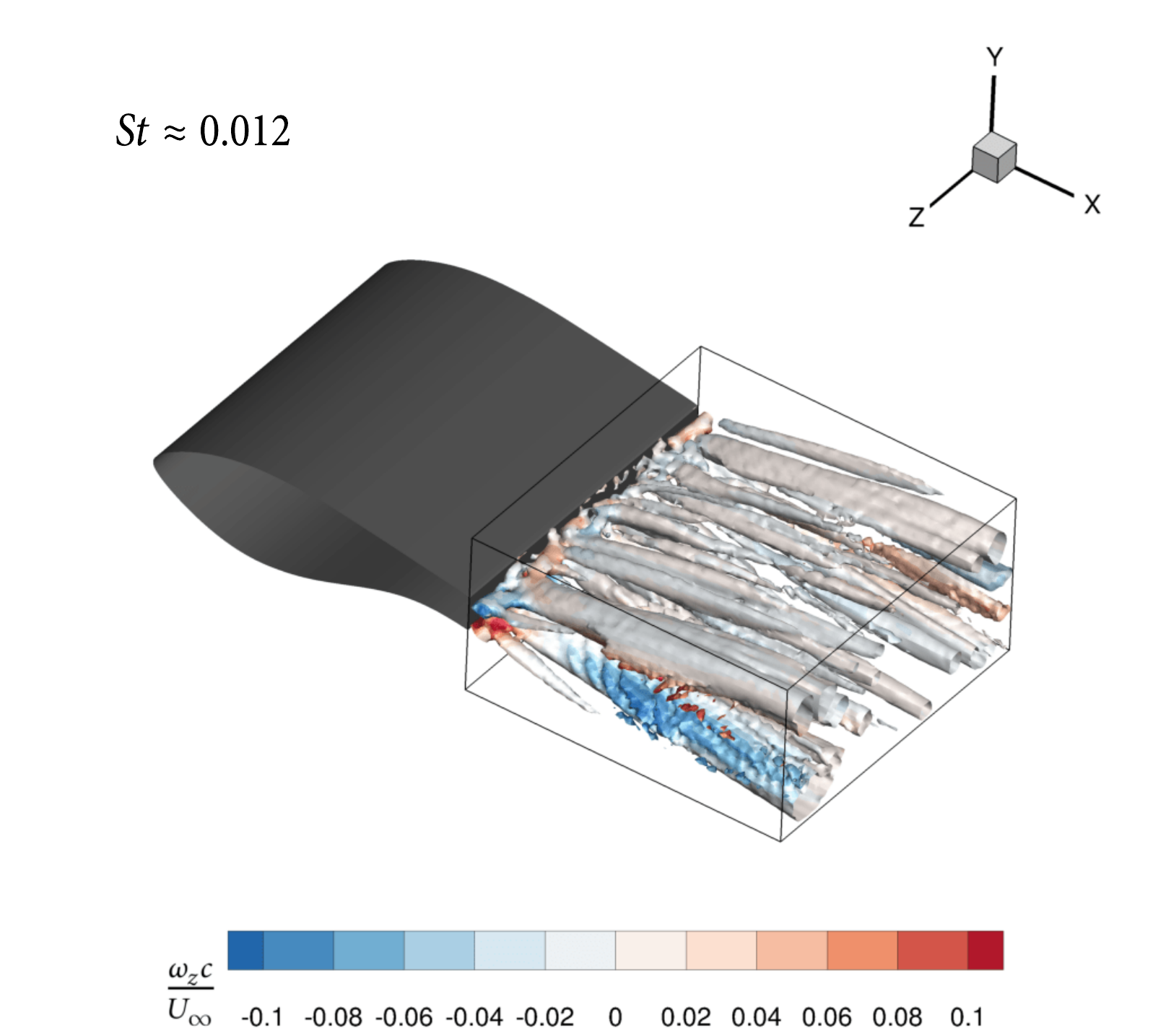}
\caption{$\alpha=0^{\circ}$ -- 11th mode}
\label{fig-mode11_0}
\end{subfigure}%
\begin{subfigure}{0.47\textwidth}
\centering
\includegraphics[width=\linewidth]{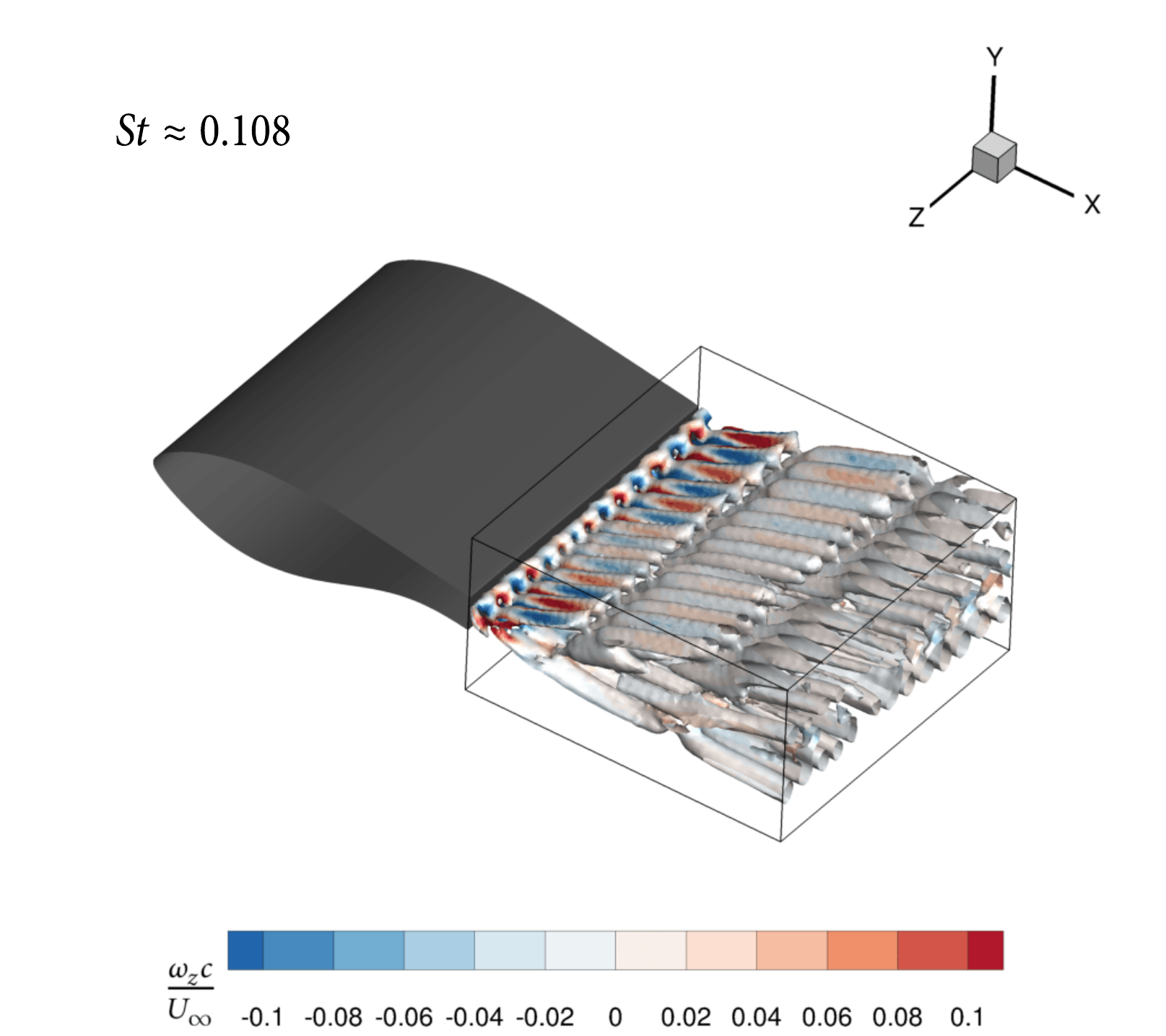}
\caption{$\alpha=12^{\circ}$ -- 7th mode}
\label{fig-mode7_12}
\end{subfigure}%
\\
\begin{subfigure}{0.47\textwidth}
\centering
\includegraphics[width=\linewidth]{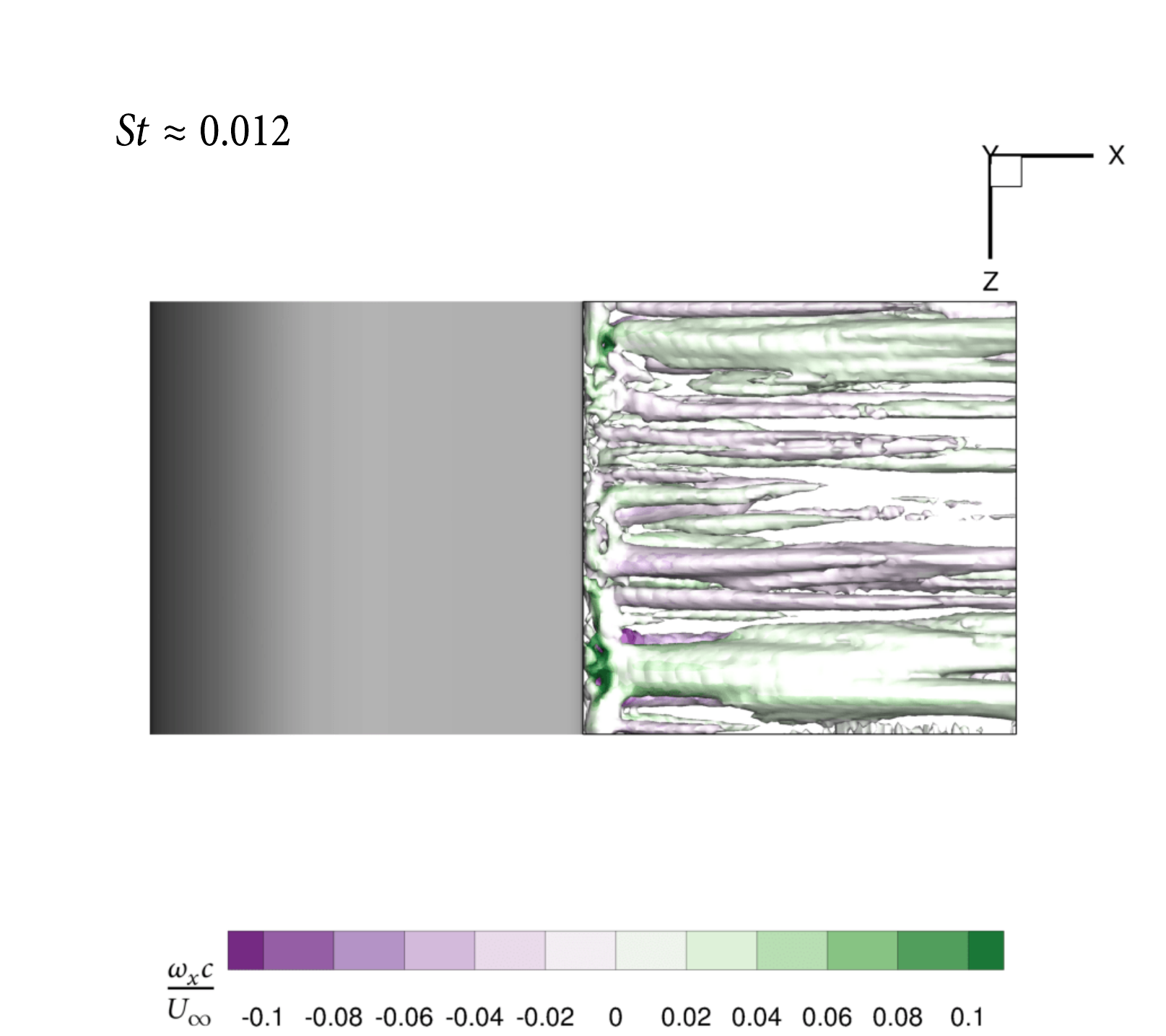}
\caption{$\alpha=0^{\circ}$ -- 11th mode}
\label{fig-mode11_0_X_top}
\end{subfigure}%
\begin{subfigure}{0.47\textwidth}
\centering
\includegraphics[width=\linewidth]{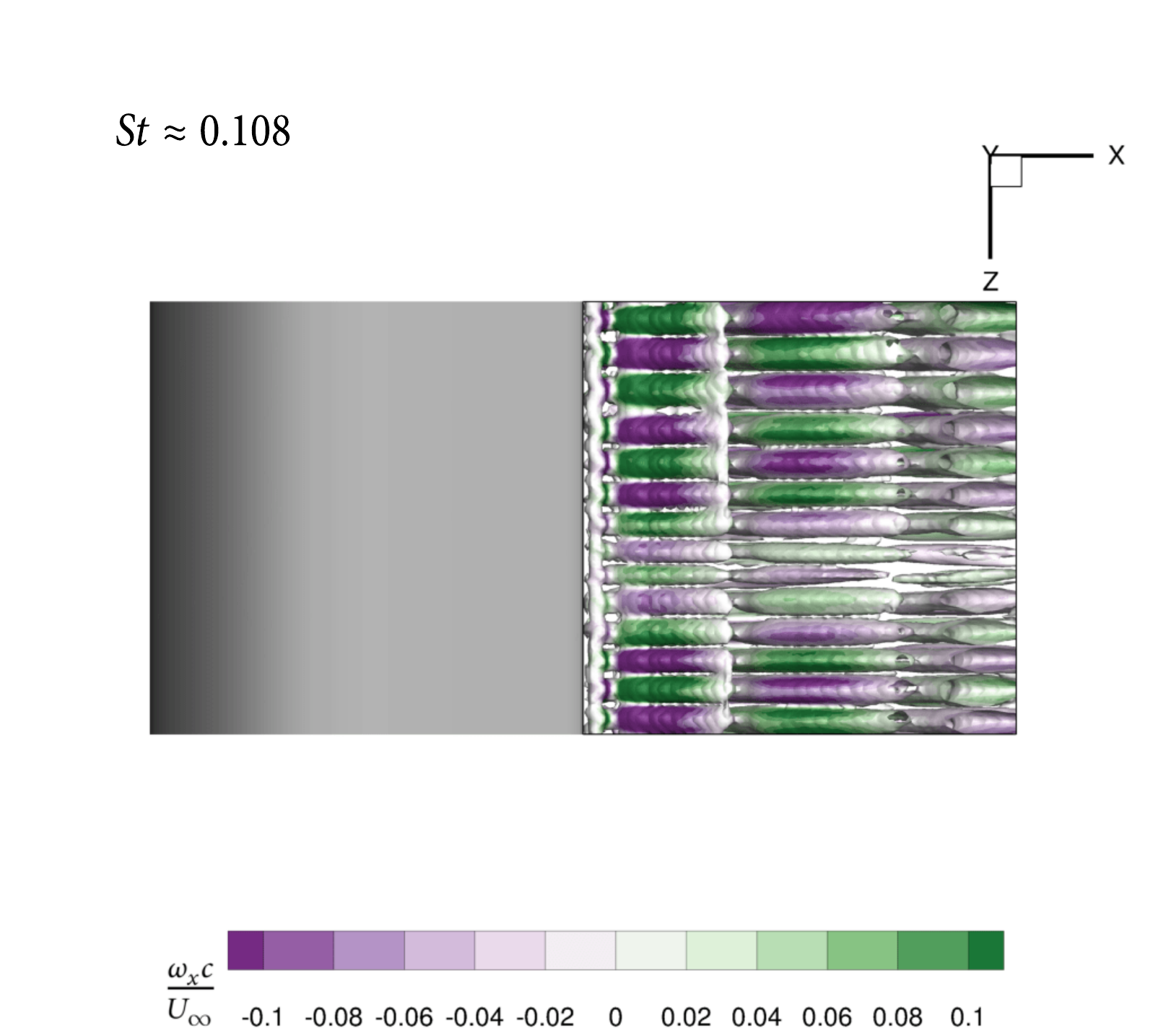}
\caption{$\alpha=12^{\circ}$ -- 7th mode}
\label{fig-mode7_12_X_top}
\end{subfigure}%
\caption{Q-criterion isosurfaces for the third group of mPOD modes. The left column shows the first mPOD mode for $\alpha=0^{\circ}$ and the right columns for $\alpha=12^{\circ}$.
The isosurfaces are colored by $\omega_{z}c/U_{\infty}$ for the first row and by $\omega_{x}c/U_{\infty}$ for the second row.}
\label{fig-modes3}
\end{figure}

The third group of mPOD modes corresponds to the secondary instability present in the wake, i.e. streamwise vortical pairs connecting the B\'enard-von K\'arm\'an vortices. 
The first modes exhibiting this behavior for both cases are shown in \autoref{fig-modes3}. 
In order to better illustrate the nature of these streamwise vortices, associated with the secondary instabilities, Q-criterion isosurfaces colored by streamwise vorticity are shown in
\autoref{fig-mode11_0_X_top} and \autoref{fig-mode7_12_X_top}.

The identified behavior is different for the two cases.
Ten counter-rotating vortex pairs can be identified for the high-drag case $\left(\alpha=0^{\circ}\right)$,
even though the vortex pairs are convoluted. This leads to a visual estimation of
$\lambda_{z}/h_{TE}\approx1.0$ and the $St$ number associated with this mode is low ($St\approx0.012$).
In addition, the rotational direction of these coherent structures
does not change along the streamwise direction.

For the low-drag case $\left(\alpha=12^{\circ}\right)$, by simple inspection of \autoref{fig-mode7_12_X_top},
seven counter-rotating vortex pairs can be identified, leading to a visual
estimation of $\lambda_{z}/h_{TE}=1.4$.
This is in agreement with the estimation in \autoref{fig-OMEGAX_TOP_12} and
the statistical analysis presented in \autoref{intra-pair-spanwise-distance-1} and \autoref{spanwise-distance-between-vortex-pairs-1}.
The dominant Strouhal number is $St\approx0.108$, i.e., a subharmonic of the dominant one. Also, it is noted that there is a streamwise change in the rotational direction
that corresponds to the different primary vortices that are connected
with the braids. The sub-harmonic Strouhal number does not
hint at some different wake dynamics but is explained by recalling
\autoref{fig-2WAKESNAP_X_12} and \autoref{fig-mode1_12}. In the
former, as discussed in \autoref{secondary-instability-characterization}, it is evident
that two complete cycles of the primary instability are required to
conclude a complete cycle of the secondary instability, i.e., shedding
of braids from the same side of TE in different directions; therefore,
the frequency corresponding to the secondary instabilities is a
subharmonic of the primary. In the latter, two fully developed vortex
pairs and one vortex shed from the TE are identified for the modes
corresponding to the primary instability. From a modal analysis point of
view, considering that the streamwise braids connect two consecutive
shed vortices, it is evident that the frequency associated with their
spatial behavior will be a subharmonic of the primary shedding frequency
in order to allow for pairs with alternating rotational directions to
appear.

\begin{figure}[h]
\begin{subfigure}{0.47\textwidth}
\centering
\includegraphics[width=\linewidth]{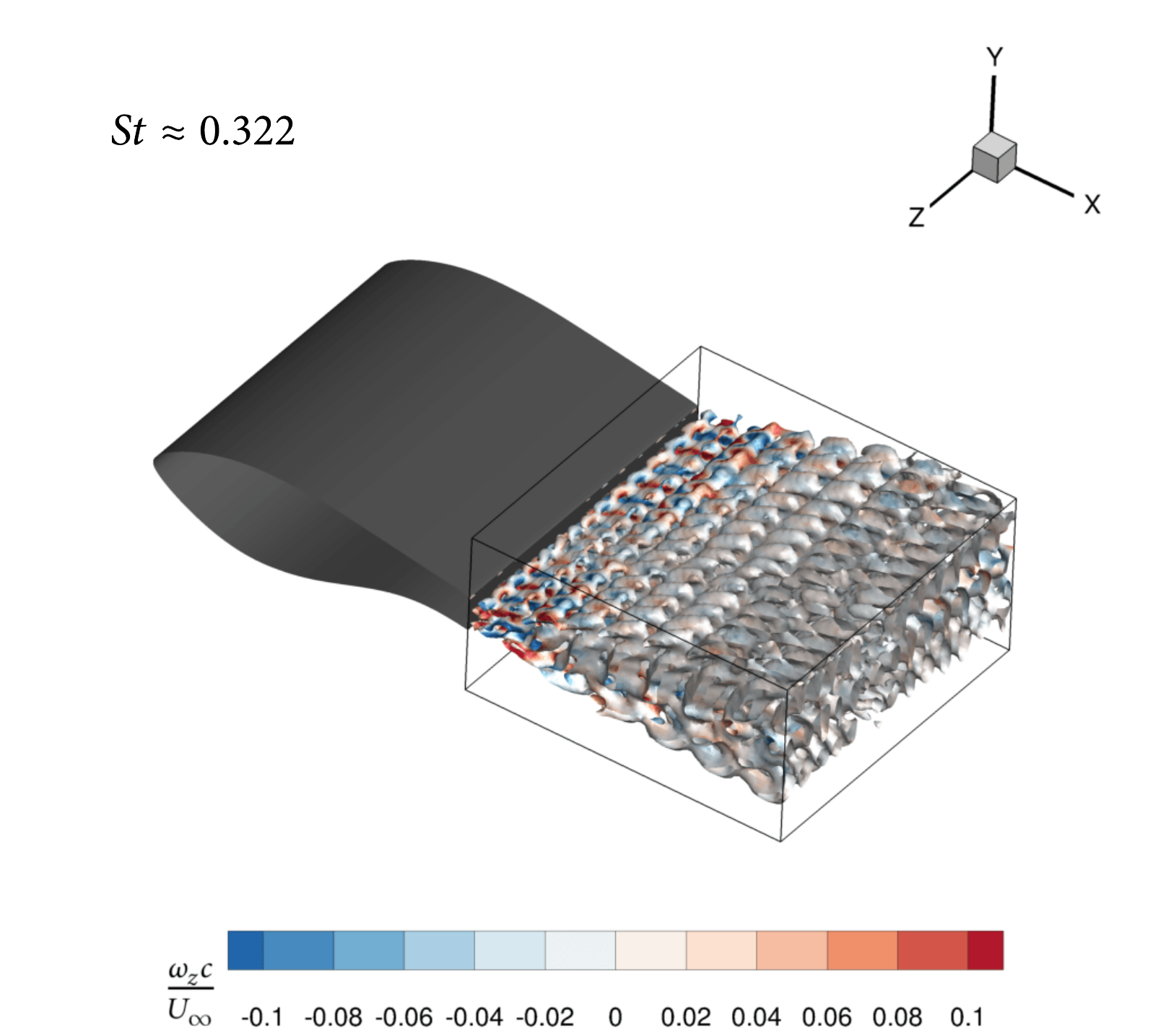}
\caption{$\alpha=12^{\circ}$ -- 13th mode}
\label{fig-mode13_12}
\end{subfigure}%
\begin{subfigure}{0.47\textwidth}
\centering
\includegraphics[width=\linewidth]{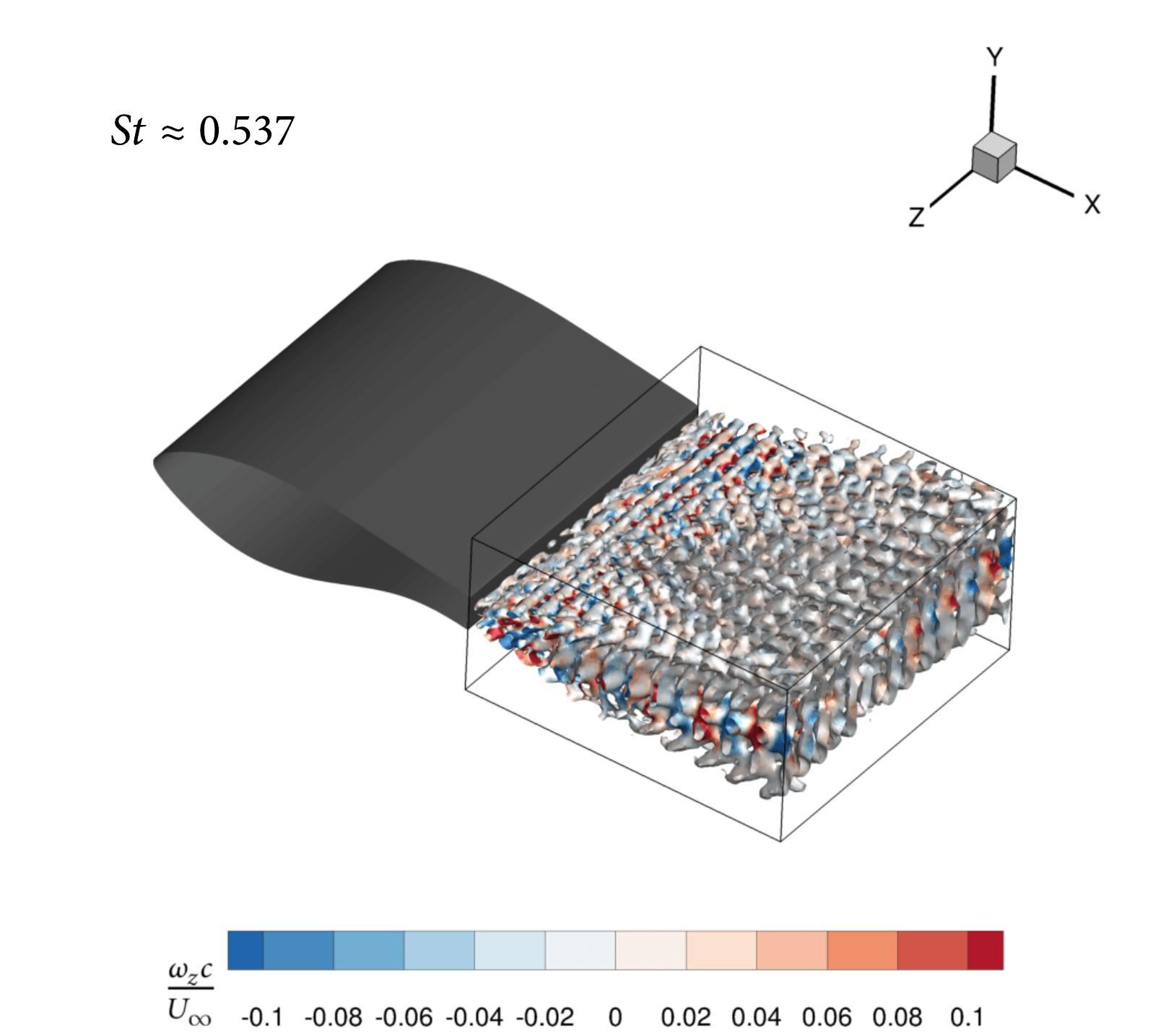}
\caption{$\alpha=12^{\circ}$ -- 25th mode}
\label{fig-mode25_12}
\end{subfigure}%
\\
\begin{subfigure}{0.47\textwidth}
   \centering
   \includegraphics[width=\linewidth]{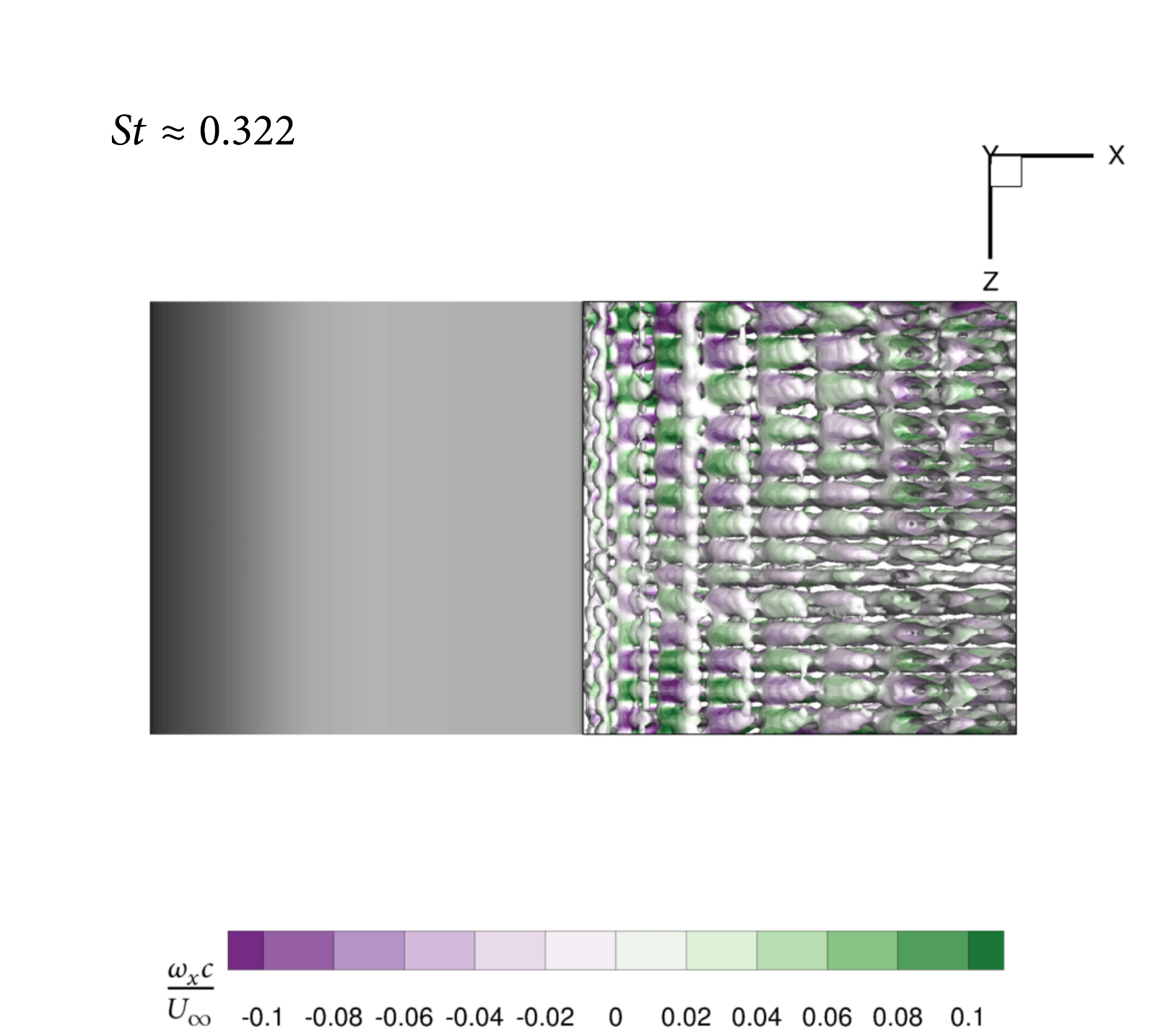}
   \caption{$\alpha=12^{\circ}$ -- 13th mode}
   \label{fig-mode13_12_X}
   \end{subfigure}%
   \begin{subfigure}{0.47\textwidth}
   \centering
   \includegraphics[width=\linewidth]{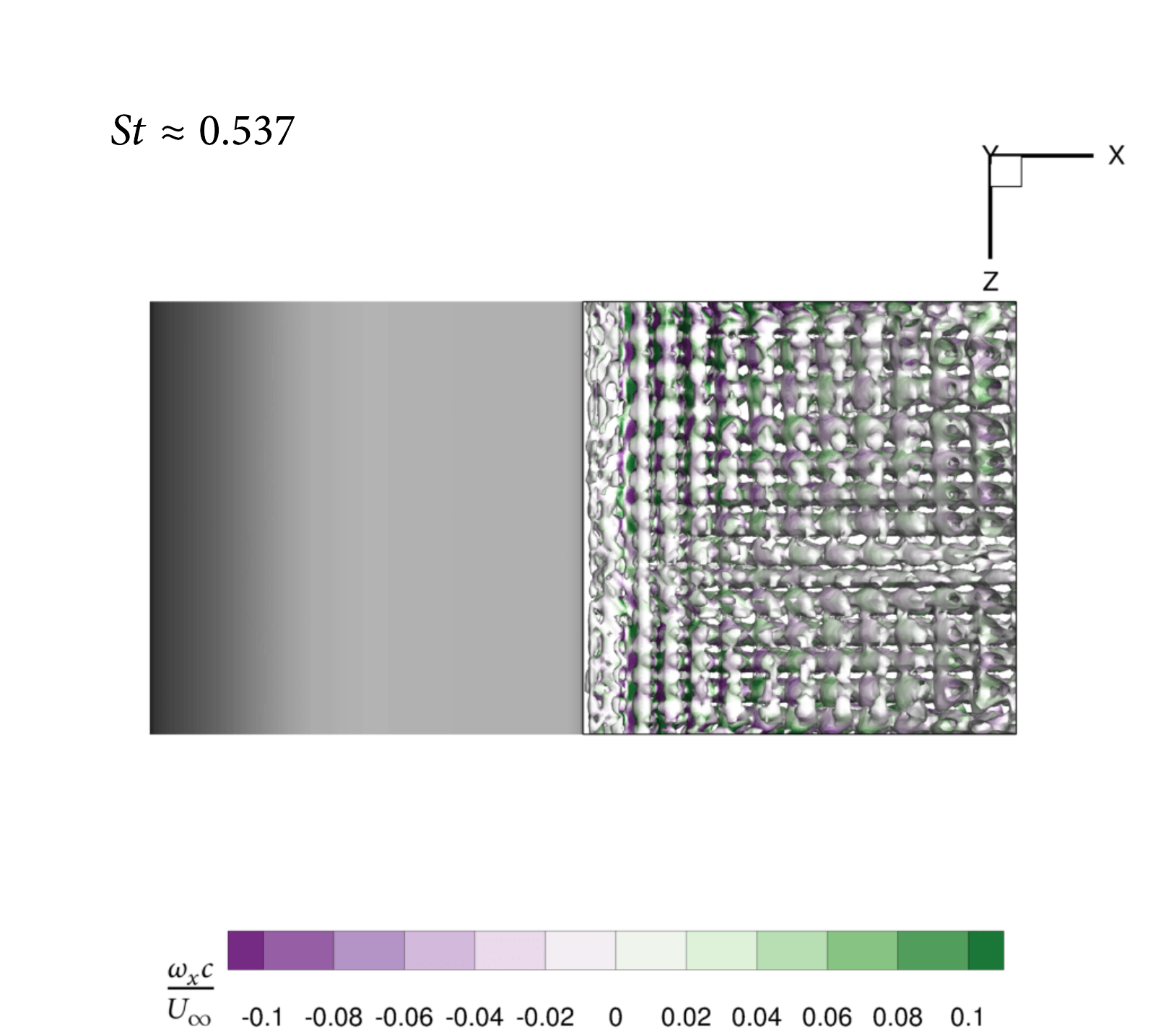}
   \caption{$\alpha=12^{\circ}$ -- 25th mode}
   \label{fig-mode25_12_X}
   \end{subfigure}%
\caption{Q-criterion isosurfaces colored by
$\omega_{z}c/U_{\infty}$ for the fourth group of mPOD modes, corresponding to the first and second harmonics of the secondary instability, only for the $\alpha=12^{\circ}$ case.
The isosurfaces in panels (a) and (c) show the first harmonic colored by $\omega_{z}c/U_{\infty}$ and $\omega_{x}c/U_{\infty}$, respectively.
Similarly, panels (b) and (d) show the second harmonic colored by $\omega_{z}c/U_{\infty}$ and $\omega_{x}c/U_{\infty}$, respectively.}
\label{fig-modes4}
\end{figure}

The final group contains the harmonics of the secondary instabilities. 
The existence of this type of modes is observed only in the low-drag case. The modes representing the first two harmonics are presented in \autoref{fig-modes4}. 
The inspection of \autoref{fig-mode13_12_X} and \autoref{fig-mode25_12_X} reveals that these modes represent the same number of streamwise counter-rotating vertical pairs (7 vertical pairs for the low-drag case) and therefore have the same spanwise wavelength as the second group of modes, i.e. the modes that correspond to the secondary instability.
The harmonic characterization of these modes stems from the number of observed changes in rotational direction along the streamwise direction.
Namely, 4 pairs with different streamwise vorticity appear as one moves downstream of the TE in \autoref{fig-mode7_12_X_top}, while for the harmonics, 8 and 16 appear in \autoref{fig-mode13_12_X} and \autoref{fig-mode25_12_X} respectively.

\section{Summary and conclusions}\label{sec-Conclusions}

In this paper, the low-drag regime for FB airfoils observed in \cite{Smith1950,Winnemoeller2007,Post2008,Baker2008,Barone2009,Manolesos2016,Soerensen2011,Xu2014} and first identified in \cite{Manolesos2021} was investigated using high-fidelity numerical simulations.
Namely, IDDES simulations were carried out for a range of
AoA at $\mathrm{Re}=1.5\times10^{6}$. Firstly, an investigation of the
BL characteristics showed that the effective BL thickness increases as
the AoA increases, leading to better base pressure recovery (up to
$44\%$ relative increase of $C_{p,b}$ compared to $\alpha=0^{\circ}$),
reducing the drag coefficient. Next, the focus was put on two AoA, one
outside $\left(\alpha=0^{\circ}\right)$ and one inside $\alpha=12^{\circ}$ the low-drag regime.
To characterize the secondary instabilities in the flow, two spanwise distances were estimated: the distance between two vortices of the same vortex pair and the distance
between adjacent pairs.
Finally, multiscale Proper Orthogonal Decomposition was used to decompose the three-dimensional wake into three-dimensional elemental coherent structures using the downsampled dataset.

The behavior of the upstream BL thickness was investigated, as it has
been linked with base pressure recovery and, consequently, drag reduction \cite{Durgesh2013}. The results show that, as the AoA increases, the BL on the pressure side of the airfoil becomes thinner; however, a significant increase was observed on the suction side due to the adverse pressure gradient. Their combination resulted in a thicker effective BL with an increase in AoA, reducing the base pressure coefficient and increasing the vortex formation length. These findings are in line with previous studies conducted for elongated bluff bodies \cite{Petrusma1994,Rowe2001,Mariotti2013,Durgesh2013}
at lower $Re_{h_{TE}}$ and provide a rudimentary framework for flow control.
Further analysis of lift force coefficients and base pressure frequency spectra showed a reduction in the shedding strength. The shedding intensity reduction with an increase in BL thickness is in line with the one discussed in
\cite{Rowe2001,ElGammal2007,Durgesh2013}, as thicker separating
boundary leads to a greater shedding organization and weaker vortex
strength and consequently to base-drag reduction \cite{Durgesh2013}.
Therefore, the observed base-drag reduction is attributed to greater shedding organization and weaker vortex strength caused by the thicker separating BL.
The change of the BL at separation represents a distinct base-drag reduction mechanism compared to approaches that aim to disrupt vortex shedding or limit vortex interactions \cite{Durgesh2013} that can be potentially utilized.    

This work also presented the statistical estimation of two metrics describing the spanwise distances between streamwise vortex pairs. Specifically, the intra-pair distance $\left(\Lambda_{z}\right)$
between the counter-rotating braids was similar for both AoA, ranging
from $\Lambda_{z}/h_{TE}=0.35-0.65$. The determination of spanwise vortex pair distance was most accurate for $\alpha=12^{\circ}$, yielding $\lambda_{z}/h_{TE}\approx1.4$, in agreement with \cite{Manolesos2021}, who arrived at similar results through the inspection of streamwise velocity
undulations.

Finally, normalized streamwise vorticity $\omega_{x}c/U_{\infty}$ isosurfaces for a time interval of 2000 timesteps --approximately eight
vortex shedding cycles-- showed strong indication that the Mode S$^{\prime}$ predicted for an elongated bluff body wake \cite{Ryan2005} is present in the wake of the FB airfoil for the low drag case $\left(\alpha=12^{\circ}\right)$. This conclusion is drawn from the observation that the rotational direction of the streamwise vortex pair alternates after each full vortex-shedding cycle of the primary B\'enard-von K\'arm\'an vortices. These isosurfaces also allows for visually estimating $\lambda_{z}/h_{TE}\approx1.4$. Furthermore, normalizing with
effective thickness $h_{TE}^{\prime}$, which accounts for BL thickness, one gets $\lambda_{z}/h_{TE}^{\prime}\approx1.0$. This is in excellent agreement with \citet{Ryan2005} and hints at the influence of the upstream BL to the behavior of the secondary instability \cite{Gibeau2020}.

The three-dimensional instantaneous Q-criterion isosurfaces show that the wake is more organized at the low-drag case $\left(\alpha=12^{\circ}\right)$ compared to the high-drag case $\left(\alpha=0^{\circ}\right)$. This behavior is consistent with findings in the literature on base drag reduction \cite{Durgesh2013}, while large-scale distortions are evident, resembling oblique shedding
behavior. The analysis of the three-dimensional mPOD modes, their associated dominant Strouhal numbers, and associated energy reinforced this observation and showed a large number of modes associated with the
primary spanwise instability for the high drag regime case. These modes indicate the complex undulations of the spanwise vortices for $\alpha=0^{\circ}$. The mPOD modes also provide a visual way to identify the
spanwise correlation length using the coherent structures, where the streamwise counter-rotating vortex pairs were observed. The results are more conclusive for $\alpha=12^{\circ}$ because of the more distinct flow organization and align well with the statistical analysis using the spanwise autocorrelation of $\Gamma_{1}$, revealing the presence of a Mode S$^{\prime}$ type secondary instability.

In summary, increasing the AoA results in a thicker the effective BL, which is subsequently linked to a drop in drag and a more organized three-dimensional wake. To the authors' knowledge, this study is the first to establish a direct link between the state of the upstream BL and the drag of a flatblack airfoil.
Additionally, the wake is more organized, and the shedding intensity decreases-- a behavior also correlated with a decrease in
base-pressure drag for elongated bluff bodies for similar Reynolds numbers $\left(\mathrm{Re}_{h_{TE}}=0.5-1.6\times10^{4}\right)$.
Finally, the upstream BL influence the secondary instability, here identified as Mode S$^{\prime}$ at the low-drag case $\left(\alpha=12^{\circ}\right)$. This leads to the conclusion that flow control strategies for drag reduction could leverage the increase in the BL momentum thickness to achieve better base pressure recovery. 

\bibliography{bibliography}% Produces the bibliography via BibTeX.

\begin{thebibliography}{66}%
\makeatletter
\providecommand \@ifxundefined [1]{%
 \@ifx{#1\undefined}
}%
\providecommand \@ifnum [1]{%
 \ifnum #1\expandafter \@firstoftwo
 \else \expandafter \@secondoftwo
 \fi
}%
\providecommand \@ifx [1]{%
 \ifx #1\expandafter \@firstoftwo
 \else \expandafter \@secondoftwo
 \fi
}%
\providecommand \natexlab [1]{#1}%
\providecommand \enquote  [1]{``#1''}%
\providecommand \bibnamefont  [1]{#1}%
\providecommand \bibfnamefont [1]{#1}%
\providecommand \citenamefont [1]{#1}%
\providecommand \href@noop [0]{\@secondoftwo}%
\providecommand \href [0]{\begingroup \@sanitize@url \@href}%
\providecommand \@href[1]{\@@startlink{#1}\@@href}%
\providecommand \@@href[1]{\endgroup#1\@@endlink}%
\providecommand \@sanitize@url [0]{\catcode `\\12\catcode `\$12\catcode
  `\&12\catcode `\#12\catcode `\^12\catcode `\_12\catcode `\%12\relax}%
\providecommand \@@startlink[1]{}%
\providecommand \@@endlink[0]{}%
\providecommand \url  [0]{\begingroup\@sanitize@url \@url }%
\providecommand \@url [1]{\endgroup\@href {#1}{\urlprefix }}%
\providecommand \urlprefix  [0]{URL }%
\providecommand \Eprint [0]{\href }%
\providecommand \doibase [0]{https://doi.org/}%
\providecommand \selectlanguage [0]{\@gobble}%
\providecommand \bibinfo  [0]{\@secondoftwo}%
\providecommand \bibfield  [0]{\@secondoftwo}%
\providecommand \translation [1]{[#1]}%
\providecommand \BibitemOpen [0]{}%
\providecommand \bibitemStop [0]{}%
\providecommand \bibitemNoStop [0]{.\EOS\space}%
\providecommand \EOS [0]{\spacefactor3000\relax}%
\providecommand \BibitemShut  [1]{\csname bibitem#1\endcsname}%
\let\auto@bib@innerbib\@empty
%</preamble>
\bibitem [{\citenamefont {Baker}\ \emph {et~al.}(2006)\citenamefont {Baker},
  \citenamefont {Mayda},\ and\ \citenamefont {Van~Dam}}]{Baker2006}%
  \BibitemOpen
  \bibfield  {author} {\bibinfo {author} {\bibfnamefont {J.~P.}\ \bibnamefont
  {Baker}}, \bibinfo {author} {\bibfnamefont {E.~A.}\ \bibnamefont {Mayda}},\
  and\ \bibinfo {author} {\bibfnamefont {C.~P.}\ \bibnamefont {Van~Dam}},\
  }\bibfield  {title} {\bibinfo {title} {Experimental analysis of thick blunt
  trailing-edge wind turbine airfoils},\ }\href
  {https://doi.org/10.1115/1.2346701} {\bibfield  {journal} {\bibinfo
  {journal} {Journal of Solar Energy Engineering}\ }\textbf {\bibinfo {volume}
  {128}},\ \bibinfo {pages} {422} (\bibinfo {year} {2006})}\BibitemShut
  {NoStop}%
\bibitem [{\citenamefont {Griffith}\ and\ \citenamefont
  {Richards}(2014)}]{Griffith2014}%
  \BibitemOpen
  \bibfield  {author} {\bibinfo {author} {\bibfnamefont {D.}~\bibnamefont
  {Griffith}}\ and\ \bibinfo {author} {\bibfnamefont {P.}~\bibnamefont
  {Richards}},\ }\href {https://www.osti.gov/servlets/purl/1159116/} {\emph
  {\bibinfo {title} {The {{SNL100-03 Blade}}: {{Design Studies}} with
  {{Flatback Airfoils}} for the {{Sandia}} 100-Meter {{Blade}}.}}},\ \bibinfo
  {type} {Tech. Rep.}\ \bibinfo {number} {SAND2014-18129, 1159116, 537751}\
  (\bibinfo  {institution} {Sandia National Lab},\ \bibinfo {year}
  {2014})\BibitemShut {NoStop}%
\bibitem [{\citenamefont {Barone}\ \emph {et~al.}(2009)\citenamefont {Barone},
  \citenamefont {Berg}, \citenamefont {Devenport},\ and\ \citenamefont
  {Burdisso}}]{Barone2009}%
  \BibitemOpen
  \bibfield  {author} {\bibinfo {author} {\bibfnamefont {M.~F.}\ \bibnamefont
  {Barone}}, \bibinfo {author} {\bibfnamefont {D.~E.}\ \bibnamefont {Berg}},
  \bibinfo {author} {\bibfnamefont {W.~J.}\ \bibnamefont {Devenport}},\ and\
  \bibinfo {author} {\bibfnamefont {R.~A.}\ \bibnamefont {Burdisso}},\ }\href
  {https://doi.org/10.2172/1504612} {\emph {\bibinfo {title} {Aerodynamic and
  Aeroacoustic Tests of a Flatback Version of the {{DU97-W-300}} Airfoil}}},\
  \bibinfo {type} {Tech. Rep.}\ \bibinfo {number} {SAND--2009-4185, 1504612}\
  (\bibinfo  {institution} {Sandia National Laboratories},\ \bibinfo {year}
  {2009})\BibitemShut {NoStop}%
\bibitem [{\citenamefont {Manolesos}\ and\ \citenamefont
  {Voutsinas}(2016)}]{Manolesos2016}%
  \BibitemOpen
  \bibfield  {author} {\bibinfo {author} {\bibfnamefont {M.}~\bibnamefont
  {Manolesos}}\ and\ \bibinfo {author} {\bibfnamefont {S.~G.}\ \bibnamefont
  {Voutsinas}},\ }\bibfield  {title} {\bibinfo {title} {Experimental study of
  drag-reduction devices on a flatback airfoil},\ }\href
  {https://doi.org/10.2514/1.j054901} {\bibfield  {journal} {\bibinfo
  {journal} {AIAA Journal}\ }\textbf {\bibinfo {volume} {54}},\ \bibinfo
  {pages} {3382} (\bibinfo {year} {2016})}\BibitemShut {NoStop}%
\bibitem [{\citenamefont {Papadakis}\ and\ \citenamefont
  {Manolesos}(2020)}]{Papadakis2020}%
  \BibitemOpen
  \bibfield  {author} {\bibinfo {author} {\bibfnamefont {G.}~\bibnamefont
  {Papadakis}}\ and\ \bibinfo {author} {\bibfnamefont {M.}~\bibnamefont
  {Manolesos}},\ }\bibfield  {title} {\bibinfo {title} {The flow past a
  flatback airfoil with flow control devices: Benchmarking numerical
  simulations against wind tunnel data},\ }\href
  {https://doi.org/10.5194/wes-5-911-2020} {\bibfield  {journal} {\bibinfo
  {journal} {Wind Energy Science}\ }\textbf {\bibinfo {volume} {5}},\ \bibinfo
  {pages} {911} (\bibinfo {year} {2020})}\BibitemShut {NoStop}%
\bibitem [{\citenamefont {Stone}\ \emph {et~al.}(2009)\citenamefont {Stone},
  \citenamefont {Barone}, \citenamefont {Smith},\ and\ \citenamefont
  {Lynch}}]{Stone2009}%
  \BibitemOpen
  \bibfield  {author} {\bibinfo {author} {\bibfnamefont {C.}~\bibnamefont
  {Stone}}, \bibinfo {author} {\bibfnamefont {M.}~\bibnamefont {Barone}},
  \bibinfo {author} {\bibfnamefont {M.}~\bibnamefont {Smith}},\ and\ \bibinfo
  {author} {\bibfnamefont {E.}~\bibnamefont {Lynch}},\ }\bibfield  {title}
  {\bibinfo {title} {A {{Computational Study}} of the {{Aerodynamics}} and
  {{Aeroacoustics}} of a {{Flatback Airfoil Using Hybrid RANS-LES}}},\ }in\
  \href {https://doi.org/10.2514/6.2009-273} {\emph {\bibinfo {booktitle} {47th
  {{AIAA Aerospace Sciences Meeting}} Including {{The New Horizons Forum}} and
  {{Aerospace Exposition}}}}}\ (\bibinfo  {publisher} {{American Institute of
  Aeronautics and Astronautics}},\ \bibinfo {address} {Orlando, Florida},\
  \bibinfo {year} {2009})\BibitemShut {NoStop}%
\bibitem [{\citenamefont {Doosttalab}\ and\ \citenamefont
  {Frommann}(2019)}]{Doosttalab2019}%
  \BibitemOpen
  \bibfield  {author} {\bibinfo {author} {\bibfnamefont {M.}~\bibnamefont
  {Doosttalab}}\ and\ \bibinfo {author} {\bibfnamefont {O.}~\bibnamefont
  {Frommann}},\ }\bibfield  {title} {\bibinfo {title} {Multidisciplinary
  {{Design}} and {{Verification}} of the {{HB Flatback Airfoil Family}}},\
  }\href {https://doi.org/10.2514/1.J057684} {\bibfield  {journal} {\bibinfo
  {journal} {AIAA Journal}\ }\textbf {\bibinfo {volume} {57}},\ \bibinfo
  {pages} {4639} (\bibinfo {year} {2019})}\BibitemShut {NoStop}%
\bibitem [{\citenamefont {Fuchs}\ \emph {et~al.}(2022)\citenamefont {Fuchs},
  \citenamefont {Weihing}, \citenamefont {Kuehn}, \citenamefont {Herr},
  \citenamefont {Suryadi}, \citenamefont {Mockett}, \citenamefont
  {Knobbe-Eschen}, \citenamefont {Kramer},\ and\ \citenamefont
  {Knacke}}]{Fuchs2022}%
  \BibitemOpen
  \bibfield  {author} {\bibinfo {author} {\bibfnamefont {M.}~\bibnamefont
  {Fuchs}}, \bibinfo {author} {\bibfnamefont {P.}~\bibnamefont {Weihing}},
  \bibinfo {author} {\bibfnamefont {T.}~\bibnamefont {Kuehn}}, \bibinfo
  {author} {\bibfnamefont {M.}~\bibnamefont {Herr}}, \bibinfo {author}
  {\bibfnamefont {A.}~\bibnamefont {Suryadi}}, \bibinfo {author} {\bibfnamefont
  {C.}~\bibnamefont {Mockett}}, \bibinfo {author} {\bibfnamefont
  {H.}~\bibnamefont {Knobbe-Eschen}}, \bibinfo {author} {\bibfnamefont
  {F.}~\bibnamefont {Kramer}},\ and\ \bibinfo {author} {\bibfnamefont
  {T.}~\bibnamefont {Knacke}},\ }\bibfield  {title} {\bibinfo {title} {Two
  computational studies of a flatback airfoil using non-zonal and embedded
  scale-resolving turbulence modelling approaches},\ }in\ \href
  {https://doi.org/10.2514/6.2022-2860} {\emph {\bibinfo {booktitle} {28th
  {{AIAA}}/{{CEAS Aeroacoustics}} 2022 {{Conference}}}}}\ (\bibinfo
  {publisher} {{American Institute of Aeronautics and Astronautics}},\ \bibinfo
  {address} {Southampton, UK},\ \bibinfo {year} {2022})\BibitemShut {NoStop}%
\bibitem [{\citenamefont {Jaffar}\ \emph {et~al.}(2023)\citenamefont {Jaffar},
  \citenamefont {Al-Sadawi}, \citenamefont {Khudhair},\ and\ \citenamefont
  {Biedermann}}]{Jaffar2023}%
  \BibitemOpen
  \bibfield  {author} {\bibinfo {author} {\bibfnamefont {H.}~\bibnamefont
  {Jaffar}}, \bibinfo {author} {\bibfnamefont {L.}~\bibnamefont {Al-Sadawi}},
  \bibinfo {author} {\bibfnamefont {A.}~\bibnamefont {Khudhair}},\ and\
  \bibinfo {author} {\bibfnamefont {T.}~\bibnamefont {Biedermann}},\ }\bibfield
   {title} {\bibinfo {title} {Aerodynamics improvement of {{DU97-W-300}} wind
  turbine flat-back airfoil using slot-induced air jet},\ }\href
  {https://doi.org/10.1016/j.ijft.2022.100267} {\bibfield  {journal} {\bibinfo
  {journal} {International Journal of Thermofluids}\ }\textbf {\bibinfo
  {volume} {17}},\ \bibinfo {pages} {100267} (\bibinfo {year}
  {2023})}\BibitemShut {NoStop}%
\bibitem [{\citenamefont {Wang}\ \emph {et~al.}(2018)\citenamefont {Wang},
  \citenamefont {Zhang},\ and\ \citenamefont {Shen}}]{Wang2018}%
  \BibitemOpen
  \bibfield  {author} {\bibinfo {author} {\bibfnamefont {G.}~\bibnamefont
  {Wang}}, \bibinfo {author} {\bibfnamefont {L.}~\bibnamefont {Zhang}},\ and\
  \bibinfo {author} {\bibfnamefont {W.~Z.}\ \bibnamefont {Shen}},\ }\bibfield
  {title} {\bibinfo {title} {{{LES}} simulation and experimental validation of
  the unsteady aerodynamics of blunt wind turbine airfoils},\ }\href
  {https://doi.org/10.1016/j.energy.2018.06.093} {\bibfield  {journal}
  {\bibinfo  {journal} {Energy}\ }\textbf {\bibinfo {volume} {158}},\ \bibinfo
  {pages} {911} (\bibinfo {year} {2018})}\BibitemShut {NoStop}%
\bibitem [{\citenamefont {Papadakis}\ \emph {et~al.}(2020)\citenamefont
  {Papadakis}, \citenamefont {Manolesos}, \citenamefont {Diakakis},\ and\
  \citenamefont {Riziotis}}]{Papadakis2020a}%
  \BibitemOpen
  \bibfield  {author} {\bibinfo {author} {\bibfnamefont {G.}~\bibnamefont
  {Papadakis}}, \bibinfo {author} {\bibfnamefont {M.}~\bibnamefont
  {Manolesos}}, \bibinfo {author} {\bibfnamefont {K.}~\bibnamefont
  {Diakakis}},\ and\ \bibinfo {author} {\bibfnamefont {V.~A.}\ \bibnamefont
  {Riziotis}},\ }\bibfield  {title} {\bibinfo {title} {{{DES}} vs {{RANS}}:
  {{The}} flatback airfoil case},\ }\href
  {https://doi.org/10.1088/1742-6596/1618/5/052062} {\bibfield  {journal}
  {\bibinfo  {journal} {Journal of Physics: Conference Series}\ }\textbf
  {\bibinfo {volume} {1618}},\ \bibinfo {pages} {052062} (\bibinfo {year}
  {2020})}\BibitemShut {NoStop}%
\bibitem [{\citenamefont {Hongpeng}\ \emph {et~al.}(2020)\citenamefont
  {Hongpeng}, \citenamefont {Yu}, \citenamefont {Rujing}, \citenamefont
  {Peng},\ and\ \citenamefont {Qing}}]{Hongpeng2020}%
  \BibitemOpen
  \bibfield  {author} {\bibinfo {author} {\bibfnamefont {L.}~\bibnamefont
  {Hongpeng}}, \bibinfo {author} {\bibfnamefont {W.}~\bibnamefont {Yu}},
  \bibinfo {author} {\bibfnamefont {Y.}~\bibnamefont {Rujing}}, \bibinfo
  {author} {\bibfnamefont {X.}~\bibnamefont {Peng}},\ and\ \bibinfo {author}
  {\bibfnamefont {W.}~\bibnamefont {Qing}},\ }\bibfield  {title} {\bibinfo
  {title} {Influence of the modification of asymmetric trailing-edge thickness
  on the aerodynamic performance of a wind turbine airfoil},\ }\href
  {https://doi.org/10.1016/j.renene.2019.09.073} {\bibfield  {journal}
  {\bibinfo  {journal} {Renewable Energy}\ }\textbf {\bibinfo {volume} {147}},\
  \bibinfo {pages} {1623} (\bibinfo {year} {2020})}\BibitemShut {NoStop}%
\bibitem [{\citenamefont {Manolesos}\ and\ \citenamefont
  {Papadakis}(2021)}]{Manolesos2021}%
  \BibitemOpen
  \bibfield  {author} {\bibinfo {author} {\bibfnamefont {M.}~\bibnamefont
  {Manolesos}}\ and\ \bibinfo {author} {\bibfnamefont {G.}~\bibnamefont
  {Papadakis}},\ }\bibfield  {title} {\bibinfo {title} {Investigation of the
  three-dimensional flow past a flatback wind turbine airfoil at high angles of
  attack},\ }\href {https://doi.org/10.1063/5.0055822} {\bibfield  {journal}
  {\bibinfo  {journal} {Physics of Fluids}\ }\textbf {\bibinfo {volume} {33}},\
  \bibinfo {pages} {085106} (\bibinfo {year} {2021})}\BibitemShut {NoStop}%
\bibitem [{\citenamefont {Bangga}\ \emph {et~al.}(2022)\citenamefont {Bangga},
  \citenamefont {Seel}, \citenamefont {Lutz},\ and\ \citenamefont
  {Kühn}}]{Bangga2022}%
  \BibitemOpen
  \bibfield  {author} {\bibinfo {author} {\bibfnamefont {G.}~\bibnamefont
  {Bangga}}, \bibinfo {author} {\bibfnamefont {F.}~\bibnamefont {Seel}},
  \bibinfo {author} {\bibfnamefont {T.}~\bibnamefont {Lutz}},\ and\ \bibinfo
  {author} {\bibfnamefont {T.}~\bibnamefont {Kühn}},\ }\bibfield  {title}
  {\bibinfo {title} {Aerodynamic and acoustic simulations of thick flatback
  airfoils employing high order {{DES}} methods},\ }\href
  {https://doi.org/10.1002/adts.202200129} {\bibfield  {journal} {\bibinfo
  {journal} {Advanced Theory and Simulations}\ }\textbf {\bibinfo {volume}
  {5}},\ \bibinfo {pages} {2200129} (\bibinfo {year} {2022})}\BibitemShut
  {NoStop}%
\bibitem [{\citenamefont {Smith}\ and\ \citenamefont
  {Schaefer}(1950)}]{Smith1950}%
  \BibitemOpen
  \bibfield  {author} {\bibinfo {author} {\bibfnamefont {H.~A.}\ \bibnamefont
  {Smith}}\ and\ \bibinfo {author} {\bibfnamefont {R.~F.}\ \bibnamefont
  {Schaefer}},\ }\href@noop {} {\emph {\bibinfo {title} {Aerodynamic
  Characteristics at Reynolds Numbers of 3.0 x 10 (Exp 6) and 6.0 x 10 (Exp 6)
  of Three Airfoil Sections Formed by Cutting off Various Amounts from the Rear
  Portion of the {{NACA}} 0012 Airfoil Section}}},\ \bibinfo {type} {Tech.
  Rep.}\ (\bibinfo  {institution} {Langley Aeronautical Laboratory},\ \bibinfo
  {year} {1950})\BibitemShut {NoStop}%
\bibitem [{\citenamefont {Post}\ \emph {et~al.}(2008)\citenamefont {Post},
  \citenamefont {Jones}, \citenamefont {Denton},\ and\ \citenamefont
  {Millard}}]{Post2008}%
  \BibitemOpen
  \bibfield  {author} {\bibinfo {author} {\bibfnamefont {M.}~\bibnamefont
  {Post}}, \bibinfo {author} {\bibfnamefont {R.}~\bibnamefont {Jones}},
  \bibinfo {author} {\bibfnamefont {A.}~\bibnamefont {Denton}},\ and\ \bibinfo
  {author} {\bibfnamefont {R.}~\bibnamefont {Millard}},\ }\bibfield  {title}
  {\bibinfo {title} {Characterization of a flatback airfoil for use in wind
  power generation},\ }in\ \href@noop {} {\emph {\bibinfo {booktitle} {46th
  {{AIAA}} Aerospace Sciences Meeting and Exhibit}}}\ (\bibinfo {year} {2008})\
  p.\ \bibinfo {pages} {1330}\BibitemShut {NoStop}%
\bibitem [{\citenamefont {Baker}\ \emph {et~al.}(2008)\citenamefont {Baker},
  \citenamefont {Mayda},\ and\ \citenamefont {Van~Dam}}]{Baker2008}%
  \BibitemOpen
  \bibfield  {author} {\bibinfo {author} {\bibfnamefont {J.~P.}\ \bibnamefont
  {Baker}}, \bibinfo {author} {\bibfnamefont {E.~A.}\ \bibnamefont {Mayda}},\
  and\ \bibinfo {author} {\bibfnamefont {C.~P.}\ \bibnamefont {Van~Dam}},\
  }\bibfield  {title} {\bibinfo {title} {Drag reduction of blunt trailing-edge
  airfoils},\ }in\ \href
  {https://citeseerx.ist.psu.edu/document?repid=rep1&type=pdf&doi=f982cc00175ee0dcdf04a3244a16f59ea3750648}
  {\emph {\bibinfo {booktitle} {{{BBAA VI International Colloquium}} on:
  {{Bluff Bodies Aerodynamics}} \& {{Applications}}}}}\ (\bibinfo {address}
  {Milano},\ \bibinfo {year} {2008})\BibitemShut {NoStop}%
\bibitem [{\citenamefont {Winnemöller}\ and\ \citenamefont
  {Van~Dam}(2007)}]{Winnemoeller2007}%
  \BibitemOpen
  \bibfield  {author} {\bibinfo {author} {\bibfnamefont {T.}~\bibnamefont
  {Winnemöller}}\ and\ \bibinfo {author} {\bibfnamefont {C.~P.}\ \bibnamefont
  {Van~Dam}},\ }\bibfield  {title} {\bibinfo {title} {Design and {{Numerical
  Optimization}} of {{Thick Airfoils Including Blunt Trailing Edges}}},\ }\href
  {https://doi.org/10.2514/1.23057} {\bibfield  {journal} {\bibinfo  {journal}
  {Journal of Aircraft}\ }\textbf {\bibinfo {volume} {44}},\ \bibinfo {pages}
  {232} (\bibinfo {year} {2007})}\BibitemShut {NoStop}%
\bibitem [{\citenamefont {Sørensen}(2011)}]{Soerensen2011}%
  \BibitemOpen
  \bibfield  {author} {\bibinfo {author} {\bibfnamefont {N.~N.}\ \bibnamefont
  {Sørensen}},\ }\bibfield  {title} {\bibinfo {title} {A {{Small Study}} of
  {{Flatback Airfoils}}},\ }in\ \href@noop {} {\emph {\bibinfo {booktitle}
  {Presentations from the {{Aeroelastic Workshop}} – Latest Results from
  {{AeroOpt}}}}},\ \bibinfo {series and number} {Denmark. {{Forskningscenter
  Risoe}}. {{Risoe-R}}},\ \bibinfo {editor} {edited by\ \bibinfo {editor}
  {\bibfnamefont {M.~H.}\ \bibnamefont {Hansen}}}\ (\bibinfo  {publisher}
  {Danmarks Tekniske Universitet, Risø Nationallaboratoriet for Bæredygtig
  Energi},\ \bibinfo {address} {Roskilde},\ \bibinfo {year} {2011})\ pp.\
  \bibinfo {pages} {153--195}\BibitemShut {NoStop}%
\bibitem [{\citenamefont {Xu}\ \emph {et~al.}(2014)\citenamefont {Xu},
  \citenamefont {Shen}, \citenamefont {Zhu}, \citenamefont {Yang},\ and\
  \citenamefont {Liu}}]{Xu2014}%
  \BibitemOpen
  \bibfield  {author} {\bibinfo {author} {\bibfnamefont {H.}~\bibnamefont
  {Xu}}, \bibinfo {author} {\bibfnamefont {W.}~\bibnamefont {Shen}}, \bibinfo
  {author} {\bibfnamefont {W.}~\bibnamefont {Zhu}}, \bibinfo {author}
  {\bibfnamefont {H.}~\bibnamefont {Yang}},\ and\ \bibinfo {author}
  {\bibfnamefont {C.}~\bibnamefont {Liu}},\ }\bibfield  {title} {\bibinfo
  {title} {Aerodynamic {{Analysis}} of {{Trailing Edge Enlarged Wind Turbine
  Airfoils}}},\ }\href {https://doi.org/10.1088/1742-6596/524/1/012010}
  {\bibfield  {journal} {\bibinfo  {journal} {Journal of Physics: Conference
  Series}\ }\textbf {\bibinfo {volume} {524}},\ \bibinfo {pages} {012010}
  (\bibinfo {year} {2014})}\BibitemShut {NoStop}%
\bibitem [{\citenamefont {Williamson}(1996{\natexlab{a}})}]{Williamson1996}%
  \BibitemOpen
  \bibfield  {author} {\bibinfo {author} {\bibfnamefont {C.~H.~K.}\
  \bibnamefont {Williamson}},\ }\bibfield  {title} {\bibinfo {title} {Vortex
  dynamics in the cylinder wake},\ }\href
  {https://doi.org/10.1146/annurev.fl.28.010196.002401} {\bibfield  {journal}
  {\bibinfo  {journal} {Annual Review of Fluid Mechanics}\ }\textbf {\bibinfo
  {volume} {28}},\ \bibinfo {pages} {477} (\bibinfo {year}
  {1996}{\natexlab{a}})}\BibitemShut {NoStop}%
\bibitem [{\citenamefont {Naghib-Lahouti}\ \emph {et~al.}(2012)\citenamefont
  {Naghib-Lahouti}, \citenamefont {Doddipatla},\ and\ \citenamefont
  {Hangan}}]{NaghibLahouti2012}%
  \BibitemOpen
  \bibfield  {author} {\bibinfo {author} {\bibfnamefont {A.}~\bibnamefont
  {Naghib-Lahouti}}, \bibinfo {author} {\bibfnamefont {L.~S.}\ \bibnamefont
  {Doddipatla}},\ and\ \bibinfo {author} {\bibfnamefont {H.}~\bibnamefont
  {Hangan}},\ }\bibfield  {title} {\bibinfo {title} {Secondary wake
  instabilities of a blunt trailing edge profiled body as a basis for flow
  control},\ }\href {https://doi.org/10.1007/s00348-012-1273-9} {\bibfield
  {journal} {\bibinfo  {journal} {Experiments in Fluids}\ }\textbf {\bibinfo
  {volume} {52}},\ \bibinfo {pages} {1547} (\bibinfo {year}
  {2012})}\BibitemShut {NoStop}%
\bibitem [{\citenamefont {Yang}\ and\ \citenamefont {Baeder}(2018)}]{Yang2018}%
  \BibitemOpen
  \bibfield  {author} {\bibinfo {author} {\bibfnamefont {S.~J.}\ \bibnamefont
  {Yang}}\ and\ \bibinfo {author} {\bibfnamefont {J.~D.}\ \bibnamefont
  {Baeder}},\ }\bibfield  {title} {\bibinfo {title} {Blunt-{{Wavy Combined
  Trailing Edge}} for {{Wind Turbine Blade Inboard Performance Improvement}}},\
  }\href {https://doi.org/10.1088/1742-6596/1037/2/022004} {\bibfield
  {journal} {\bibinfo  {journal} {Journal of Physics: Conference Series}\
  }\textbf {\bibinfo {volume} {1037}},\ \bibinfo {pages} {022004} (\bibinfo
  {year} {2018})}\BibitemShut {NoStop}%
\bibitem [{\citenamefont {Ryan}\ \emph {et~al.}(2005)\citenamefont {Ryan},
  \citenamefont {Thompson},\ and\ \citenamefont {Hourigan}}]{Ryan2005}%
  \BibitemOpen
  \bibfield  {author} {\bibinfo {author} {\bibfnamefont {K.}~\bibnamefont
  {Ryan}}, \bibinfo {author} {\bibfnamefont {M.~C.}\ \bibnamefont {Thompson}},\
  and\ \bibinfo {author} {\bibfnamefont {K.}~\bibnamefont {Hourigan}},\
  }\bibfield  {title} {\bibinfo {title} {Three-dimensional transition in the
  wake of bluff elongated cylinders},\ }\href
  {https://doi.org/10.1017/s0022112005005082} {\bibfield  {journal} {\bibinfo
  {journal} {Journal of Fluid Mechanics}\ }\textbf {\bibinfo {volume} {538}},\
  \bibinfo {pages} {1} (\bibinfo {year} {2005})}\BibitemShut {NoStop}%
\bibitem [{\citenamefont {Naghib-Lahouti}\ \emph {et~al.}(2014)\citenamefont
  {Naghib-Lahouti}, \citenamefont {Lavoie},\ and\ \citenamefont
  {Hangan}}]{NaghibLahouti2014}%
  \BibitemOpen
  \bibfield  {author} {\bibinfo {author} {\bibfnamefont {A.}~\bibnamefont
  {Naghib-Lahouti}}, \bibinfo {author} {\bibfnamefont {P.}~\bibnamefont
  {Lavoie}},\ and\ \bibinfo {author} {\bibfnamefont {H.}~\bibnamefont
  {Hangan}},\ }\bibfield  {title} {\bibinfo {title} {Wake instabilities of a
  blunt trailing edge profiled body at intermediate {{Reynolds}} numbers},\
  }\bibfield  {journal} {\bibinfo  {journal} {Experiments in Fluids}\ }\textbf
  {\bibinfo {volume} {55}},\ \href {https://doi.org/10.1007/s00348-014-1779-4}
  {10.1007/s00348-014-1779-4} (\bibinfo {year} {2014})\BibitemShut {NoStop}%
\bibitem [{\citenamefont {Gibeau}\ \emph {et~al.}(2018)\citenamefont {Gibeau},
  \citenamefont {Koch},\ and\ \citenamefont {Ghaemi}}]{Gibeau2018}%
  \BibitemOpen
  \bibfield  {author} {\bibinfo {author} {\bibfnamefont {B.}~\bibnamefont
  {Gibeau}}, \bibinfo {author} {\bibfnamefont {C.~R.}\ \bibnamefont {Koch}},\
  and\ \bibinfo {author} {\bibfnamefont {S.}~\bibnamefont {Ghaemi}},\
  }\bibfield  {title} {\bibinfo {title} {Secondary instabilities in the wake of
  an elongated two-dimensional body with a blunt trailing edge},\ }\href
  {https://doi.org/10.1017/jfm.2018.285} {\bibfield  {journal} {\bibinfo
  {journal} {Journal of Fluid Mechanics}\ }\textbf {\bibinfo {volume} {846}},\
  \bibinfo {pages} {578} (\bibinfo {year} {2018})}\BibitemShut {NoStop}%
\bibitem [{\citenamefont {Gibeau}\ and\ \citenamefont
  {Ghaemi}(2020)}]{Gibeau2020}%
  \BibitemOpen
  \bibfield  {author} {\bibinfo {author} {\bibfnamefont {B.}~\bibnamefont
  {Gibeau}}\ and\ \bibinfo {author} {\bibfnamefont {S.}~\bibnamefont
  {Ghaemi}},\ }\bibfield  {title} {\bibinfo {title} {The mode {{B}} structure
  of streamwise vortices in the wake of a two-dimensional blunt trailing
  edge},\ }\href {https://doi.org/10.1017/jfm.2019.931} {\bibfield  {journal}
  {\bibinfo  {journal} {Journal of Fluid Mechanics}\ }\textbf {\bibinfo
  {volume} {884}},\ \bibinfo {pages} {A12} (\bibinfo {year}
  {2020})}\BibitemShut {NoStop}%
\bibitem [{\citenamefont {Forouzi~Feshalami}\ \emph {et~al.}(2022)\citenamefont
  {Forouzi~Feshalami}, \citenamefont {He}, \citenamefont {Scarano},
  \citenamefont {Gan},\ and\ \citenamefont {Morton}}]{ForouziFeshalami2022}%
  \BibitemOpen
  \bibfield  {author} {\bibinfo {author} {\bibfnamefont {B.}~\bibnamefont
  {Forouzi~Feshalami}}, \bibinfo {author} {\bibfnamefont {S.}~\bibnamefont
  {He}}, \bibinfo {author} {\bibfnamefont {F.}~\bibnamefont {Scarano}},
  \bibinfo {author} {\bibfnamefont {L.}~\bibnamefont {Gan}},\ and\ \bibinfo
  {author} {\bibfnamefont {C.}~\bibnamefont {Morton}},\ }\bibfield  {title}
  {\bibinfo {title} {A review of experiments on stationary bluff body wakes},\
  }\href {https://doi.org/10.1063/5.0077323} {\bibfield  {journal} {\bibinfo
  {journal} {Physics of Fluids}\ }\textbf {\bibinfo {volume} {34}},\ \bibinfo
  {pages} {011301} (\bibinfo {year} {2022})}\BibitemShut {NoStop}%
\bibitem [{\citenamefont {Hoerner}(1950)}]{Hoerner1950}%
  \BibitemOpen
  \bibfield  {author} {\bibinfo {author} {\bibfnamefont {S.~F.}\ \bibnamefont
  {Hoerner}},\ }\bibfield  {title} {\bibinfo {title} {Base {{Drag}} and {{Thick
  Trailing Edges}}},\ }\href {https://doi.org/10.2514/8.1750} {\bibfield
  {journal} {\bibinfo  {journal} {Journal of the Aeronautical Sciences}\
  }\textbf {\bibinfo {volume} {17}},\ \bibinfo {pages} {622} (\bibinfo {year}
  {1950})}\BibitemShut {NoStop}%
\bibitem [{\citenamefont {Bearman}(1965)}]{Bearman1965}%
  \BibitemOpen
  \bibfield  {author} {\bibinfo {author} {\bibfnamefont {P.~W.}\ \bibnamefont
  {Bearman}},\ }\bibfield  {title} {\bibinfo {title} {Investigation of the flow
  behind a two-dimensional model with a blunt trailing edge and fitted with
  splitter plates},\ }\href {https://doi.org/10.1017/S0022112065000162}
  {\bibfield  {journal} {\bibinfo  {journal} {Journal of Fluid Mechanics}\
  }\textbf {\bibinfo {volume} {21}},\ \bibinfo {pages} {241} (\bibinfo {year}
  {1965})}\BibitemShut {NoStop}%
\bibitem [{\citenamefont {Bearman}(1967)}]{Bearman1967}%
  \BibitemOpen
  \bibfield  {author} {\bibinfo {author} {\bibfnamefont {P.~W.}\ \bibnamefont
  {Bearman}},\ }\bibfield  {title} {\bibinfo {title} {The {{Effect}} of {{Base
  Bleed}} on the {{Flow}} behind a {{Two-Dimensional Model}} with a {{Blunt
  Trailing Edge}}},\ }\href {https://doi.org/10.1017/S0001925900004212}
  {\bibfield  {journal} {\bibinfo  {journal} {Aeronautical Quarterly}\ }\textbf
  {\bibinfo {volume} {18}},\ \bibinfo {pages} {207} (\bibinfo {year}
  {1967})}\BibitemShut {NoStop}%
\bibitem [{\citenamefont {Petrusma}\ and\ \citenamefont
  {Gai}(1994)}]{Petrusma1994}%
  \BibitemOpen
  \bibfield  {author} {\bibinfo {author} {\bibfnamefont {M.}~\bibnamefont
  {Petrusma}}\ and\ \bibinfo {author} {\bibfnamefont {S.}~\bibnamefont {Gai}},\
  }\bibfield  {title} {\bibinfo {title} {The effect of geometry on the base
  pressure recovery of segmented blunt trailing edges},\ }\href
  {https://www.semanticscholar.org/paper/The-effect-of-geometry-on-the-base-pressure-of-Petrusma-Gai/92beec7f588c75a698cfbd2649b371343c6f6510}
  {\bibfield  {journal} {\bibinfo  {journal} {Aeronautical Journal}\ }
  (\bibinfo {year} {1994})}\BibitemShut {NoStop}%
\bibitem [{\citenamefont {Rowe}\ \emph {et~al.}(2001)\citenamefont {Rowe},
  \citenamefont {Fry},\ and\ \citenamefont {Motallebi}}]{Rowe2001}%
  \BibitemOpen
  \bibfield  {author} {\bibinfo {author} {\bibfnamefont {A.}~\bibnamefont
  {Rowe}}, \bibinfo {author} {\bibfnamefont {A.~L.~A.}\ \bibnamefont {Fry}},\
  and\ \bibinfo {author} {\bibfnamefont {F.}~\bibnamefont {Motallebi}},\
  }\bibfield  {title} {\bibinfo {title} {Influence of {{Boundary-Layer
  Thickness}} on {{Base Pressure}} and {{Vortex Shedding Frequency}}},\ }\href
  {https://doi.org/10.2514/2.1377} {\bibfield  {journal} {\bibinfo  {journal}
  {AIAA Journal}\ }\textbf {\bibinfo {volume} {39}},\ \bibinfo {pages} {754}
  (\bibinfo {year} {2001})}\BibitemShut {NoStop}%
\bibitem [{\citenamefont {Mariotti}\ and\ \citenamefont
  {Buresti}(2013)}]{Mariotti2013}%
  \BibitemOpen
  \bibfield  {author} {\bibinfo {author} {\bibfnamefont {A.}~\bibnamefont
  {Mariotti}}\ and\ \bibinfo {author} {\bibfnamefont {G.}~\bibnamefont
  {Buresti}},\ }\bibfield  {title} {\bibinfo {title} {Experimental
  investigation on the influence of boundary layer thickness on the base
  pressure and near-wake flow features of an axisymmetric blunt-based body},\
  }\href {https://doi.org/10.1007/s00348-013-1612-5} {\bibfield  {journal}
  {\bibinfo  {journal} {Experiments in Fluids}\ }\textbf {\bibinfo {volume}
  {54}},\ \bibinfo {pages} {1612} (\bibinfo {year} {2013})}\BibitemShut
  {NoStop}%
\bibitem [{\citenamefont {Durgesh}\ \emph {et~al.}(2013)\citenamefont
  {Durgesh}, \citenamefont {Naughton},\ and\ \citenamefont
  {Whitmore}}]{Durgesh2013}%
  \BibitemOpen
  \bibfield  {author} {\bibinfo {author} {\bibfnamefont {V.}~\bibnamefont
  {Durgesh}}, \bibinfo {author} {\bibfnamefont {J.~W.}\ \bibnamefont
  {Naughton}},\ and\ \bibinfo {author} {\bibfnamefont {S.~A.}\ \bibnamefont
  {Whitmore}},\ }\bibfield  {title} {\bibinfo {title} {Experimental
  {{Investigation}} of {{Base-Drag Reduction}} via {{Boundary-Layer
  Modification}}},\ }\href {https://doi.org/10.2514/1.J051825} {\bibfield
  {journal} {\bibinfo  {journal} {AIAA Journal}\ }\textbf {\bibinfo {volume}
  {51}},\ \bibinfo {pages} {416} (\bibinfo {year} {2013})}\BibitemShut
  {NoStop}%
\bibitem [{\citenamefont {Park}\ \emph {et~al.}(2006)\citenamefont {Park},
  \citenamefont {Lee}, \citenamefont {Jeon}, \citenamefont {Hahn},
  \citenamefont {Kim}, \citenamefont {Kim}, \citenamefont {Choi},\ and\
  \citenamefont {Choi}}]{Park2006}%
  \BibitemOpen
  \bibfield  {author} {\bibinfo {author} {\bibfnamefont {H.}~\bibnamefont
  {Park}}, \bibinfo {author} {\bibfnamefont {D.}~\bibnamefont {Lee}}, \bibinfo
  {author} {\bibfnamefont {W.-P.}\ \bibnamefont {Jeon}}, \bibinfo {author}
  {\bibfnamefont {S.}~\bibnamefont {Hahn}}, \bibinfo {author} {\bibfnamefont
  {J.}~\bibnamefont {Kim}}, \bibinfo {author} {\bibfnamefont {J.}~\bibnamefont
  {Kim}}, \bibinfo {author} {\bibfnamefont {J.}~\bibnamefont {Choi}},\ and\
  \bibinfo {author} {\bibfnamefont {H.}~\bibnamefont {Choi}},\ }\bibfield
  {title} {\bibinfo {title} {Drag reduction in flow over a two-dimensional
  bluff body with a blunt trailing edge using a new passive device},\ }\href
  {https://doi.org/10.1017/S0022112006001364} {\bibfield  {journal} {\bibinfo
  {journal} {Journal of Fluid Mechanics}\ }\textbf {\bibinfo {volume} {563}},\
  \bibinfo {pages} {389} (\bibinfo {year} {2006})}\BibitemShut {NoStop}%
\bibitem [{\citenamefont {Trip}\ and\ \citenamefont
  {Fransson}(2014)}]{Trip2014}%
  \BibitemOpen
  \bibfield  {author} {\bibinfo {author} {\bibfnamefont {R.}~\bibnamefont
  {Trip}}\ and\ \bibinfo {author} {\bibfnamefont {J.~H.~M.}\ \bibnamefont
  {Fransson}},\ }\bibfield  {title} {\bibinfo {title} {Boundary layer
  modification by means of wall suction and the effect on the wake behind a
  rectangular forebody},\ }\href {https://doi.org/10.1063/1.4904376} {\bibfield
   {journal} {\bibinfo  {journal} {Physics of Fluids}\ }\textbf {\bibinfo
  {volume} {26}},\ \bibinfo {pages} {125105} (\bibinfo {year}
  {2014})}\BibitemShut {NoStop}%
\bibitem [{\citenamefont {Trip}\ and\ \citenamefont
  {Fransson}(2017)}]{Trip2017}%
  \BibitemOpen
  \bibfield  {author} {\bibinfo {author} {\bibfnamefont {R.}~\bibnamefont
  {Trip}}\ and\ \bibinfo {author} {\bibfnamefont {J.~H.~M.}\ \bibnamefont
  {Fransson}},\ }\bibfield  {title} {\bibinfo {title} {Bluff body
  boundary-layer modification and its effect on the near-wake topology},\
  }\href {https://doi.org/10.1063/1.5003383} {\bibfield  {journal} {\bibinfo
  {journal} {Physics of Fluids}\ }\textbf {\bibinfo {volume} {29}},\ \bibinfo
  {pages} {095105} (\bibinfo {year} {2017})}\BibitemShut {NoStop}%
\bibitem [{\citenamefont {El-Gammal}\ and\ \citenamefont
  {Hangan}(2007)}]{ElGammal2007}%
  \BibitemOpen
  \bibfield  {author} {\bibinfo {author} {\bibfnamefont {M.}~\bibnamefont
  {El-Gammal}}\ and\ \bibinfo {author} {\bibfnamefont {H.}~\bibnamefont
  {Hangan}},\ }\bibfield  {title} {\bibinfo {title} {Three-dimensional wake
  dynamics of a blunt and divergent trailing edge airfoil},\ }\href
  {https://doi.org/10.1007/s00348-007-0428-6} {\bibfield  {journal} {\bibinfo
  {journal} {Experiments in Fluids}\ }\textbf {\bibinfo {volume} {44}},\
  \bibinfo {pages} {705} (\bibinfo {year} {2007})}\BibitemShut {NoStop}%
\bibitem [{\citenamefont {Mendez}\ \emph {et~al.}(2019)\citenamefont {Mendez},
  \citenamefont {Balabane},\ and\ \citenamefont {Buchlin}}]{Mendez2019}%
  \BibitemOpen
  \bibfield  {author} {\bibinfo {author} {\bibfnamefont {M.~A.}\ \bibnamefont
  {Mendez}}, \bibinfo {author} {\bibfnamefont {M.}~\bibnamefont {Balabane}},\
  and\ \bibinfo {author} {\bibfnamefont {J.-M.}\ \bibnamefont {Buchlin}},\
  }\bibfield  {title} {\bibinfo {title} {Multi-scale proper orthogonal
  decomposition of complex fluid flows},\ }\href
  {https://doi.org/10.1017/jfm.2019.212} {\bibfield  {journal} {\bibinfo
  {journal} {Journal of Fluid Mechanics}\ }\textbf {\bibinfo {volume} {870}},\
  \bibinfo {pages} {988} (\bibinfo {year} {2019})}\BibitemShut {NoStop}%
\bibitem [{\citenamefont {Mendez}(2023)}]{Mendez2023}%
  \BibitemOpen
  \bibfield  {author} {\bibinfo {author} {\bibfnamefont {M.}~\bibnamefont
  {Mendez}},\ }\bibinfo {title} {Generalized and multiscale modal analysis},\
  in\ \href {https://doi.org/10.1017/9781108896214.013} {\emph {\bibinfo
  {booktitle} {Data-Driven Fluid Mechanics}}}\ (\bibinfo  {publisher}
  {Cambridge University Press},\ \bibinfo {year} {2023})\ pp.\ \bibinfo {pages}
  {153--181}\BibitemShut {NoStop}%
\bibitem [{\citenamefont {Lumley}\ and\ \citenamefont
  {Panofsky}(1964)}]{Lumley1964}%
  \BibitemOpen
  \bibfield  {author} {\bibinfo {author} {\bibfnamefont {J.~L.}\ \bibnamefont
  {Lumley}}\ and\ \bibinfo {author} {\bibfnamefont {H.~A.}\ \bibnamefont
  {Panofsky}},\ }\href@noop {} {\emph {\bibinfo {title} {The Structure of
  Atmospheric Turbulence, Monographs and Texts in Physics and Astronomy Vol.
  {{XII}}}}}\ (\bibinfo  {publisher} {John Wiley \& Sons Inc.},\ \bibinfo
  {year} {1964})\ p.\ \bibinfo {pages} {244}\BibitemShut {NoStop}%
\bibitem [{\citenamefont {Schmid}(2010)}]{Schmid2010}%
  \BibitemOpen
  \bibfield  {author} {\bibinfo {author} {\bibfnamefont {P.~J.}\ \bibnamefont
  {Schmid}},\ }\bibfield  {title} {\bibinfo {title} {Dynamic mode decomposition
  of numerical and experimental data},\ }\href
  {https://doi.org/10.1017/s0022112010001217} {\bibfield  {journal} {\bibinfo
  {journal} {Journal of Fluid Mechanics}\ }\textbf {\bibinfo {volume} {656}},\
  \bibinfo {pages} {5} (\bibinfo {year} {2010})}\BibitemShut {NoStop}%
\bibitem [{\citenamefont {Papadakis}(2015)}]{Papadakis2015}%
  \BibitemOpen
  \bibfield  {author} {\bibinfo {author} {\bibfnamefont {G.}~\bibnamefont
  {Papadakis}},\ }\emph {\bibinfo {title} {Development of a Hybrid Compressible
  Vortex Particle Method and Application to External Problems Including
  Helicopter Flows}},\ \href {https://doi.org/10.26240/HEAL.NTUA.1582} {Ph.D.
  thesis},\ \bibinfo  {school} {National Technical University of Athens}
  (\bibinfo {year} {2015})\BibitemShut {NoStop}%
\bibitem [{\citenamefont {Ntouras}\ and\ \citenamefont
  {Papadakis}(2020)}]{Ntouras2020}%
  \BibitemOpen
  \bibfield  {author} {\bibinfo {author} {\bibfnamefont {D.}~\bibnamefont
  {Ntouras}}\ and\ \bibinfo {author} {\bibfnamefont {G.}~\bibnamefont
  {Papadakis}},\ }\bibfield  {title} {\bibinfo {title} {A coupled artificial
  compressibility method for free surface flows},\ }\href
  {https://doi.org/10.3390/jmse8080590} {\bibfield  {journal} {\bibinfo
  {journal} {Journal of Marine Science and Engineering}\ }\textbf {\bibinfo
  {volume} {8}},\ \bibinfo {pages} {590} (\bibinfo {year} {2020})}\BibitemShut
  {NoStop}%
\bibitem [{\citenamefont {Roe}(1981)}]{Roe1981}%
  \BibitemOpen
  \bibfield  {author} {\bibinfo {author} {\bibfnamefont {P.~L.}\ \bibnamefont
  {Roe}},\ }\bibfield  {title} {\bibinfo {title} {Approximate {{Riemann}}
  solvers, parameter vectors, and difference schemes},\ }\href
  {https://doi.org/10.1016/0021-9991(81)90128-5} {\bibfield  {journal}
  {\bibinfo  {journal} {Journal of Computational Physics}\ }\textbf {\bibinfo
  {volume} {43}},\ \bibinfo {pages} {357} (\bibinfo {year} {1981})}\BibitemShut
  {NoStop}%
\bibitem [{\citenamefont {Papadakis}\ and\ \citenamefont
  {Voutsinas}(2019)}]{Papadakis2019}%
  \BibitemOpen
  \bibfield  {author} {\bibinfo {author} {\bibfnamefont {G.}~\bibnamefont
  {Papadakis}}\ and\ \bibinfo {author} {\bibfnamefont {S.~G.}\ \bibnamefont
  {Voutsinas}},\ }\bibfield  {title} {\bibinfo {title} {A strongly coupled
  {{Eulerian Lagrangian}} method verified in {{2D}} external compressible
  flows},\ }\href {https://doi.org/10.1016/j.compfluid.2019.104325} {\bibfield
  {journal} {\bibinfo  {journal} {Computers \& Fluids}\ }\textbf {\bibinfo
  {volume} {195}},\ \bibinfo {pages} {104325} (\bibinfo {year}
  {2019})}\BibitemShut {NoStop}%
\bibitem [{\citenamefont {Biedron}\ \emph {et~al.}(2005)\citenamefont
  {Biedron}, \citenamefont {Vatsa},\ and\ \citenamefont
  {Atkins}}]{Biedron2005}%
  \BibitemOpen
  \bibfield  {author} {\bibinfo {author} {\bibfnamefont {R.}~\bibnamefont
  {Biedron}}, \bibinfo {author} {\bibfnamefont {V.}~\bibnamefont {Vatsa}},\
  and\ \bibinfo {author} {\bibfnamefont {H.}~\bibnamefont {Atkins}},\
  }\bibfield  {title} {\bibinfo {title} {Simulation of {{Unsteady Flows Using}}
  an {{Unstructured Navier-Stokes Solver}} on {{Moving}} and {{Stationary
  Grids}}},\ }in\ \href {https://doi.org/10.2514/6.2005-5093} {\emph {\bibinfo
  {booktitle} {23rd {{AIAA Applied Aerodynamics Conference}}}}}\ (\bibinfo
  {publisher} {{American Institute of Aeronautics and Astronautics}},\ \bibinfo
  {address} {Toronto, Ontario, Canada},\ \bibinfo {year} {2005})\BibitemShut
  {NoStop}%
\bibitem [{\citenamefont {Diakakis}(2019)}]{Diakakis2019}%
  \BibitemOpen
  \bibfield  {author} {\bibinfo {author} {\bibfnamefont {K.}~\bibnamefont
  {Diakakis}},\ }\emph {\bibinfo {title} {Computational Analysis of
  Transitional and Massively Separated Flows with Application to Wind
  Turbines}},\ \href {https://doi.org/10.12681/eadd/46180} {Ph.D. thesis},\
  \bibinfo  {school} {National Technical University of Athens} (\bibinfo {year}
  {2019})\BibitemShut {NoStop}%
\bibitem [{\citenamefont {Shur}\ \emph {et~al.}(2008)\citenamefont {Shur},
  \citenamefont {Spalart}, \citenamefont {Strelets},\ and\ \citenamefont
  {Travin}}]{Shur2008}%
  \BibitemOpen
  \bibfield  {author} {\bibinfo {author} {\bibfnamefont {M.~L.}\ \bibnamefont
  {Shur}}, \bibinfo {author} {\bibfnamefont {P.~R.}\ \bibnamefont {Spalart}},
  \bibinfo {author} {\bibfnamefont {M.~K.}\ \bibnamefont {Strelets}},\ and\
  \bibinfo {author} {\bibfnamefont {A.~K.}\ \bibnamefont {Travin}},\ }\bibfield
   {title} {\bibinfo {title} {A hybrid {{RANS-LES}} approach with
  delayed-{{DES}} and wall-modelled {{LES}} capabilities},\ }\href
  {https://doi.org/10.1016/j.ijheatfluidflow.2008.07.001} {\bibfield  {journal}
  {\bibinfo  {journal} {International Journal of Heat and Fluid Flow}\ }\textbf
  {\bibinfo {volume} {29}},\ \bibinfo {pages} {1638} (\bibinfo {year}
  {2008})}\BibitemShut {NoStop}%
\bibitem [{\citenamefont {Spalart}\ and\ \citenamefont
  {Allmaras}(1992)}]{Spalart1992}%
  \BibitemOpen
  \bibfield  {author} {\bibinfo {author} {\bibfnamefont {P.}~\bibnamefont
  {Spalart}}\ and\ \bibinfo {author} {\bibfnamefont {S.}~\bibnamefont
  {Allmaras}},\ }\bibfield  {title} {\bibinfo {title} {A one-equation
  turbulence model for aerodynamic flows},\ }in\ \href
  {https://doi.org/10.2514/6.1992-439} {\emph {\bibinfo {booktitle} {30th
  Aerospace Sciences Meeting and Exhibit}}}\ (\bibinfo  {publisher} {{American
  Institute of Aeronautics and Astronautics}},\ \bibinfo {year}
  {1992})\BibitemShut {NoStop}%
\bibitem [{\citenamefont {Poletti}\ \emph {et~al.}(2023)\citenamefont
  {Poletti}, \citenamefont {Schena}, \citenamefont {Ninni},\ and\ \citenamefont
  {Mendez}}]{Poletti_MODULO_2023}%
  \BibitemOpen
  \bibfield  {author} {\bibinfo {author} {\bibfnamefont {R.}~\bibnamefont
  {Poletti}}, \bibinfo {author} {\bibfnamefont {L.}~\bibnamefont {Schena}},
  \bibinfo {author} {\bibfnamefont {D.}~\bibnamefont {Ninni}},\ and\ \bibinfo
  {author} {\bibfnamefont {M.~A.}\ \bibnamefont {Mendez}},\ }\bibfield  {title}
  {\bibinfo {title} {{MODULO: a python toolbox for data-driven modal
  decomposition}},\ }\href@noop {} {\bibfield  {journal} {\bibinfo  {journal}
  {submitted to Journal of Open Software (JOSS)}\ } (\bibinfo {year}
  {2023})}\BibitemShut {NoStop}%
\bibitem [{\citenamefont {Ninni}\ and\ \citenamefont
  {Mendez}(2020)}]{Ninni2020}%
  \BibitemOpen
  \bibfield  {author} {\bibinfo {author} {\bibfnamefont {D.}~\bibnamefont
  {Ninni}}\ and\ \bibinfo {author} {\bibfnamefont {M.~A.}\ \bibnamefont
  {Mendez}},\ }\bibfield  {title} {\bibinfo {title} {Modulo: A software for
  multiscale proper orthogonal decomposition of data},\ }\href
  {https://doi.org/10.1016/j.softx.2020.100622} {\bibfield  {journal} {\bibinfo
   {journal} {SoftwareX}\ }\textbf {\bibinfo {volume} {12}},\ \bibinfo {pages}
  {100622} (\bibinfo {year} {2020})}\BibitemShut {NoStop}%
\bibitem [{\citenamefont {Sirovich}(1987)}]{Sirovich1987}%
  \BibitemOpen
  \bibfield  {author} {\bibinfo {author} {\bibfnamefont {L.}~\bibnamefont
  {Sirovich}},\ }\bibfield  {title} {\bibinfo {title} {Turbulence and the
  dynamics of coherent structures. {{I}}. {{Coherent}} structures},\ }\href
  {https://doi.org/10.1090/qam/910462} {\bibfield  {journal} {\bibinfo
  {journal} {Quarterly of Applied Mathematics}\ }\textbf {\bibinfo {volume}
  {45}},\ \bibinfo {pages} {561} (\bibinfo {year} {1987})}\BibitemShut
  {NoStop}%
\bibitem [{\citenamefont {Kennel}(2004)}]{Kennel2004}%
  \BibitemOpen
  \bibfield  {author} {\bibinfo {author} {\bibfnamefont {M.~B.}\ \bibnamefont
  {Kennel}},\ }\bibfield  {title} {\bibinfo {title} {{{KDTREE}} 2: {{Fortran}}
  95 and {{C}}++ software to efficiently search for near neighbors in a
  multi-dimensional {{Euclidean}} space},\ }\Eprint
  {https://arxiv.org/abs/physics/0408067} {arXiv:physics/0408067}  (\bibinfo
  {year} {2004})\BibitemShut {NoStop}%
\bibitem [{\citenamefont {Spalart}(1988)}]{Spalart1988}%
  \BibitemOpen
  \bibfield  {author} {\bibinfo {author} {\bibfnamefont {P.~R.}\ \bibnamefont
  {Spalart}},\ }\bibfield  {title} {\bibinfo {title} {Direct simulation of a
  turbulent boundary layer up to {{Re}}{\textsubscript{θ}} = 1410},\ }\href
  {https://doi.org/10.1017/s0022112088000345} {\bibfield  {journal} {\bibinfo
  {journal} {Journal of Fluid Mechanics}\ }\textbf {\bibinfo {volume} {187}},\
  \bibinfo {pages} {61} (\bibinfo {year} {1988})}\BibitemShut {NoStop}%
\bibitem [{\citenamefont {Hunt}\ \emph {et~al.}(1988)\citenamefont {Hunt},
  \citenamefont {Wray},\ and\ \citenamefont {Moin}}]{Hunt1988}%
  \BibitemOpen
  \bibfield  {author} {\bibinfo {author} {\bibfnamefont {J.~C.~R.}\
  \bibnamefont {Hunt}}, \bibinfo {author} {\bibfnamefont {A.~A.}\ \bibnamefont
  {Wray}},\ and\ \bibinfo {author} {\bibfnamefont {P.}~\bibnamefont {Moin}},\
  }\bibfield  {title} {\bibinfo {title} {Eddies, streams, and convergence zones
  in turbulent flows},\ }in\ \href
  {https://ntrs.nasa.gov/citations/19890015184} {\emph {\bibinfo {booktitle}
  {Center for Turbulence Research. Proceedings of the 1988 Summer Program}}}\
  (\bibinfo {year} {1988})\BibitemShut {NoStop}%
\bibitem [{\citenamefont {Raffel}\ \emph {et~al.}(2018)\citenamefont {Raffel},
  \citenamefont {Willert}, \citenamefont {Scarano}, \citenamefont {Kähler},
  \citenamefont {Wereley},\ and\ \citenamefont {Kompenhans}}]{Raffel2018}%
  \BibitemOpen
  \bibfield  {author} {\bibinfo {author} {\bibfnamefont {M.}~\bibnamefont
  {Raffel}}, \bibinfo {author} {\bibfnamefont {C.~E.}\ \bibnamefont {Willert}},
  \bibinfo {author} {\bibfnamefont {F.}~\bibnamefont {Scarano}}, \bibinfo
  {author} {\bibfnamefont {C.~J.}\ \bibnamefont {Kähler}}, \bibinfo {author}
  {\bibfnamefont {S.~T.}\ \bibnamefont {Wereley}},\ and\ \bibinfo {author}
  {\bibfnamefont {J.}~\bibnamefont {Kompenhans}},\ }\href
  {https://doi.org/10.1007/978-3-319-68852-7} {\emph {\bibinfo {title}
  {Particle Image Velocimetry}}}\ (\bibinfo  {publisher} {Springer
  International Publishing},\ \bibinfo {year} {2018})\BibitemShut {NoStop}%
\bibitem [{\citenamefont {Graftieaux}\ \emph {et~al.}(2001)\citenamefont
  {Graftieaux}, \citenamefont {Michard},\ and\ \citenamefont
  {Grosjean}}]{Graftieaux2001}%
  \BibitemOpen
  \bibfield  {author} {\bibinfo {author} {\bibfnamefont {L.}~\bibnamefont
  {Graftieaux}}, \bibinfo {author} {\bibfnamefont {M.}~\bibnamefont
  {Michard}},\ and\ \bibinfo {author} {\bibfnamefont {N.}~\bibnamefont
  {Grosjean}},\ }\bibfield  {title} {\bibinfo {title} {Combining {{PIV}},
  {{POD}} and vortex identification algorithms for the study of unsteady
  turbulent swirling flows},\ }\href
  {https://doi.org/10.1088/0957-0233/12/9/307} {\bibfield  {journal} {\bibinfo
  {journal} {Measurement Science and Technology}\ }\textbf {\bibinfo {volume}
  {12}},\ \bibinfo {pages} {1422} (\bibinfo {year} {2001})}\BibitemShut
  {NoStop}%
\bibitem [{\citenamefont {Soto-Valle}\ \emph {et~al.}(2022)\citenamefont
  {Soto-Valle}, \citenamefont {Cioni}, \citenamefont {Bartholomay},
  \citenamefont {Manolesos}, \citenamefont {Nayeri}, \citenamefont
  {Bianchini},\ and\ \citenamefont {Paschereit}}]{SotoValle2022}%
  \BibitemOpen
  \bibfield  {author} {\bibinfo {author} {\bibfnamefont {R.}~\bibnamefont
  {Soto-Valle}}, \bibinfo {author} {\bibfnamefont {S.}~\bibnamefont {Cioni}},
  \bibinfo {author} {\bibfnamefont {S.}~\bibnamefont {Bartholomay}}, \bibinfo
  {author} {\bibfnamefont {M.}~\bibnamefont {Manolesos}}, \bibinfo {author}
  {\bibfnamefont {C.~N.}\ \bibnamefont {Nayeri}}, \bibinfo {author}
  {\bibfnamefont {A.}~\bibnamefont {Bianchini}},\ and\ \bibinfo {author}
  {\bibfnamefont {C.~O.}\ \bibnamefont {Paschereit}},\ }\bibfield  {title}
  {\bibinfo {title} {Vortex identification methods applied to wind turbine tip
  vortices},\ }\href {https://doi.org/10.5194/wes-7-585-2022} {\bibfield
  {journal} {\bibinfo  {journal} {Wind Energy Science}\ }\textbf {\bibinfo
  {volume} {7}},\ \bibinfo {pages} {585} (\bibinfo {year} {2022})}\BibitemShut
  {NoStop}%
\bibitem [{\citenamefont {Williamson}(1989)}]{Williamson1989}%
  \BibitemOpen
  \bibfield  {author} {\bibinfo {author} {\bibfnamefont {C.~H.~K.}\
  \bibnamefont {Williamson}},\ }\bibfield  {title} {\bibinfo {title} {Oblique
  and parallel modes of vortex shedding in the wake of a circular cylinder at
  low {{Reynolds}} numbers},\ }\href
  {https://doi.org/10.1017/S0022112089002429} {\bibfield  {journal} {\bibinfo
  {journal} {Journal of Fluid Mechanics}\ }\textbf {\bibinfo {volume} {206}},\
  \bibinfo {pages} {579} (\bibinfo {year} {1989})}\BibitemShut {NoStop}%
\bibitem [{\citenamefont {Williamson}(1996{\natexlab{b}})}]{Williamson1996b}%
  \BibitemOpen
  \bibfield  {author} {\bibinfo {author} {\bibfnamefont {C.~H.~K.}\
  \bibnamefont {Williamson}},\ }\bibfield  {title} {\bibinfo {title}
  {Three-dimensional vortex dynamics in bluff body wakes},\ }\href
  {https://doi.org/10.1016/0894-1777(95)00085-2} {\bibfield  {journal}
  {\bibinfo  {journal} {Experimental Thermal and Fluid Science}\ }\textbf
  {\bibinfo {volume} {12}},\ \bibinfo {pages} {150} (\bibinfo {year}
  {1996}{\natexlab{b}})}\BibitemShut {NoStop}%
\bibitem [{\citenamefont {Luo}\ \emph {et~al.}(2007)\citenamefont {Luo},
  \citenamefont {Tong},\ and\ \citenamefont {Khoo}}]{Luo2007}%
  \BibitemOpen
  \bibfield  {author} {\bibinfo {author} {\bibfnamefont {S.~C.}\ \bibnamefont
  {Luo}}, \bibinfo {author} {\bibfnamefont {X.~H.}\ \bibnamefont {Tong}},\ and\
  \bibinfo {author} {\bibfnamefont {B.~C.}\ \bibnamefont {Khoo}},\ }\bibfield
  {title} {\bibinfo {title} {Transition phenomena in the wake of a square
  cylinder},\ }\href {https://doi.org/10.1016/j.jfluidstructs.2006.08.012}
  {\bibfield  {journal} {\bibinfo  {journal} {Journal of Fluids and
  Structures}\ }\textbf {\bibinfo {volume} {23}},\ \bibinfo {pages} {227}
  (\bibinfo {year} {2007})}\BibitemShut {NoStop}%
\bibitem [{\citenamefont {Mendez}\ \emph {et~al.}(2020)\citenamefont {Mendez},
  \citenamefont {Hess}, \citenamefont {Watz},\ and\ \citenamefont
  {Buchlin}}]{Mendez2020}%
  \BibitemOpen
  \bibfield  {author} {\bibinfo {author} {\bibfnamefont {M.~A.}\ \bibnamefont
  {Mendez}}, \bibinfo {author} {\bibfnamefont {D.}~\bibnamefont {Hess}},
  \bibinfo {author} {\bibfnamefont {B.~B.}\ \bibnamefont {Watz}},\ and\
  \bibinfo {author} {\bibfnamefont {J.-M.}\ \bibnamefont {Buchlin}},\
  }\bibfield  {title} {\bibinfo {title} {Multiscale proper orthogonal
  decomposition ({{mPOD}}) of {{TR-PIV}} data—a case study on stationary and
  transient cylinder wake flows},\ }\href
  {https://doi.org/10.1088/1361-6501/ab82be} {\bibfield  {journal} {\bibinfo
  {journal} {Measurement Science and Technology}\ }\textbf {\bibinfo {volume}
  {31}},\ \bibinfo {pages} {094014} (\bibinfo {year} {2020})}\BibitemShut
  {NoStop}%
\bibitem [{\citenamefont {Towne}\ \emph {et~al.}(2018)\citenamefont {Towne},
  \citenamefont {Schmidt},\ and\ \citenamefont {Colonius}}]{Towne2018}%
  \BibitemOpen
  \bibfield  {author} {\bibinfo {author} {\bibfnamefont {A.}~\bibnamefont
  {Towne}}, \bibinfo {author} {\bibfnamefont {O.~T.}\ \bibnamefont {Schmidt}},\
  and\ \bibinfo {author} {\bibfnamefont {T.}~\bibnamefont {Colonius}},\
  }\bibfield  {title} {\bibinfo {title} {Spectral proper orthogonal
  decomposition and its relationship to dynamic mode decomposition and
  resolvent analysis},\ }\href {https://doi.org/10.1017/jfm.2018.283}
  {\bibfield  {journal} {\bibinfo  {journal} {Journal of Fluid Mechanics}\
  }\textbf {\bibinfo {volume} {847}},\ \bibinfo {pages} {821} (\bibinfo {year}
  {2018})}\BibitemShut {NoStop}%
\bibitem [{\citenamefont {Williamson}(1992)}]{Williamson1992}%
  \BibitemOpen
  \bibfield  {author} {\bibinfo {author} {\bibfnamefont {C.~H.~K.}\
  \bibnamefont {Williamson}},\ }\bibfield  {title} {\bibinfo {title} {The
  natural and forced formation of spot-like ‘vortex dislocations’ in the
  transition of a wake},\ }\href {https://doi.org/10.1017/S0022112092002763}
  {\bibfield  {journal} {\bibinfo  {journal} {Journal of Fluid Mechanics}\
  }\textbf {\bibinfo {volume} {243}},\ \bibinfo {pages} {393} (\bibinfo {year}
  {1992})}\BibitemShut {NoStop}%
\end{thebibliography}%
\begin{acknowledgments}
The research work was supported by the Hellenic Foundation for Research and Innovation (HFRI) under the 5th Call for HFRI PhD Fellowships (Fellowship Number: 20455).
Additionally, K. Kellaris and M. Manolesos kindly acknowledge the financial support of the European project
‘TWEET-IE’ funded by the European Union’s Horizon 2020 Research and Innovation Programme (Grant Agreement
101079125).
Furthermore, this work was supported by computational time granted from the Greek Research \& Technology Network (GRNET) in the National HPC facility - ARIS - under project “HPC4FAIR” with ID pr014018\textunderscore thin. The computational grids were generated using the ANSA pre-processor of BETA-CAE Systems.
\end{acknowledgments}

\end{document}